\documentclass[seceqn]{elsart}
\usepackage[pagebackref]{hyperref}
\usepackage{amssymb,amsmath,graphicx,mathptm,epsfig,natbib,bm,lscape}

\journal{Physics Reports}

\newcommand{\eprint}[1]{\href{http://arxiv.org/abs/#1}{#1}}

\newcommand{\aap}[3]{{\em A\&A\/},{\bf #2},(#1),#3.}

\newcommand{\aarv}[3]{{\em A\&ARv\/},{\bf #2},(#1),#3.}
\newcommand{\aj}[3]{{\em AJ\/},{\bf #2},(#1),#3.}
\newcommand{\annp}[3]{{\em  Ann. Phys.\/},{\bf #2},(#1),#3.}
\newcommand{\anyas}[3]{{\em Ann. NY. Acad. Sci.\/},{\bf #2},(#1),#3.}
\newcommand{\apj}[3]{{\em ApJ\/},{\bf #2},(#1),#3.}
\newcommand{\apjl}[3]{{\em ApJL\/},{\bf #2},(#1),#3.}

\newcommand{\araa}[3]{{\em ARA\&A\/},{\bf #2},(#1),#3.}
\newcommand{\ass}[3]{{\em Astrophys. and Space Sci.\/},{\bf #2},(#1),#3.}
\newcommand{\casp}[3]{{\em Comm. Astrophys. Sp. Phys.\/},{\bf #2},(#1),#3.}
\newcommand{\coa}[3]{{\em Comments on Astrophysics\/},{\bf #2},(#1),#3.}
\newcommand{\cqg}[3]{{\em Class. Quant. Grav.\/},{\bf #2},(#1),#3.}
\newcommand{\ijmpd}[3]{{\em Int. J. Mod. Phy. D\/},{\bf #2},(#1),#3.}
\newcommand{\jcap}[3]{{\em JCAP\/},{\bf #2},(#1),#3.}
\newcommand{\mnras}[3]{{\em MNRAS\/},{\bf #2},(#1),#3.}
\newcommand{\nar}[3]{{\em New Astronomy Reviews\/},{\bf #2},(#1),#3.}
\newcommand{\nas}[3]{{\em New Astronomy\/},{\bf #2},(#1),#3.}
\newcommand{\nat}[3]{{\em Nature\/},{\bf #2},(#1),#3.}
\newcommand{\phil}[3]{{\em Phil. Trans. Roy. Soc.\/},{\bf #2},(#1),#3.}

\newcommand{\phlb}[3]{{\em Phys. Lett. B\/},{\bf #2},(#1),#3.}
\newcommand{\phr}[3]{{\em Phys. Rep.\/},{\bf #2},(#1),#3.}
\newcommand{\prd}[3]{{\em Phys. Rev. D\/},{\bf #2},(#1),#3.}
\newcommand{\prl}[3]{{\em Phys. Rev. Lett.\/},{\bf #2},(#1),#3.}
\newcommand{\prs}[3]{{\em Proc. R. Soc. Lond.\/},{\bf #2},(#1),#3.}
\newcommand{\scam}[3]{{\em Sci. Am.\/},{\bf #2},(#1),#3.}
\newcommand{\sci}[3]{{\em Science\/},{\bf #2},(#1),#3.}

\newcommand{\cal}{\mathcal}
\newcommand{\pl}{\partial}
\renewcommand{\d}{{\rm d}}
\newcommand{\beq}{\begin{equation}}
\newcommand{\eeq}{\end{equation}}
\newcommand{\beqa}{\begin{eqnarray}}
\newcommand{\eeqa}{\end{eqnarray}}
\newcommand{\bea}{\begin{array}}
\newcommand{\ea}{\end{array}}
\newcommand{\lag}{\langle}
\newcommand{\rag}{\rangle}
\newcommand{\cP}{{\cal P}}

\newcommand{\xib}{\overline{\xi}}
\newcommand{\dum}{s}
\newcommand{\xidum}{\xi_{\dum}}
\newcommand{\phidum}{\varphi_{\dum}}

\newcommand{\cA}{{\cal A}}
\newcommand{\bc}{{\bf c}}

\newcommand{\Om}{\Omega_{\rm m}}
\newcommand{\Ol}{\Omega_{\Lambda}}
\newcommand{\Ode}{\Omega_{\rm de}}
\newcommand{\rhob}{\overline{\rho}}

\newcommand{\chirad}{{\chi}}
\newcommand{\De}{{\cal D}}
\newcommand{\wh}{\hat{w}}
\renewcommand{\r}{{\bf r}}
\newcommand{\bx}{{\bf x}}

\newcommand{\bv}{{\bf v}}
\newcommand{\vperp}{\vec{v}_{\perp}}
\newcommand{\vperpdot}{\dot{\vec{v}}_{\perp}}

\newcommand{\bk}{{\bf k}}
\newcommand{\kpar}{k_{\parallel}}
\newcommand{\kperp}{{\vec{k}}_{\perp}}
\newcommand{\kperpj}{{\vec{k}}_{\perp j}}
\newcommand{\vell}{\vec{\ell}}

\newcommand{\valpha}{\vec{\alpha}}
\newcommand{\vtheta}{\vec{\theta}}
\newcommand{\vex}{{\vec{e}}_x}
\newcommand{\vey}{{\vec{e}}_y}

\newcommand{\Map}{M_{\rm ap}}
\newcommand{\vgamma}{\vec{\gamma}}
\newcommand{\gammat}{\gamma_{t}}
\newcommand{\gammap}{\gamma_+}
\newcommand{\gammac}{\gamma_{\times}}
\newcommand{\barX}{\bar{X}}
\newcommand{\barkappa}{\bar{\kappa}}
\newcommand{\bargamma}{\bar{\gamma}}

\newcommand{\edth}{\,\eth\,}

\newcommand{\vt}{\vartheta}

\begin{document}

\begin{frontmatter}

\title{Cosmology with Weak Lensing Surveys}

\author{Dipak Munshi}
\address{Institute of Astronomy,  Madingley Road, Cambridge, CB3 OHA, UK \\
Astrophysics Group, Cavendish Laboratory, Madingley Road, \\
Cambridge CB3 OHE, UK}
\author{Patrick Valageas}
\address{Service de Physique Th\'eorique,
CEA Saclay,\\ 91191 Gif-sur-Yvette, France}
\author{Ludovic van Waerbeke}
\address{University of British Columbia,  Department of Physics \& Astronomy,\\ 6224
Agricultural Road,  Vancouver, B.C. V6T 1Z1, Canada.}
\author{Alan Heavens}
\address{SUPA (Scottish Universities Physics Alliance), Institute for Astronomy,\\University of Edinburgh,
Blackford Hill,  Edinburgh EH9 3HJ, UK }

\begin{abstract}

Weak gravitational lensing is responsible for the shearing and
magnification of the images of high-redshift sources due to the
presence of intervening matter. The distortions are due to
fluctuations in the gravitational potential, and are directly
related to the distribution of matter and to the geometry and
dynamics of the Universe. As a consequence, weak gravitational
lensing offers unique possibilities for probing the Dark Matter and
Dark Energy in the Universe.  In this review, we summarise the
theoretical and observational state of the subject, focussing on the
statistical aspects of weak lensing, and consider the prospects for
weak lensing surveys in the future.

Weak gravitational lensing surveys are complementary to both galaxy
surveys and cosmic microwave background (CMB) observations as they
probe the unbiased non-linear matter power spectrum at modest
redshifts.  Most of the cosmological parameters are accurately
estimated from CMB and large-scale galaxy surveys, so the focus of
attention is shifting to understanding the nature of Dark Matter and
Dark Energy.  On the theoretical side, recent advances in the use of
3D information of the sources from photometric redshifts promise
greater statistical power, and these are further enhanced by the use
of statistics beyond two-point quantities such as the power
spectrum.  The use of 3D information also alleviates difficulties
arising from physical effects such as the intrinsic alignment of
galaxies, which can mimic weak lensing to some extent.  On the
observational side, in the next few years weak lensing surveys such
as CFHTLS, VST-KIDS and Pan-STARRS, and the planned Dark Energy
Survey, will provide the first weak lensing surveys covering very
large sky areas and depth. In the long run even more ambitious
programmes such as DUNE, the Supernova Anisotropy Probe (SNAP) and
Large-aperture Synoptic Survey Telescope (LSST) are planned. Weak
lensing of diffuse components such as the CMB and 21cm emission can
also provide valuable cosmological information. Finally, we consider
the prospects for joint analysis with other probes, such as (1) the
CMB to probe background cosmology (2) galaxy surveys to probe
large-scale bias and (3) Sunyaev-Zeldovich surveys to study
small-scale baryonic physics, and consider the lensing effect on
cosmological supernova observations.

\end{abstract}


\end{frontmatter}

\newpage

\tableofcontents

\section{Introduction and notations}

Gravitational lensing refers to the deflection of light rays from
distant sources by the gravitational force arising from massive
bodies present along the line of sight. Such an effect was already
raised by Newton in 1704 and computed by Cavendish around 1784.  As
is well known, General Relativity put lensing on a firm theoretical
footing, and yields twice the Newtonian value for the deflection
angle \cite{Einstein1915}. The agreement of this prediction with the
deflection of light from distant stars by the Sun measured during
the solar eclipse of 1919 \cite{Dyson_Eddington_Davidson1920} was a
great success for Einstein's theory and brought General Relativity
to the general attention. The eclipse was necessary to allow one to
detect stars with a line of sight which comes close to the Sun.

In a similar fashion, light rays emitted by a distant galaxy are
deflected by the matter distribution along the line of sight toward
the observer. This creates a distortion of the image of this galaxy,
which is both sheared and amplified (or attenuated). It is possible
to distinguish two fields of study which make use of these
gravitational lensing effects. First, strong-lensing studies
correspond to strongly non-linear perturbations (which can lead to
multiple images of distant objects) produced by highly non-linear
massive objects (e.g. clusters of galaxies). In this case, the
analysis of the distortion of the images of background sources can
be used to extract some information on the properties of the
well-identified foreground lens (e.g. its mass).  Second, cosmic
shear, or weak gravitational lensing not associated with a
particular intervening lens, corresponds to the small distortion (of
the order of $1\%$) of the images of distant galaxies by all density
fluctuations along typical lines of sight. Then, one does not use
gravitational lensing to obtain the characteristics of a single
massive object but tries to derive the statistical properties of the
density field as well as the geometrical properties of the Universe
(as described by the cosmological parameters, such as the mean
density or the curvature). To this order, one computes the mean
shear over a rather large region on the sky (a few arcmin$^2$ or
more) from the ellipticities of many galaxies (one hundred or more).
Indeed, since galaxies are not exactly spherical one needs to
average over many galaxies and cross-correlate their observed
ellipticity in order to extract a meaningful signal.  Putting
together many such observations one obtains a large survey (a few to
many thousands of square degrees) which may have an intricate
geometry (as observational constraints may produce many holes).
Then, by performing various statistical measures one can derive from
such observations some constraints on the cosmological parameters as
well as on the statistical properties of the density field over
scales between a few arcmin to one degree, see for instance
Refs.~\cite{Miralda-Escude1991,Mellier1999,Bartelmann_Schneider2001,Refregier2003,LVWM2003,Mellier04,Schneider2005}.

Traditionally, the study of large scale structures has been done by
analyzing galaxy catalogues. However, this method is plagued by the
problem of the galaxy bias (i.e. the distribution of light may not
exactly follow the distribution of mass). The advantage of weak
lensing is its ability to probe directly the matter distribution,
through the gravitational potential, which is much more easily
related to theory.  In this way, one does not need to involve less
well-understood processes like galaxy or star formation.

In the last few years many studies have managed to detect
cosmological shear in random patches of the sky
\cite{Bacon_Refregier_Ellis2000,Bacon_et_al2003,Brown_et_al2003,Hamana_et_al2003,Hammerle_et_al2002,Hoekstra_Yee_Gladders2002,Hoekstra_et_al2002,Jarvis_et_al2002,Kaiser_Wilson_Luppino2000,Maoli_et_al2001,Refregier_Rhodes_Groth2002,Rhodes_Refregier_Groth2001,vanWaerbeke_et_al2001a,vanWaerbeke_et_al2002,Wittman_et_al2000}.
While early studies were primarily concerned with the detection of a
non-zero weak lensing signal, present weak lensing studies are
already putting constraints on cosmological parameters such as the
matter density parameter $\Om$ and the amplitude $\sigma_8$ of
the power-spectrum of matter density fluctuations. These works also
help to lift parameter degeneracies when used along with other
cosmological probes such as Cosmic Microwave Background (CMB)
observations. In combination with galaxy redshift surveys they can
be used to study the bias associated with various galaxies which
will be useful for galaxy formation scenarios thereby providing much
needed clues to the galaxy formation processes. 
For cosmological purposes, perhaps most exciting is the 
possibility that weak lensing will determine the properties of the 
dominant contributor to the Universe's energy budget: Dark 
Energy. Indeed, the recent acceleration of the Universe detected from the
magnitude-redshift relation of supernovae (SNeIa) occurs at too late 
redshifts to be probed by the CMB fluctuations. On the other hand,
weak lensing surveys offer a detailed probe of the dynamics of the 
Universe at low redshifts $z < 3$.
Thus weak lensing is among the best independent techniques to 
confirm this 
acceleration and to analyze in greater details the equation of state of 
this dark energy component which may open a window on new physics beyond 
the standard model (such as extra dimensions).  

In this review we describe the recent progress that has been made and various
prospects of future weak lensing surveys. We first describe in Section~\ref{ch:2}
the basic elements of the deflection of light rays by gravity and the various
observables associated with cosmological weak gravitational lensing.
In Section~\ref{ch2:2Pointstatistics} we review the 2-point statistics of these
observables (power-spectra and 2-point correlations) and the problem of mass
reconstruction from observed shear maps. Next, we explain in 
Section~\ref{ch4:3Dweaklensing} how the knowledge of the redshift of background 
sources can be used to improve constraints on theoretical cosmological models
or to perform fully 3-dimensional analysis (3D weak lensing). 
Then, we describe in Section~\ref{Non-Gaussianities} how to extract further 
information from weak lensing surveys by studying higher-order correlations
which can tighten the constraints on cosmological parameters or provide some
information on non-Gaussianities associated with non-linear dynamics or primordial
physics. We turn to the determination of weak lensing shear maps from actual 
observations of galaxy images and to the correction techniques which have been
devised to this order in Section~\ref{ch:6}. In Section~\ref{Simulations} 
we discuss the numerical simulations which are essential to compare theoretical
predictions with observational data. We describe in Section~\ref{Otherwavelengths}
how weak lensing surveys can also be performed at other wavelengths than the
common optical range, using for instance the 21cm emission of first generation
protogalaxies as distant sources. We present in greater detail the weak lensing
distortion of the CMB radiation in Section~\ref{CosmicMicrowaveBackground}.
In Section~\ref{External-data-sets} we also discuss how weak lensing can be
combined or cross-correlated with other data sets, such as the CMB or galaxy 
surveys, to help constrain cosmological models or derive some information on the
matter distribution (e.g. mass-to-light relationships). Finally, we conclude
in Section~\ref{Conclusion}. To help the reader, we also give in 
Tables~\ref{table:map1}--\ref{table:map3} below
our notations for most coordinate systems and variables used in this review.

\newpage

\begin{table}
\caption{Notation for cosmological variables}
\vspace{.2 cm}
\label{table:map1}
\begin{tabular}{lc}
\hline
total matter density in units of critical density & $\Om$ \\
reduced cosmological constant & $\Ol$ \\
reduced dark energy density & $\Ode$ \\
mean comoving density of the Universe & $\rhob$ \\
Hubble constant at present time & $H_0$ \\
Hubble constant at present time in units & $h$ \\ [-0.3cm]
of $100$ km.s$^{-1}$.Mpc$^{-1}$ & \\
rms linear density contrast in a sphere & $\sigma_8$ \\ [-0.3cm]
of radius $8 h^{-1}$ Mpc & \\
\hline
\end{tabular}
\end{table}

\begin{table}
\caption{Notation for coordinates}
\vspace{.2 cm}
\label{table:map2}
\begin{tabular}{lc}
\hline
metric & $\d s^2 = c^2 \d t^2 - a^2(t) [\d\chirad^2+\De^2(\d\theta^2+\sin^2\!\theta\,\d\varphi^2)]$ \\
speed of light & $c$ \\
scale factor & $a$ \\
comoving radial coordinate & $\chirad , r$ \\
comoving angular diameter distance & $\De$ \\
comoving position in 3D real space & $\bx, \r, (\chirad,\De\vtheta)$ \\
comoving wavenumber in 3D Fourier space & $\bk, (\kpar,\kperp), (\kpar,\vell/\De)$ \\
bend angle & $\valpha$ \\
deflection angle & $\delta\vtheta$ \\
image position on the sky & $\vtheta$ \\
flat-sky angle & $(\theta_1,\theta_2)$ \\
2D angular wavenumber & $\vell, (\ell_x,\ell_y), (\ell_1,\ell_2)$ \\
\hline
\end{tabular}
\end{table}

\begin{table}
\caption{Notation for fields and weak-lensing variables}
\vspace{.2 cm}
\label{table:map3}
\begin{tabular}{lc}
\hline
gravitational potential & $\Phi$ \\
lensing potential & $\phi$ \\
shear matrix & $\Psi$ \\
amplification matrix & $\cA$ \\
weak-lensing convergence & $\kappa$ \\
complex weak-lensing shear & $\gamma=\gamma_1+i\,\gamma_2$ \\
shear pseudo-vector & $\vgamma=\gamma_1\vex+\gamma_2\vey$ \\
tangential component of shear & $\gammat , \gammap$ \\
cross component of shear & $\gammac$ \\
weak-lensing magnification & $\mu$ \\
angular filter radius & $\theta_s$ \\
smoothed convergence, smoothed shear & $\barkappa , \bargamma$ \\
weak-lensing aperture mass & $\Map$ \\
3D matter density power spectrum & $P(k)$ \\
2D convergence power spectrum & $P_{\kappa}(\ell)$ \\
2D shear power spectrum & $P_{\gamma}(\ell)$ \\
two-point correlation & $\xi$ \\
3D density contrast bispectrum & $B(k_1,k_2,k_3)$ \\
2D convergence bispectrum & $B_{\kappa}(\ell_1,\ell_2,\ell_3)$ \\
probability distribution function of the smoothed convergence & $\cP_{\kappa}(\barkappa)$ \\
\hline
\end{tabular}
\end{table}

\section{\label{ch:2} Weak Lensing Theory}

\subsection{Deflection of light rays}
\label{Deflection}

We briefly describe here the basic idea behind weak gravitational
lensing as we present a simple heuristic derivation of the
first-order result for the deflection of light rays by gravity. For
a rigorous derivation using General Relativity the reader can
consult references
~\cite{Kaiser1998,Bartelmann_Schneider2001,Schneider_et_al1992,Seitz_et_al1994}.

\begin{figure}
\begin{center}
\epsfig{file=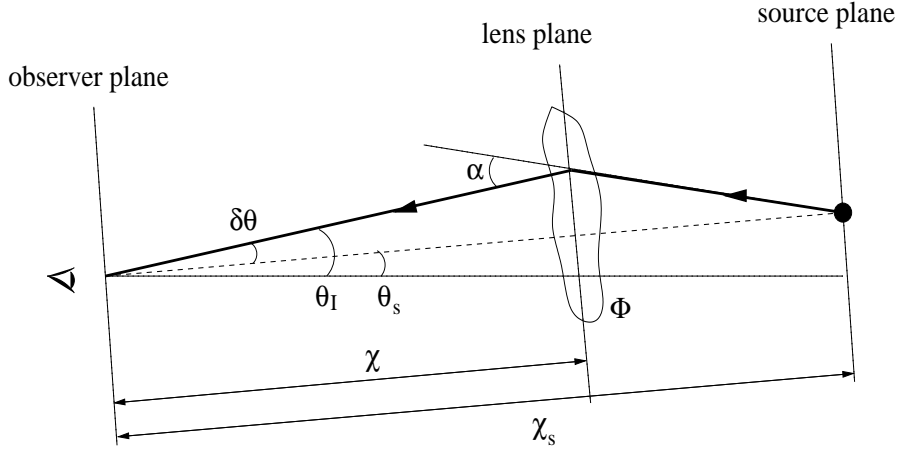,height=6cm,width=12cm}
\end{center}
\caption[]{Deflection of light rays from a distant source at comoving radial
distance $\chirad_s$ by a gravitational potential fluctuation $\Phi$ at distance
$\chirad$. For a thin lens the deflection by the angle $\alpha$ is taken as
instantaneous. This changes the observed position of the source by the
angle $\delta\theta$, from the intrinsic source direction $\theta_s$ to the
image direction $\theta_I$ on the sky.}
\label{deflec.fig}
\end{figure}

We assume in the following that deflections angles are small so that
we only consider first-order terms. This is sufficient for most
applications of weak lensing since by definition the latter
corresponds to the case of small perturbations of light rays by the
large-scale structures of the universe. Let us consider within
Newtonian theory the deflection of a photon with velocity $\bv$ that
passes through a small region of space where the gravitational
potential $\Phi$ is non-zero. The acceleration perpendicular to the
unperturbed trajectory, $\vperpdot=-\nabla_{\perp}\Phi$,
yields a small transverse velocity $\vperp= -\int \d t
\nabla_{\perp}\Phi$. This gives a deflection angle
$\valpha=\vperp/c=-\int \d l \nabla_{\perp}\Phi/c^2$
for a constant velocity $|\bv|=c$. As is well-known, General
Relativity simply yields this Newtonian result multiplied by a
factor two. This deflection changes the observed position on the sky
of the radiation source by a small angle $\delta\vtheta$. For an
extended source (e.g. a galaxy) this also leads to both a
magnification and a shear of the image of the source from which one
can extract some information on the gravitational potential $\Phi$.
If the deflection takes place within a small distance it can be
taken as instantaneous which corresponds to the thin lens
approximation (as in geometrical optics) as displayed in
Fig.~\ref{deflec.fig}. Besides, in cosmology transverse distances
are related to angles through the comoving angular diameter distance $\De$ 
given by:
\begin{equation}
\De(\chirad) = \frac{c\sin_K \left(|1-\Om-\Ol|^{1/2}H_0\;\chirad/c\right)} 
{H_0 |1-\Om-\Ol|^{1/2}} , 
\label{De}
\end{equation}
where $\sin_K$ means the hyperbolic sine, sinh, if $(1-\Om-\Ol)> 0$, 
or sine if $(1-\Om-\Ol) < 0$; if $(1-\Om-\Ol) = 0$, then
$\De(\chirad) = \chirad$ (case of a flat Universe). 
The radial comoving distance $\chirad$ measured by a light ray which travels
from a source at redshift $z$ to the observer at $z=0$ is given by:
\beq 
\chirad = \frac{c}{H_0} \int_0^z \frac{\d z'}{\sqrt{\Ol+(1-\Om-\Ol)(1+z')^2+
\Om(1+z')^3}} , 
\label{chi} 
\eeq 
where $z'$ is the redshift along the line of sight. Note that $\chi$ measures
both a spatial coordinate distance and a travel time. 
Here we also introduced the Hubble constant $H_0$ and
the cosmological parameters $\Om$ (matter density parameter) and
$\Ol$ (dark energy in the form of a cosmological constant).
Therefore, the source appears to have moved in the source plane over
a comoving distance
$\De(\chirad_s)\delta\vtheta=-\De(\chirad_s-\chirad){\vec\alpha}$ as can be
seen from Fig.~\ref{deflec.fig}, where ${\vec\alpha}$ and
$\delta\vtheta$ are 2D vectors in the plane perpendicular to the
unperturbed light ray.  Summing up the deflections arising from all
potential gradients between the observer and the source gives the
total shift on the sky: 
\beq 
\delta\vtheta = \vtheta_I - \vtheta_s =
\frac{2}{c^2} \int_0^{\chirad_s} \d\chirad \;
\frac{\De(\chirad_s-\chirad)}{\De(\chirad_s)} \; \nabla_{\perp} \Phi(\chirad) ,
\label{dtheta} 
\eeq

where $\vtheta_s$ is the intrinsic position of the source on the sky
and $\vtheta_I$ is the observed position. However, generally we do
not know the true position of the source but only the position of
the observed image. Thus the observable quantities are not the
displacements $\delta\vtheta$ themselves but the distortions induced
by these deflections. They are given at lowest order by the
symmetric shear matrix $\Psi_{ij}$
\cite{Kaiser1992,Bartelmann_Schneider2001,Jain_Seljak_White2000}
which we define as: 
\beq 
\Psi_{ij} = \frac{\pl\delta\theta_{i}}{\pl\theta_{sj}} 
=  \frac{2}{c^2} \int_0^{\chirad_s} \d\chirad \;
\frac{\De(\chirad)\De(\chirad_s-\chirad)}{\De(\chirad_s)} \; \nabla_i
\nabla_{\!j} \Phi(\chirad) , 
\label{Psi} 
\eeq 
Eq.(\ref{Psi}) follows
from eq.(\ref{dtheta}) if we note that a change of angle $\d\vtheta$
for the unperturbed light ray corresponds to a transverse distance
$\De(\chirad)\d\vtheta$ in the lens plane where the gravitational
potential $\Phi$ produces the gravitational lensing. The reasoning
presented above clearly shows that Eq.(\ref{Psi}) uses the weak
lensing approximation; the derivatives
$\nabla_i\nabla_{\!j}\Phi(\chirad)$ of the gravitational potential are
computed along the unperturbed trajectory of the photon. This
assumes that the components of the shear tensor are small but the
density fluctuations $\delta$ can be large \cite{Kaiser1992}. We can
also express the shear matrix $\Psi_{ij}$ in terms of a lensing
potential $\phi(\vtheta;\chirad_s)$ (also called the deflection
potential) as: 
\beq 
\Psi_{ij} =  \phi_{,ij}  \;\;\;\;\; \mbox{with}
\;\;\;\;\; \phi(\vtheta;\chirad_s) = \frac{2}{c^2}\int_0^{\chirad_s}
\d\chirad \; \frac{\De(\chirad_s-\chirad)}{\De(\chirad_s)\De(\chirad)} \;
\Phi(\chirad,\De(\chirad)\vtheta) . 
\label{Phi} 
\eeq
The expression (\ref{Phi}) is formally divergent because of the term 
$1/\De(\chirad)$ near $\chirad=0$, but this only affects the monopole
term which does not contribute to the shear matrix $\Psi_{ij}$
(indeed derivatives with respect to angles yield powers of $\De$
as in Eq.(\ref{Psi})). Therefore, we may set the constant term to zero
so that $\phi(\vtheta)$ is well defined. Eq.(\ref{Phi})
clearly shows how the weak lensing distortions are related to the
gravitational potential projected onto the sky and can be fully
described at this order by the 2D lensing potential $\phi(\vtheta)$.
Thus, in this approximation lensing by the 3D matter distribution
from the observer to the redshift $z_s$ of the source plane is
equivalent to a thin lens plane with the same deflection potential
$\phi(\vtheta)$. However, from the dependence of
$\phi(\vtheta;\chirad_s)$ on the redshift $z_s$ of the source plane we
can recover the 3D matter distribution as discussed below in
Section~\ref{ch4:3Dweaklensing}. Note that weak lensing effects grow with
the redshift of the source as the line of sight is more extended. However,
since distant galaxies are fainter and more difficult to observe weak lensing
surveys mainly probe redshifts $z_s\sim 1$. On the other hand, this range of
redshifts of order unity is of great interest to probe the dark energy
component of the Universe. Next, one can also introduce the
amplification matrix $\cA$ of image flux densities which is simply
given by the ratio of image areas, that is by the Jacobian:
\begin{equation}
\cA = \frac{\pl\vtheta_s}{\pl\vtheta_I} = \left(\delta_{ij} +
\Psi_{ij}\right)^{-1}
= \left(\bea{cc} 1-\kappa-\gamma_1 & -\gamma_2 \\
-\gamma_2 & 1-\kappa+\gamma_1 \ea\right) , 
\label{defA}
\end{equation}
which defines the {\em convergence} $\kappa$ and the {\em complex
shear} $\gamma=\gamma_1+i\,\gamma_2$. At linear order the convergence
gives the magnification of the source as 
$\mu=[{\rm det}(\cA)]^{-1}\simeq 1+2\kappa$.
The shear describes the area-preserving distortion of amplitude
given by $|\gamma|$ and of direction given by its phase, see also
Section~\ref{Shape-measurement} and Fig.~\ref{lensedgal.ps}. In general
the matrix $\cA$ also contains an antisymmetric part associated with
a rotation of the image but this term vanishes at linear order as
can be seen from Eq.(\ref{Psi}). From Eq.(\ref{defA}) the
convergence $\kappa$ and the shear components $\gamma_1$, $\gamma_2$,
can be written at linear order in terms of the shear tensor as:
\begin{equation}
\kappa = \frac{\Psi_{11}+\Psi_{22}}{2}, \;\;\;\; \gamma = \gamma_1 + i
\; \gamma_2 \;\;\; \mbox{with} \;\;\; \gamma_1 =
\frac{\Psi_{11}-\Psi_{22}}{2} , \;\; \gamma_2 = \Psi_{12} .
\label{defkappagam}
\end{equation}
On the other hand, the gravitational potential, $\Phi$, is related
to the fluctuations of the density contrast, $\delta$, by Poisson's
equation:
\begin{equation}
 \nabla^2 \Phi = \frac{3}{2} \Om H_0^2
(1+z) \; \delta \hspace{0.4cm} \mbox{with} \hspace{0.4cm}
\delta(\bx) = \frac{\rho(\bx)-\rhob}{\rhob} , 
\label{Poisson}
\end{equation}
where $\rhob$ is the mean density of the universe. Note that since
the convergence $\kappa$ and the shear components $\gamma_i$ can be
expressed in terms of the scalar lensing potential $\phi$ they are
not independent. For instance, one can check from the first Eq.(\ref{Phi})
and Eq.(\ref{defkappagam}) that we have 
$\kappa_{,1}=\gamma_{1,1}+\gamma_{2,2}$ \cite{Kaiser95}.
This allows one to derive consistency relations
satisfied by weak lensing distortions (e.g.~\cite{Schneider2003})
and deviations from these relations in the observed shear fields can
be used to estimate the observational noise or systematics. Of
course such relations also imply interrelations between correlation
functions, see \cite{Schneider2005} and Section~\ref{ch2:2Pointstatistics} 
below. For a rigorous derivation of
Eqs.(\ref{Psi})-(\ref{defkappagam}) one needs to compute the paths
of light rays (null geodesics) through the perturbed metric of
spacetime using General Relativity \cite{Kaiser1998}. An alternative
approach is to follow the distortion of the cross-section of an
infinitesimal light beam
\cite{Sachs1961,Blandford_et_al1991,Seitz_et_al1994,Bartelmann_Schneider2001}.
Both methods give back the results (\ref{Psi})-(\ref{defkappagam})
obtained in a heuristic manner above.

\subsection{Convergence, shear and aperture mass}
\label{Convergence}

Thanks to the radial integration over $\chirad$ in (\ref{Psi})
gradients of the gravitational potential along the radial direction
give a negligible contribution as compared with transverse
fluctuations \cite{Kaiser1992,Jain_Seljak_White2000,Limber1954}
since positive and negative fluctuations cancel along the line of
sight. In other words, the radial integration selects Fourier radial
modes of order $|\kpar| \sim H/c$ (inverse of cosmological distances
over which the effective lensing weight $\wh(\chirad)$ varies, see
eq.(\ref{kappanz}) below) whereas transverse modes are of order $|\kperp|
\sim 1/\De\theta_s \gg |\kpar|$ where $\theta_s\ll 1$ is the typical
angular scale (a few arcmin) probed by the weak-lensing observable.
Therefore, within this small-angle approximation the 2D Laplacian
(\ref{defkappagam}) associated with $\kappa$ can be expressed in
terms of the 3D Laplacian (\ref{Poisson}) at each point along the
line of sight. This yields for the convergence along a given line of
sight up to $z_s$: 
\beq 
\kappa(z_s) \simeq \int_0^{\chirad_s} \d\chirad \;
w(\chirad,\chirad_s) \delta(\chirad) \;\; \mbox{with} \;\; w(\chirad,\chirad_s) =
\frac{3\Om H_0^2 \De(\chirad) \De(\chirad_s-\chirad)}{2 c^2 \De(\chirad_s)}
(1+z) . 
\label{kappa} 
\eeq 
Thus the convergence, $\kappa$, can be
expressed very simply as a function of the density field; it is
merely an average of the local density contrast along the line of
sight. Therefore, weak lensing observations allow one to measure the
projected density field $\kappa$ on the sky (note that by looking at
sources located at different redshifts one may also probe the radial
direction). In practice the sources have a broad redshift
distribution which needs to be taken into account. Thus, the
quantity of interest is actually: 
\beq 
\kappa= \int_0^{\infty}\d z_s
\; n(z_s)\kappa(z_s) = \int_0^{\chirad_{\rm max}} \d\chirad \; \wh(\chirad)
\delta(\chirad) \;\; \mbox{with} \;\; \wh(\chirad) = \int_z^{z_{\rm max}}
\d z_s \; n(z_s) \; w(\chirad,\chirad_s) , 
\label{kappanz} 
\eeq 
where $n(z_s)$ is the mean redshift distribution of the sources (e.g.
galaxies) normalized to unity and $z_{\rm max}$ is the depth of the
survey. Eq.(\ref{kappanz}) neglects the discrete effects due to the
finite number of galaxies, which can be obtained by taking into
account the discrete nature of the distribution $n(z_s)$. This gives
corrections of order $1/N$ to higher-order moments of weak-lensing
observables, where $N$ is the number of galaxies within the field of
interest. In practice $N$ is much larger than unity (for a circular
window of radius 1 arcmin we expect $N > 100$ for the SNAP mission)
therefore it is usually sufficient to work with Eq.(\ref{kappanz}).

In order to measure weak-lensing observables such as $\kappa$ or the
shear $\gamma$ one measures for instance the brightness or the shape
of galaxies located around a given direction $\vtheta$ on the sky.
Therefore, one is led to consider weak-lensing quantities smoothed
over a non-zero angular radius $\theta_s$ around the direction
$\vtheta$. More generally, one can define any smoothed weak-lensing
quantity $\barX(\vtheta)$ from its angular filter $U_X(\Delta
\vtheta)$ by: 
\beq 
\barX(\vtheta) = \int \d\vtheta' \;
U_X(\vtheta'-\vtheta) \kappa(\vtheta') = \int\d\chirad \; \wh \int
\d\vtheta' \; U_X(\vtheta'-\vtheta) \delta(\chirad,\De \vtheta') ,
\label{Xx} 
\eeq 
where $\vtheta'$ is the angular vector in the plane
perpendicular to the line of sight (we restrict ourselves to small
angular windows) and $\De\vtheta'$ is the two-dimensional vector of
transverse coordinates. Thus, it is customary to define the smoothed
convergence by a top-hat $U_{\kappa}$ of angular radius $\theta_s$
but this quantity is not very convenient for practical purposes
since it is easier to measure the ellipticity of galaxies (related
to the shear $\gamma$) than their magnification (related to
$\kappa$). This leads one to consider compensated filters $U_{\Map}$
with polar symmetry which define the ``aperture-mass'' $\Map$, that
is with $\int d\vtheta\, U_{\Map}(\vtheta)=0$. Then $\Map$ can be
expressed in terms of the tangential component $\gamma_t$ of the
shear \cite{Schneider1996} so that it is not necessary to build a
full convergence map from observations: 
\beq 
\Map(\vtheta) \equiv
\int \d\vtheta' U_{\Map}(|\vtheta'-\vtheta|) \kappa(\vtheta') = \int
\d\vtheta' Q_{\Map}(|\vtheta'-\vtheta|) \gamma_t(\vtheta')
\label{Map} 
\eeq 
where we introduce \cite{Schneider1996}: 
\beq
Q_{\Map}(\theta)=-U_{\Map}(\theta)+{2\over \theta^2}\int_0^\theta
\d\theta'\,\theta' \,U_{\Map}(\theta') . 
\label{QU} 
\eeq Besides,
the aperture-mass provides a useful separation between $E$ and $B$
modes, as discussed below in Section~\ref{E/B decomposition}.

For analytical and data analysis purposes it is often useful to work
in Fourier space. Thus, we write for the 3D matter density contrast 
$\delta(\bx)$ and the 2D lensing potential $\phi(\vtheta)$:
\beq
\delta({\bx}) = \int \frac{\d\bk}{(2\pi)^3} \, e^{-i \bk.\bx} \, \delta(\bk) 
\;\;\; \mbox{and} \;\;\; \phi(\vtheta)= \int \frac{\d\vell}{(2\pi)^2} 
\, e^{-i \vell.\vtheta} \, \phi(\vell) ,
\label{def_deltak_phil}
\eeq
where we use a flat-sky approximation for 2D fields. This is sufficient
for most weak lensing purposes where we consider angular scales of the
order of $1-10$ arcmin, but we shall describe in Section~\ref{3Danalysis}
the more general expansion over spherical harmonics.
From Eq.(\ref{Phi}) and Poisson's equation (\ref{Poisson}) we obtain:
\beq
\phi(\vell) = -2 \int\d\chirad \, \wh(\chirad)
\int\frac{\d\kpar}{2\pi} \, e^{-i\kpar\chirad} \, \frac{1}{k^2\De(\chirad)^4}
\, \delta\left(\kpar,\frac{\vell}{\De(\chirad)};\chi\right) ,
\label{phil}
\eeq
where $\kpar$ is the component parallel to the line of sight of the 3D 
wavenumber $\bk=(\kpar,\kperp)$, with $\kperp=\vell/\De$, and 
$\delta(\bk;\chirad)$ is the matter density contrast in Fourier space 
at redshift $z(\chirad)$. The weight $\wh(\chirad)$ along the line of sight
was defined in Eqs.(\ref{kappa})-(\ref{kappanz}). Then, from 
Eq.(\ref{defkappagam}) we obtain for the convergence $\kappa$:
\beq
\kappa(\vell) = -\frac{1}{2}(\ell_x^2+\ell_y^2) \, \phi(\vell)
\simeq \int\d\chirad \, \frac{\wh(\chirad)}{\De^2} \int\frac{\d\kpar}{2\pi} \,
e^{-i\kpar\chirad} \, 
\delta\left(\kpar,\frac{\vell}{\De(\chirad)};\chi\right) .
\label{kappal}
\eeq
In the last expression we used as for Eq.(\ref{kappa}) Limber's approximation
$k^2\simeq k_{\perp}^2$ as the integration along the line of sight associated
with the projection on the sky suppresses radial modes as compared with
transverse wavenumbers (i.e. $|\kpar| \ll k_{\perp}$). In a similar fashion,
we obtain from Eq.(\ref{defkappagam}) for the complex shear $\gamma$:
\beq
\gamma(\vell) = -\frac{1}{2}(\ell_x+i\ell_y)^2 \, \phi(\vell) 
= \frac{\ell_x^2-\ell_y^2+2i\ell_x\ell_y}{\ell_x^2+\ell_y^2} \, \kappa(\vell)
= e^{i2\alpha} \, \kappa(\vell) ,
\label{gammal} 
\eeq
where $\alpha$ is the polar angle of the wavenumber $\vell=(\ell_x,\ell_y)$.
This expression clearly shows that the
complex shear $\gamma$ is a spin-2 field: it transforms as
$\gamma\rightarrow \gamma \, e^{-i2\psi}$ under a rotation of
transverse coordinates axis of angle $\psi$. This comes from the fact
that an ellipse transforms into itself through a rotation of $180$ degrees
and so does the shear which measures the area-preserving distortion,
see Fig.~\ref{lensedgal.ps}.

For smoothed weak-lensing observables $\bar X$ as defined in
Eq.(\ref{Xx}) we obtain:
\beq
\bar X(\vell) = W_X(-\vell\theta_s) \kappa(\vell) \;\;\; \mbox{with} \;\;\;
W_X(\vell\theta_s) = \int\d\vtheta \, e^{i \vell.\vtheta} U_X(\vtheta)  ,
\label{WX} 
\eeq
where we introduced the Fourier transform $W_X$ of the real-space filter 
$U_X$ of angular scale $\theta_s$. 
This gives for the convergence and the shear smoothed with a top-hat of 
angular radius $\theta_s$:
\beq
W_{\kappa}(\vell\theta_s) = \frac{2 J_1(\ell\theta_s)}{\ell\theta_s} , \;\;\;
W_{\gamma}(\vell\theta_s) = W_{\kappa}(\ell\theta_s) \; e^{i2\alpha} ,
\label{Wkappa_gamma}
\eeq
where $J_1$ is the Bessel function of the first kind of order 1.
In real space this gives back (with $\theta=|\vtheta|$):
\beq
U_{\kappa}(\vtheta) = \frac{\Theta(\theta_s-\theta)}{\pi\theta_s^2}
, \;\;\;
U_{\gamma}(\vtheta) = - \frac{\Theta(\theta-\theta_s)}{\pi\theta^2}
\; e^{i2\beta} ,
\label{Ukappa_gamma}
\eeq
where $\Theta$ is the Heaviside function and $\beta$ is the polar angle of
the angular vector $\vtheta$. Note that Eq.(\ref{Ukappa_gamma})
clearly shows that the smoothed convergence is an average of the density
contrast over the cone of angular radius $\theta_s$ whereas the smoothed
shear can be written as an average of the density contrast outside of this
cone.

One drawback of the shear components is that they are even
quantities (their sign can be changed through a rotation of axis,
see Eq.(\ref{gammal})), hence their third-order moment vanishes
by symmetry and one must measure the fourth-order moment $\lag
\bargamma_i^4\rag$ (i.e. the kurtosis) in order to probe the
deviations from Gaussianity. Therefore it is more convenient to use
the aperture-mass defined in Eq.(\ref{Map}) which can be derived
from the shear but is not even, so that deviations from Gaussianity
can be detected through the third-order moment $\lag\Map^3\rag$. A
simple example is provided by the pair of filters
\cite{Schneider1996}: 
\beq 
U_{\Map}(\vtheta) =
\frac{\Theta(\theta_s-\theta)}{\pi\theta_s^2} \; 9
\left(1-\frac{\theta^2}{\theta_s^2}\right) \left(\frac{1}{3} -
\frac{\theta^2}{\theta_s^2}\right) , \label{UMap} \eeq 
and: 
\beq
W_{\Map}(\vell\theta_s) = \frac{24 J_4(\ell\theta_s)}{(\ell\theta_s)^2}. 
\label{WMap} 
\eeq

\subsection{Approximations}

The derivation of Eq.(\ref{Psi}) does not assume that
the density fluctuations $\delta$ are small but it assumes that
deflection angles $\delta\vtheta$ are small so that the relative
deflection $\Psi_{ij}$ of neighboring light rays can be computed
from the gravitational potential gradients along the unperturbed
trajectory (Born approximation). This may not be a good
approximation for individual light beams, but in cosmological
weak-lensing studies considered in this review one is only
interested in the statistical properties of the gravitational
lensing distortions. Since the statistical properties of the tidal
field $\Phi_{ij}$ are, to an excellent approximation, identical
along the perturbed and unperturbed paths, the use of Eq.(\ref{Psi})
is well-justified to compute statistical quantities such as the
correlation functions of the shear field
\cite{Kaiser1992,Bernardeau_et_al1997}.

Apart from the higher-order corrections to the Born approximation discussed
above (multiple lens couplings), other higher-order terms are produced by
the observational procedure.
Indeed, in Eq.(\ref{kappanz}) we neglected the fluctuations of the galaxy
distribution $n(z_s)$ which can be coupled to the matter density fluctuations
along the line of sight. This source-lens correlation effect is more important
as the overlapping area between the distributions of sources and lenses
increases. On the other hand, source density fluctuations themselves can
lead to spurious small-scale power (as the average distance to the sources
can vary with the direction on the sky). Using analytical methods
Ref.~\cite{Bernardeau1998} found that both these effects are negligible for the
skewness and kurtosis of the convergence provided the source redshift
dispersion is less than about $0.15$. These source clustering effects were
further discussed in \cite{Hamana_et_al2002} who found that numerical
simulations agree well with semi-analytical estimates and that the amplitude
of such effects strongly depends on the redshift distribution of the sources.
A recent study of the source-lens clustering \cite{Forero-Romero}, 
using numerical simulations coupled to realistic semi-analytical models 
for the distribution of galaxies, finds that this effect can bias the 
estimation of $\sigma_8$ by $2\%-5\%$. 
Therefore, accurate photometric redshifts will be needed for future missions
such as SNAP or LSST to handle this effect.

\section{\label{ch2:2Pointstatistics}Statistics of 2D Cosmic Shear}

For statistical analysis of cosmic shear, it is most common to use
2-point quantities, i.e. those which are quadratic in the shear, and
calculated either in real or harmonic space.  For this Section, we
will restrict the discussion to 2D fields, where we consider the
statistics of the shear pattern on the sky only, and not in 3D.  The
shear field will be treated as a 3D field in Section~\ref{ch4:3Dweaklensing}.
Examples of real-space 2-point statistics are the
average shear variance and various shear correlation functions. In
general there are advantages for cosmological parameter estimation
in using harmonic-space statistics, as their correlation properties
are more convenient, but for surveys with complicated geometry, such
as happens with removal of bright stars and artifacts, there can be
practical advantages to using real-space measures, as they can be
easier to estimate.  All the 2-point statistics can be related to
the underlying 3D matter power spectrum via the (2D) convergence
power spectrum $P_\kappa(\ell)$, and inspection of the relationship
between the two point statistic and $P_\kappa(\ell)$ can be
instructive, as it shows which wavenumbers are picked out by each
statistic.  In general, a narrow window in $\ell$ space may be
desirable if the power spectrum is to be estimated.

\subsection{Convergence and shear power spectra}

We define the power spectra $P(k)$ of the 3D matter density contrast
and $P_\kappa(\ell)$ of the 2D convergence as:
\beq
\lag\delta(\bk_1)\delta(\bk_2)\rag = (2\pi)^3 \delta_D(\bk_1+\bk_2) P(k_1)
\label{Powerk}
\eeq
and:
\beq
\lag\kappa(\vell_1)\kappa(\vell_2)\rag = (2\pi)^2 \delta_D(\vell_1+\vell_2)
P_{\kappa}(\ell_1) .
\label{powerl}
\eeq
The Dirac functions $\delta_D$ express statistical homogeneity whereas
statistical isotropy implies that $P(\bk)$ and $P_{\kappa}(\vell)$ only
depend on $k=|\bk|$ and $\ell=|\vell|$. In Eq.(\ref{powerl}) we used
a flat-sky approximation which is sufficient for most weak-lensing purposes.
We shall discuss in Section~\ref{ch4:3Dweaklensing} the expansion over
spherical harmonics (instead of plane waves as in Eq.(\ref{powerl})) which
is necessary for instance for full-sky studies.
Then, from Eqs.(\ref{kappal})-(\ref{gammal}) we obtain:
\begin{equation}
P_\kappa(\ell) = P_\gamma(\ell) = \frac{1}{4}\ell^4\,P_\phi(\ell)
\end{equation}
and:
\begin{equation}
P_\kappa(\ell) = \int_0^\chirad d\chirad'
\frac{\wh^2(\chirad')}{\De^2(\chirad')}
P\left({\ell\over \De(\chi')};\chi'\right).
\label{Powerkappa}
\end{equation}
Thus this expression gives the 2D convergence power spectrum in terms of the
3D matter power spectrum $P(k;\chirad)$ integrated along the line of sight,
using Limber's approximation.

\subsection{2-point statistics in real space}

As an example of a real-space 2-point statistic, consider the {\em
shear variance}, defined as the variance of the average shear
$\bargamma$ evaluated in circular patches of varying radius
$\theta_s$.  The averaging is a convolution, so the power is
multiplied (see Eqs.(\ref{WX}),(\ref{Wkappa_gamma})):
\begin{equation}
\langle |\bargamma|^2\rangle = \int {d\ell\over 2\pi} \ell P_\kappa(\ell)
{4J_1^2(\ell\theta_s)\over(\ell\theta_s)^2},
\end{equation}
where $J_n$ is a Bessel function of order $n$.

The {\em shear correlation functions} can either be defined with
reference to the coordinate axes,
\begin{equation}
\xi_{ij}(\theta) \equiv \langle \gamma_i(\vtheta')\gamma_j(\vtheta'+\vtheta)\rangle
\end{equation}
where $i,j=1,2$ and the averaging is done over pairs of galaxies
separated by angle $\theta=|\vtheta|$.  By parity $\xi_{12}=0$, and
by isotropy $\xi_{11}$ and $\xi_{22}$ are functions only of
$|\vtheta|$. The correlation function of the complex shear is
\begin{eqnarray}
\langle \gamma \gamma^*\rangle_\theta &=& \int {d^2\ell\over
(2\pi)^2}\, P_\gamma(\ell)\, e^{i\vell.\vtheta}\\\nonumber
 &=& \int {\ell d\ell\over (2\pi)^2}\, P_\kappa(\ell)
 e^{i\ell\theta\cos\varphi} d\varphi\\\nonumber
 &=& \int {d\ell\over 2\pi} \ell P_\kappa(\ell) J_0(\ell\theta).
 \label{gammacorrn}
\end{eqnarray}
Alternatively, the shears may be referred to axes oriented
tangentially ($t$) and at 45 degrees to the radius ($\times$),
defined with respect to each pair of galaxies used in the averaging.
The rotations $\gamma \rightarrow \gamma'=\gamma \, e^{-2i\psi}$, where
$\psi$ is the position angle of the pair, give tangential and cross
components of the rotated shear as
$\gamma'=-\gamma_t-i\gamma_\times$, where the components have
correlation functions $\xi_{tt}$ and $\xi_{\times\times}$
respectively.  It is common to define a pair of correlations
\begin{equation}
\xi_{\pm}(\theta) = \xi_{tt}\pm \xi_{\times\times},
\end{equation}
which can be related to the convergence power spectrum by (see
\cite{Kaiser1992})
\begin{eqnarray}\label{chipm}
\xi_+(\theta) &=& \int_0^\infty {d\ell\over 2\pi} \ell P_\kappa(\ell) J_0(\ell\theta)\nonumber\\
\xi_-(\theta) &=& \int_0^\infty {d\ell\over 2\pi} \ell
P_\kappa(\ell) J_4(\ell\theta).
\end{eqnarray}
Finally, let us consider the class of statistics referred to as {\em aperture
masses} associated with {\em compensated filters}, which we defined
in equation (\ref{Map}). This allows $\Map$ to be related to the
tangential shear \cite{Schneider1996} as in equation (\ref{Map}).
Several forms of $U_{\Map}$ have been suggested, which trade
locality in real space with locality in $\ell$ space.
Ref.~\cite{schneideretal1998} considers the filter $U_{\Map}$ of equation
(\ref{UMap}) which cuts off at some scale, $\theta_s$. From equation
(\ref{WMap}) this gives a two-point statistic
\begin{equation}
\langle \Map^2(\theta_s) \rangle = \int {d\ell\over 2\pi}\, \ell
P_\kappa(\ell)\, {576J_4^2(\ell\theta_s)\over (\ell\theta_s)^4}.
\label{Map_Pkappa}
\end{equation}
Other forms have been suggested \cite{crittendenetal2002}, which
are broader in real space, but pick up a narrower range of $\ell$
power for a given $\theta$. As we have seen, all of these two-point
statistics can be written as integrals over $\ell$ of the
convergence power spectrum $P_\kappa(\ell)$ multiplied by some
kernel function, since weak-lensing distortions can be expressed in
terms of the lensing potential $\phi$, see Eqs.(\ref{Phi}) and
(\ref{kappal}).

\begin{figure}
\begin{center}
\epsfxsize=7.12 cm \epsfysize=7.4 cm {\epsfbox[27 424 302 712]{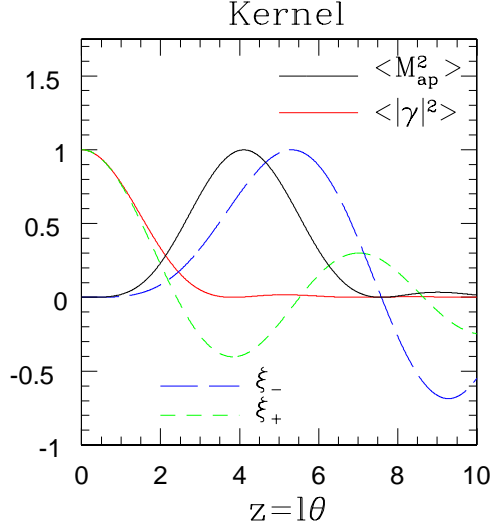}}
\end{center}
\caption{Kernel functions for the two-point statistics discussed in
this Section.  $z=\ell\theta_s$.  The thin solid line peaking at $z=0$
is corresponds with the shear variance, the thick solid line is the
aperture mass, with filter given in equation (\ref{UMap}), the short
dashed line is the kernel for $\xi_+$ and the long dashed to
$\xi_-$.}
\label{kernels}
\end{figure}

If one wants to estimate the matter power
spectrum, then there are some advantages in having a narrow kernel
function, but the uncertainty principle then demands that the
filtering is broad on the sky.  This can lead to practical
difficulties in dealing with holes, edges etc.  Filter functions for
the 2-point statistics mentioned here are shown in Fig.~\ref{kernels}.

\subsection{E/B decomposition}
\label{E/B decomposition}

Weak gravitational lensing does not produce the full range of locally
linear distortions possible.  These are characterised by translation,
rotation, dilation and shear, with six free parameters.  Translation
is not readily observable, but weak lensing is specified by three
parameters rather than the four remaining degrees of freedom permitted
by local affine transformations. This restriction is manifested in a
number of ways: for example, the transformation of angles involves a
$2 \times 2$ matrix which is symmetric, so not completely general, see
equation (\ref{defA}). Alternatively, a general spin-weight 2 field
can be written in terms of second derivatives of a {\em complex}
potential, whereas the lensing potential is real.  As noticed below
equation (\ref{Poisson}) and in equation (\ref{Powerkappa2}), this also
implies that there are many other consistency relations which have to
hold if lensing is responsible for the observed shear field. In
practice the observed ellipticity field may not satisfy the expected
relations, if it is contaminated by distortions not associated with
weak lensing.  The most obvious of these is optical distortions of the
telescope system, but could also involve physical effects such as
intrinsic alignment of galaxy ellipticities, which we will consider in
Section~\ref{ch4:3Dweaklensing}.

\begin{figure}
\begin{center}
\epsfxsize=7.12 cm \epsfysize=7.4 cm {\epsfbox[0 -40 400 400]{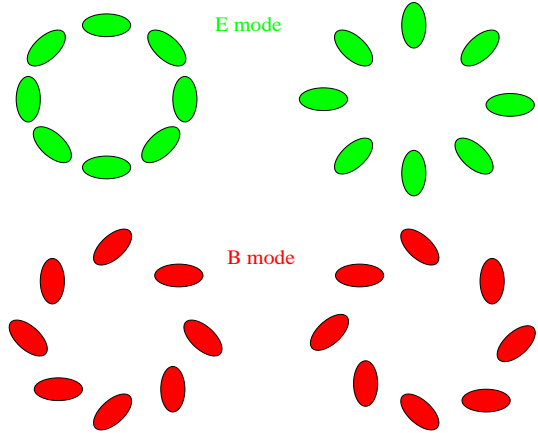}}
\end{center}
\caption{Illustrative E and B modes: the E modes show what is
expected around overdensities (left) and underdensities (right). The
B mode patterns should not be seen (from van Waerbeke \& Mellier
2003).}
\label{EBmodes}
\end{figure}

A convenient way to characterise the distortions is via E/B
decomposition, where the shear field is described in terms of an
`E-mode', which is allowed by weak lensing, and a `B-mode', which is
not.  These terms are borrowed from similar decompositions in
polarisation fields.  In fact weak lensing can generate B-modes, but
they are expected to be very small \cite{schneider2002a}, so the
existence of a significant B-mode in the observed shear pattern is
indicative of some non-lensing contamination.  Illustrative examples
of E- and B-modes are shown in Fig.\ref{EBmodes} (from \cite{LVWM2003}).
The easiest way to introduce a B-mode mathematically is to make the
lensing potential complex:
\begin{equation}
\phi=\phi_E+i\phi_B.
\end{equation}
There are various ways to determine whether a B-mode is present. A
neat way is to generalise the aperture mass to a complex
$M=\Map+iM_\perp$, where the real part picks up the E modes, and the
imaginary part the B modes.  Alternatively, the $\xi_\pm$ can be
used \cite{crittendenetal2002,schneider2002b}:
\begin{equation}
P_{\kappa\pm}(\ell) = \pi \int_0^\infty d\theta\,
\theta\,[J_0(\ell\theta)\xi_+(\theta)\pm
J_4(\ell\theta)\xi_-(\theta)]
\end{equation}
where the $\pm$ power spectra refer to E and B mode powers. In
principle this requires the correlation functions to be known over
all scales from $0$ to $\infty$. Variants of this
\cite{crittendenetal2002} allow the E/B-mode correlation functions to
be written in terms of integrals of $\xi_\pm$ over a finite range:
\begin{eqnarray}
\xi_E(\theta)&=&{1\over
2}\left[\xi_-(\theta)+\xi'_+(\theta)\right]\\\nonumber
\xi_B(\theta)&=&-\frac{1}{2}\left[\xi_-(\theta)-\xi'_+(\theta)\right],
\end{eqnarray}
where
\begin{equation}
\xi'_+(\theta)=\xi_+(\theta)+4\int_0^\theta
\frac{d\vt}{\vt}\xi_+(\vt) -12\theta^2\int_0^\theta
\frac{d\vt}{\vt^3}\xi_+(\vt).
\end{equation}
This avoids the need to know the correlation functions on large
scales, but needs the observed correlation functions to be
extrapolated to small scales; this was one of the approaches taken
in the analysis of the CFHTLS data \cite{hoekstra2006}. Difficulties
with estimating the correlation functions on small scales have led
others to prefer to extrapolate to large scales, such as in the
analysis of the GEMS \cite{heymans2004} and William Herschel data
\cite{massey2005}. Note that without full sky coverage, the
decomposition into E and B modes is ambiguous, although for scales
much smaller than the survey it is not an issue.

\subsection{Estimators and their covariance}
\label{Estimators_and_their_covariance}

The most common estimate of the cosmic shear comes from measuring
the ellipticities of individual galaxies.  We will consider the
practicalities in Section~\ref{ch:6}.  For weak gravitational
lensing, these estimates are very noisy, since the galaxies as a
population have intrinsic ellipticities $e_S$ with a dispersion of
about $0.4$, whereas the typical cosmic shear is around
$\gamma\simeq 0.01$. Therefore, one needs a large number $N$ of galaxies
to decrease the noise $\sim e_S/\sqrt{N}$ associated with these
intrinsic ellipticities (hence one needs to observe the more numerous
faint galaxies). The observed ellipticity is related to the
shear by \cite{schneiderseitz1995}
\begin{equation}
e_S = \frac{e-2g+g^2e^*}{1+g^2-2 {\rm Re}(ge^*)}
\label{eS_e}
\end{equation}
where $g=\gamma/(1-\kappa)$ is the reduced shear. Here we defined
the ellipticity $e$ such that for an elliptical image of axis ratio
$r<1$ we have:
\beq
|e|= \frac{1-r^2}{1+r^2} .
\label{def_e}
\eeq
Other definitions are also used in the literature such as $|e'|=(1-r)/(1+r)$,
see \cite{Schneider2005}. To linear order in $\gamma$ or $\kappa$,
we obtain from Eq.(\ref{eS_e}):
\beq
e\simeq e_S+2\gamma,
\label{e_2gamma}
\eeq
with a small
correction term when averaged.  $e$ is therefore dominated by the
intrinsic ellipticity, and many source galaxies are needed to get a
robust measurement of cosmic shear.  This results in estimators of
averaged quantities, such as the average shear in an aperture, or a
weighted average in the case of $\Map$.  Any analysis of these
quantities needs to take account of their noise properties, and more
generally in their covariance properties.  We will look only at a
couple of examples here; a more detailed discussion of covariance of
estimators, including non-linear cumulants, appears in
\cite{munshivalageas2005a}.

\subsubsection{Linear estimators}

Perhaps the simplest average statistic to use is the average (of
$N$) galaxy ellipticities in a 2D aperture on the sky:
\begin{equation}
\bar\gamma \equiv {1\over N}\sum_{i=1}^N {e_i\over 2}.
\end{equation}
The covariance of two of these estimators $\bar\gamma_\alpha$ and
$\bar\gamma_\beta^*$ is
\begin{equation}
\langle \bar\gamma_\alpha \bar\gamma_\beta^*\rangle = {1\over
4NM}\sum_{i=1}^N \sum_{j=1}^M \langle
(e_{Si}+2\gamma_i)(e_{Sj}^*+2\gamma_j^*)\rangle,
\end{equation}
where the apertures have $N$ and $M$ galaxies respectively.  If we
assume (almost certainly incorrectly; see Section~\ref{intrinsic})
that the source ellipticities are uncorrelated with each other, and
with the shear, then for distinct apertures the estimator is an
unbiased estimator of the shear correlation function averaged over
the pair separations.  If the apertures overlap, then this is not
the case. For example, in the shear variance, the apertures are the
same, and
\begin{equation}
\langle |\bar \gamma^2|\rangle = {1\over N^2}\sum_{i=1}^N
\sum_{j=1}^N \langle
\frac{|e_{Si}|^2}{4}\delta_{ij}+\gamma_i\gamma_j^*\rangle,
\end{equation}
which is dominated by the presence of the intrinsic ellipticity
variance, $\sigma_e^2 \equiv \langle |e_{S}|^2 \rangle \simeq
0.3^2-0.4^2$. The average shear therefore has a variance of
$\sigma_e^2/4N$. If we use the (quadratic) shear variance itself as a
statistic, then it is estimated by omitting the diagonal terms:
\begin{equation}
|\bar \gamma^2| = {1\over 4N(N-1)}\sum_{i=1}^N \sum_{j\ne i} e_i
e_j^*.
\end{equation}
For aperture masses (equation (\ref{Map}), the intrinsic ellipticity
distribution leads to a shot noise term from the finite number of
galaxies.  Again we simplify the discussion here by neglecting
correlations of source ellipticities. The shot noise can be
calculated by the standard method \cite{peebles1980} of dividing the
integration solid angle into cells $i$ of size $\Delta^2\theta_i$
containing $n_i=0$ or 1 galaxy:
\begin{equation}
\Map\simeq \sum_{i} \Delta^2\theta_i\,n_i\, Q(|\vtheta_i|)
\,(e_{Si}/2+\gamma_i)_t.
\end{equation}
Squaring and taking the ensemble average, noting that $\langle e_{Si,
t}e_{Sj,t}\rangle = \sigma_e^2 \delta_{ij}/2$, $n_i^2=n_i$, and
rewriting as a continuous integral gives
\begin{equation}
\langle \Map^2\rangle_{SN} = {\sigma_e^2\over 8}\int d^2\theta
\,Q^2(|\vtheta|).
\end{equation}
Shot noise terms for other statistics are calculated in similar
fashion.  In addition to the covariance from shot noise, there can
be signal covariance, for example from samples of different depths
in the same area of sky; both samples are affected by the lensing by
the common low-redshift foreground structure
\cite{munshivalageas2005b}.

\subsubsection{Quadratic estimators}

We have already seen how to estimate in an unbiased way the shear
variance.  The shear correlation functions can similarly be
estimated:
\begin{equation}
\hat\xi_\pm(\theta) = {\sum_{ij}w_i w_j (e_{it}e_{jt}\pm
e_{i\times}e_{j\times})\over 4\sum_{ij} w_i w_j}
\end{equation}
where the $w_i$ are arbitrary weights, and the sum extends over all
pairs of source galaxies with separations close to $\theta$.  Only in the
absence of intrinsic correlations, $\langle e_{it}e_{jt}\pm
e_{i\times}e_{j\times} \rangle = \sigma_e^2 \delta_{ij} +
4\xi_\pm(|\vtheta_i-\vtheta_j|)$, are these estimators unbiased.
The variance of the shear $\lag|\bargamma|^2\rag$ and of the aperture-mass
$\lag\Map^2\rag$ can also be obtained from the shear correlation functions
(as may be seen for instance from Eq.(\ref{Map_Pkappa})). This avoids the
need to place circular apertures on the sky which is hampered by the gaps
and holes encountered in actual weak lensing surveys.

As with any quadratic quantity, the covariance of these estimators
depends on the 4-point function of the source ellipticities and the
shear.  These expressions can be evaluated if the shear field is
assumed to be Gaussian, but the expressions for this (and the
squared aperture mass covariance) are too cumbersome to be given
here, so the reader is directed to \cite{schneider2002b}.
At small angular scales (below $\sim 10'$), which are sensitive to
the non-linear regime of gravitational clustering, the fields can no longer
be approximated as Gaussian and one must use numerical simulations to
calibrate the non-Gaussian contributions to the covariane, as described
in \cite{Semboloni07}.

In harmonic space, the convergence power spectrum may be estimated
from either $\xi_+$ or $\xi_-$ (or both), using Eq.~(\ref{chipm}).
From the orthonormality of the Bessel functions,
\begin{equation}
P_\kappa(\ell) = \int_0^\infty d\theta \, \theta \, \xi_\pm(\theta)
\,J_{0,4}(\ell\theta)\label{Powerkappa2}
\end{equation}
where the $0,4$ correspond to the $+/-$ cases.  In practice,
$\xi_\pm(\theta)$ is not known for all $\theta$, and the integral is
truncated on both small and large scales.  This can lead to
inaccuracies in the estimation of $P_\kappa(\ell)$ (see
\cite{schneider2002b}). An alternative method is to parametrise
$P_\kappa(\ell)$ in band-powers, and to use parameter estimation
techniques to estimate it from the shear correlation functions
\cite{huwhite,brown}.

\subsubsection{2-point statistics measurement}

\begin{figure}
\begin{center}
\epsfig{file=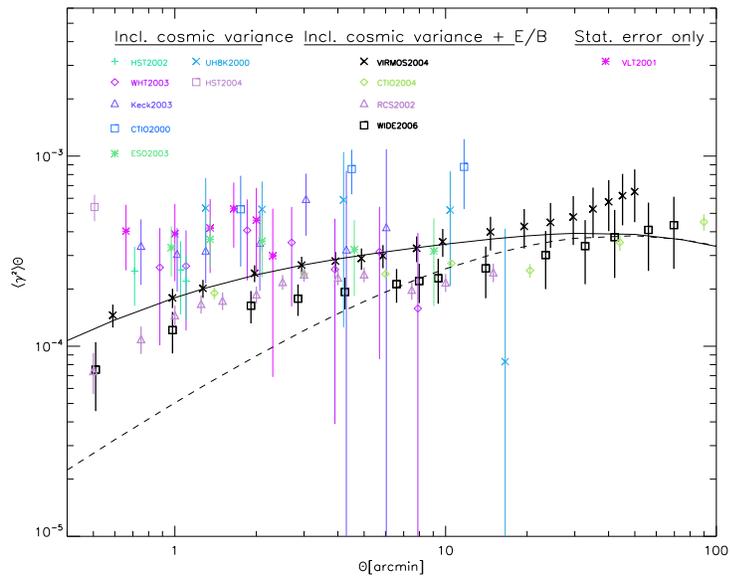,height=8cm,width=10cm}
\end{center}
\caption[]{Compilation of most of the shear measurements listed in
Table \ref{constraints}. The vertical axis is the shear top-hat
variance multiplied by the angular scale in arcminutes. The
horizontal axis is the radius of the smoothing window in arcminutes.
The positioning along the y-axis is only approximate given that the
different surveys have a slightly different source redshift
distribution. The RCS result (mean source redshift of $0.6$) was
rescaled to a mean source redshift of one.} \label{compileshear}
\end{figure}

Since the first measurements of weak lensing by large scale
structures
\cite{vwetal2000,Bacon_Refregier_Ellis2000,Wittman_et_al2000,Kaiser_Wilson_Luppino2000},
all ideas discussed above have been put in practice on real data.
Figure~\ref{compileshear} shows the measurement of shear top-hat variance
as function of scale. Some groups performed a E/B separation, which
lead to a more accurate measurement of residual systematics. Table
\ref{constraints} shows all measurements of the mass power spectrum
$\sigma_8$ to date. It is interesting to note, except for COMBO-17
\cite{heymans2004}, none of the measurements are using a source
redshift distribution obtained from the data, they all use the
Hubble Deep Fields with different prescriptions regarding galaxy
weighing. \cite{vwetal06} have shown that the Hubble Deep Field
photometric redshift distribution can lead to a source mean redshift
error of $\sim 10\%$ due to cosmic variance. The relative tension
between different measures of $\sigma_8$ in Table \ref{constraints}
comes in part from this problem, and also from an uncertainty
regarding how to treat the residual systematics, the $B$-mode, in
the cosmic shear signal. Recently, \cite{ilbert06} have released the
largest photometric redshift catalogue, obtained from the
CFHTLS-DEEP data. The most recent 2-point statitics analysis
involves the combination of this photometric redshift sample with
the largest weak lensing surveys described in Table
\ref{constraints}. In this analysis the relative tension is gone and
points towards a value of the power spectrum normalisation
$\sigma_8=0.75\pm0.05$ if one considers all possible issues with
uncertainties in the source redshift distribution \cite{jonben06}.
From Table \ref{constraints} and Figure \ref{compileshear} one can
say that the amplitude of weak lensing by large scale structure was
successfully measured and the main uncertainty remains the source
redshift distribution. The systematics due to the Point Spread
Function correction (see Section~\ref{ch:6}) seems to be much better
understood than a few years ago, and many promising techniques have
been proposed to solve it. The proper calibration of the redshift
distribution is likely to remain the major limitation for the use of
weak lensing in precision cosmology.

\begin{landscape}
\begin{table*}[t]
\begin{center}
{\small \caption{Reported constraints on the power spectrum
normalization ``$\sigma_8$" for $\Om=0.3$ for a flat Universe,
obtained from a given ``statistic" (from \cite{LVWM2003} and
extended). ``CosVar" tells us whether or not the cosmic variance has
been included, ``E/B" tells us whether or not a mode decomposition
has been used in the likelihood analysis. $z_s$ and $\Gamma$ are the
priors used for the different surveys identified with ``ID".
\label{constraints} } \label{tabcs}
\bigskip
\begin{tabular}{lcccccccc}\hline
\\
ID & $\sigma_8$ $(\Omega=0.3)$ & Statistic & Field & $m_{\rm lim}$& CosVar & E/B & $z_s$ & $\Gamma$ \\
\hline
Maoli  et al. 01 & $1.03\pm 0.05$ & $\langle\gamma^2\rangle$ & VLT+CTIO+WHT+CFHT & - & no & no & - & 0.21 \\
\hline
Van Waerbeke et al. 01& $0.88\pm 0.11$ & $\langle\gamma^2\rangle$, $\xi(r)$, $\langle M_{\rm ap}^2\rangle$  & CFHT 8 sq.deg.& I=24.5 & no & no (yes) & 1.1 & 0.21 \\
\hline
Rhodes  et al. 01& $0.91^{+0.25}_{-0.29}$ & $\xi(r)$ & HST 0.05 sq.deg.& I=26 & yes & no & 0.9-1.1 & 0.25 \\
\hline
Hoekstra et al. 02& $0.81\pm 0.08$ & $\langle\gamma^2\rangle$ & CFHT+CTIO 24 sq.deg. & R=24 & yes & no & 0.55 & 0.21 \\
\hline
Bacon et al. 03& $0.97\pm 0.13$ & $\xi(r)$ & Keck+WHT 1.6 sq.deg. & R=25 & yes & no & 0.7-0.9 & 0.21 \\
\hline
R\'efr\'egier et al. 02 & $0.94\pm 0.17$ & $\langle\gamma^2\rangle$ & HST 0.36 sq.deg. & I=23.5 & yes & no & 0.8-1.0 & 0.21 \\
\hline
Van Waerbeke et al. 02& $0.94\pm 0.12$ & $\langle M_{\rm ap}^2\rangle$ & CFHT 12 sq.deg. & I=24.5 & yes & yes & 0.78-1.08 & 0.1-0.4 \\
\hline
Hoekstra et al. 02& $0.91^{+0.05}_{-0.12}$ & $\langle\gamma^2\rangle$, $\xi(r)$, $\langle M_{\rm ap}^2\rangle $ & CFHT+CTIO 53 sq.deg. & R=24 & yes & yes & 0.54-0.66 & 0.05-0.5 \\
\hline
Brown et al. 03&  $0.74\pm 0.09$ & $\langle\gamma^2\rangle$, $\xi(r)$  & COMBO17 1.25 sq.deg.& R=25.5 & yes & no (yes) & 0.8-0.9 & - \\
\hline
Hamana et al. 03 & $(2\sigma) 0.69^{+0.35}_{-0.25}$ & $\langle M_{\rm ap}^2\rangle$, $\xi(r)$  & Subaru 2.1 sq.deg. & R=26 & yes & yes & 0.8-1.4 & 0.1-0.4 \\
\hline
Jarvis et al. 03 & $(2\sigma) 0.71^{+0.12}_{-0.16}$ & $\langle\gamma^2\rangle$, $\xi(r)$, $\langle M_{\rm ap}^2\rangle$  & CTIO 75 sq.deg. & R=23 & yes & yes & 0.66 & 0.15-0.5 \\
\hline
Rhodes et al. 04&  $1.02\pm 0.16$ & $\langle\gamma^2\rangle$, $\xi(r)$  & STIS 0.25 sq.deg.& $\langle I\rangle=24.8$ & yes & no & 1.0 $\pm$ 0.1 & - \\
\hline
Heymans et al. 05&  $0.68\pm 0.13$ & $\langle\gamma^2\rangle$, $\xi(r)$  & GEMS 0.3 sq.deg.& $\langle m_{606}\rangle=25.6$ & yes & no (yes) & $\sim 1$ & - \\
\hline
Massey et al. 05&  $1.02\pm 0.15$ & $\langle\gamma^2\rangle$, $\xi(r)$  & WHT 4 sq.deg.& R=25.8 & yes & yes & $\sim$ 0.8 & - \\
\hline
Van Waerbeke et al. 05&  $0.83\pm 0.07$ & $\langle\gamma^2\rangle$, $\xi(r)$  & CFHT 12 sq.deg.& I=24.5 & yes & no (yes) & 0.9 $\pm $ 0.1 & 0.1-0.3 \\
\hline
Heitterscheidt et al. 06&  $0.8\pm 0.1$ & $\langle\gamma^2\rangle$, $\xi(r)$  & GaBoDS 13 sq.deg.& R=[21.5,24.5] & yes & yes & $\sim$ 0.78 &  $h \in [0.63,0.77]$ \\
\hline
Semboloni et al. 06 & $0.90 \pm 0.14$ & $\langle M_{\rm ap}^2\rangle$, $\xi(r)$  & CFHTLS-DEEP 2.3 sq.deg. & i=25.5 & yes & yes & $\sim 1$ & $\Gamma =\Omega h$ \\
\hline
Hoekstra et al. 06 & $0.85 \pm 0.06$ & $\langle\gamma^2\rangle$, $\xi(r)$, $\langle M_{\rm ap}^2\rangle$  & CFHTLS-WIDE 22 sq.deg. & i=24.5 & yes & yes & 0.8 $\pm $0.1 & $\Gamma =\Omega h$ \\
\hline
\end{tabular}
}
\end{center}
\end{table*}
\end{landscape}

\subsection{Mass Reconstruction}

The problem of mass reconstruction is a central topic in weak
lensing. Historically this is because the early measurements of weak
gravitational lensing were obtained in clusters of galaxies, and
this led to the very first maps of dark matter
\cite{BM1995,fahlman94}. These maps were the very first
demonstrations that we could {\it see} the dark side of the Universe
without any assumption regarding the light-mass relation, which, of
course, was a major breakthrough in our exploration of the Universe.
Reconstructing mass maps is also the only way to perform a complete
comparison of the dark matter distribution to the Universe as seen
in other wavelengths. For these reasons, mass reconstruction is also
part of the shear measurement process. A recent example of the power
of mass reconstruction is shown by the {\it bullet cluster}
\cite{clowe2006,bradac2006}, which clearly indicates the presence of
dark matter at a location different from where most of the baryons are.
This is a clear demonstration that, at least for the extreme cases
where light and baryons do not trace the mass, weak lensing is the
only method that can probe the matter distribution.

A mass map is a convergence, $\kappa$ map (projected mass), which
can be reconstructed from the shear field $\gamma_i$:

\begin{equation}
\kappa=\frac{1}{2}\left(\partial^2_x +\partial^2_y \right) \phi \ \
; \ \ \gamma_1=\frac{1}{2}\left(\partial^2_x -\partial^2_y
\right)\phi \ \ ; \ \ \gamma_2=\partial_x\partial_y \phi,
\end{equation}
where $\phi$ is the projected gravitational potential
\cite{Kaiser1992}, see Eqs.(\ref{Phi})-(\ref{defkappagam}).
Assuming that the reduced shear and shear are
equal to first approximation, i.e. $g_i\simeq \gamma_i$ (which is
true only in the weak lensing regime when $|\gamma | \ll$ 1 and
$\kappa \ll$ 1), \cite{KS93} have shown that the Fourier transform
of the smoothed convergence $\barkappa(\vell)$ can be obtained from the
Fourier transform of the smoothed shear map $\bargamma(\vell)$:
\begin{equation}
\barkappa(\vell)= \frac{\ell_x^2+\ell_y^2}{\ell_x^2-\ell_y^2+2i\ell_x\ell_y}
 \bargamma (\vell).
\label{kappa_F_gamma}
\end{equation}
This follows from Eq.(\ref{gammal}) if smoothing is a mere convolution
as in Eq.(\ref{Xx}) which writes in Fourier space as a mere product,
see Eq.(\ref{WX}).
This relation explicitly shows that mass reconstruction must be
performed with a given smoothing window, otherwise the variance of
the mass map becomes infinite \cite{KS93}. Indeed, the random galaxy
intrinsic ellipticities $e_{Si}$ introduce a white noise which gives
a large-$\ell$ divergence when we transform back to real space for
the variance $\lag\barkappa^2\rag_c$. The Fourier transform
method is a fast $N\log N$ process, but the non-linear regions
$\kappa\sim\gamma_i\sim 1$ are not accurately reconstructed. A
likelihood reconstruction method works in the intermediate and
strong lensing regimes \cite{bartelmann1996}. A $\chi^2$ function of
the reduced shear $g_i$ is minimised by finding the best
gravitational potential $\phi_{ij}$ calculated on a grid $ij$:

\begin{equation}
\chi^2=\sum_{ij} \left[\left| g^{\rm obs}_{ij}-g^{\rm
guess}_{ij}(\phi_{ij})\right|^2 \right].
\end{equation}
%

\begin{figure}
\begin{center}
\epsfig{file=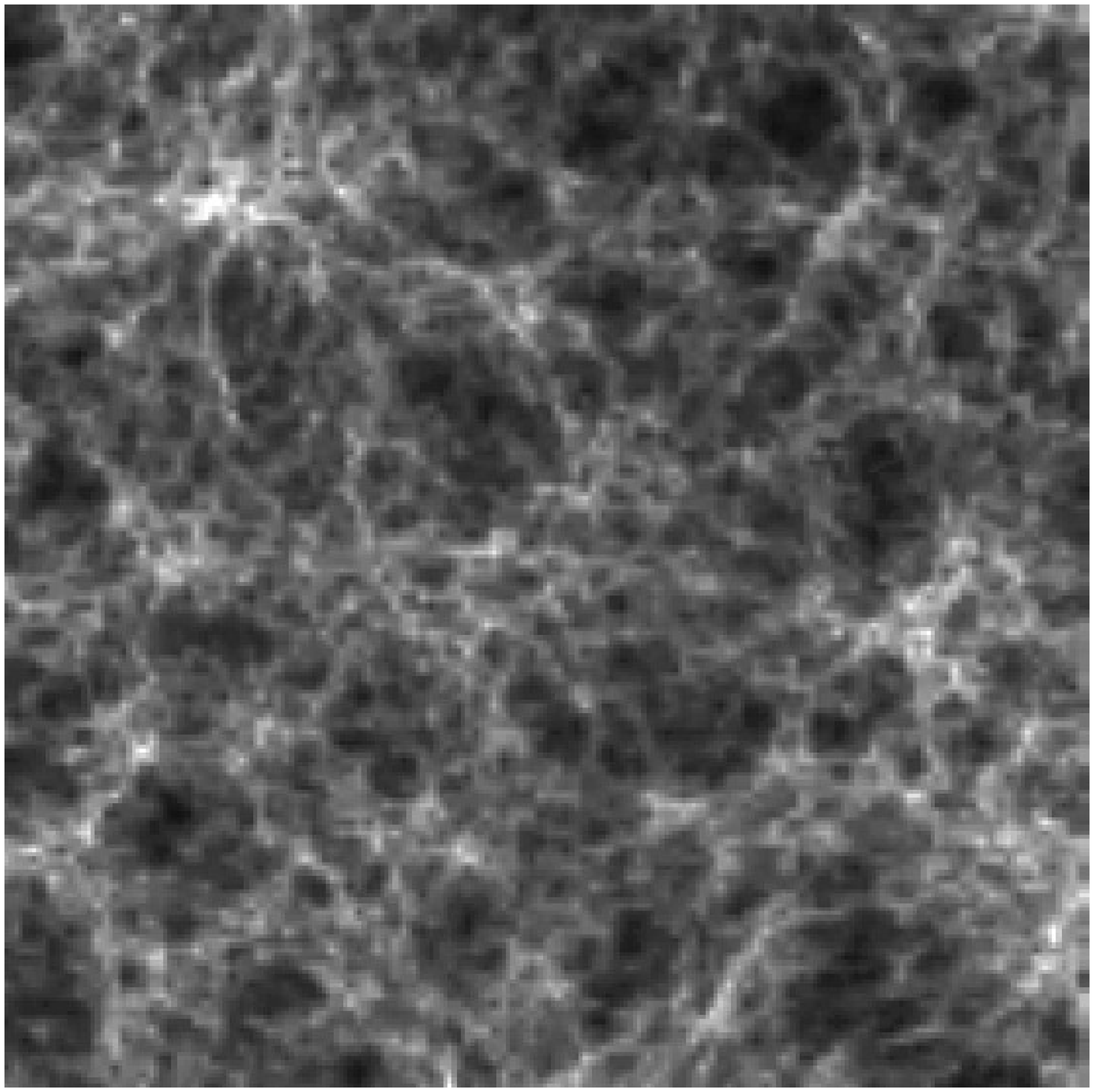,height=7cm,width=7cm}
\epsfig{file=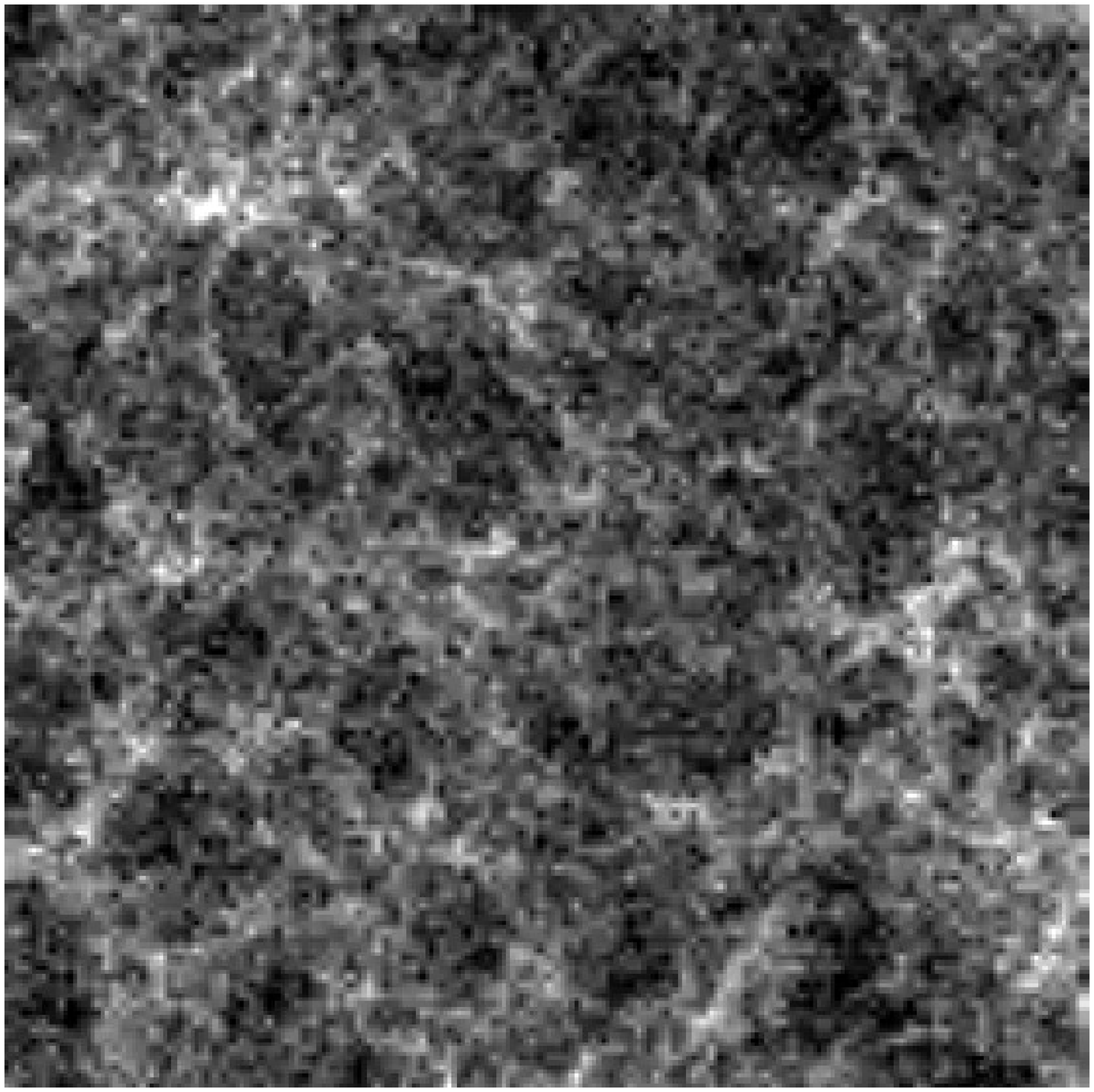,height=7cm,width=7cm}
\end{center}
\caption[]{Left: simulated noise-free $\kappa$ map. Field-of-view is
$49$ square degrees in LCDM cosmology. Right: reconstructed $\kappa$
field with realistic noise level (give details). Van Waerbeke et al.
(1999) have shown that two and three-point statistics can be
accurately measured from such mass reconstructions.}
\label{massrecon}
\end{figure}

The shot noise of the reconstructed mass map depends on the
smoothing window, the intrinsic ellipticity of the galaxies and the
number density of galaxies \cite{vw2000}. The two methods outlined
above provide an accurate description of the shot noise: it was
shown \cite{vw1999} that two and three-points statistics can be
measured accurately from reconstructed mass maps using these methods
(see Figure \ref{massrecon}). Important for cluster lensing, the
non-linear version of \cite{KS93} has been developed in \cite{KS96},
and \cite{schneiderseitz1995} have developed an alternative which
also conserves the statistical properties of the noise. The
advantage of a reconstruction method that leaves intact the shot
noise is that a statistical analysis of the mass map is relatively
straightforward (e.g. the peak statistics in \cite{JVW2000}). Mass
reconstruction has proven to be reasonably successful in blind
cluster searches \cite{miya2002,clowe2006b,GS2006}
A radically different approach in map making consists in reducing
the noise in order to identify the highest signal-to-noise peaks.
Such an approach has been developed by
\cite{SSB98,bridle98,marshall}. More recently \cite{starck} proposed
a wavelet approach, where the size of the smoothing kernel is
optimized as a function of the local noise amplitude. An application
of this method on the COSMOS data for cluster detection is shown in
\cite{massey2006}.

Mass maps are essential for some specific cosmological studies such
as morphology analysis like Minkowski functionals and Euler
characteristics \cite{sch98} and for global statistics (Probability
distribution function of the convergence, e.g. \cite{ZP06}. The
reconstruction processes is non local, and this is why it is
difficult to have a perfect control of the error propagation and
systematics in the $\kappa$ maps. In particular, note that one can only
reconstruct the convergence $\kappa$ up to a constant $\kappa_0$, since
Eq.(\ref{kappa_F_gamma}) is undetermined for $\ell=0$.
Indeed, a constant matter surface density in the lens plane does not create
any shear (since it selects no prefered direction) but it leads to a non-zero
constant convergence (which may only be eliminated if we observe a wide-enough
field where the mean convergence should vanish, or use complementary
information such as number counts which are affected by the associated
magnification). This is the well-known
mass-sheet degeneracy. The aperture mass statistics $\Map$
\cite{K94,schneideretal1998} introduced in Eq.(\ref{Map})
has been invented to enforce locality
of the mass reconstruction, therefore it might provide an
alternative to the inversion problem, although it is not yet clear
that it can achieve a signal-to-noise as good as top-hat or gaussian
smoothing windows.

The aperture mass statistic can also be used to provide unbiased
estimates of the power spectrum \cite{BS99}, galaxy biasing
\cite{vw98,sch98}, high order statistics
\cite{schneideretal1998,vw2001}, and peak statistics
\cite{Hetal2006}.

\section{\label{ch4:3Dweaklensing}3D Weak Lensing}

\subsection{What is 3D weak lensing?}

The way in which weak lensing surveys have been analysed to date has
been to look for correlations of shapes of galaxies on the sky; this
can be done even if there is no distance information available for
individual sources.  However, as we have seen, the interpretation of
observed correlations depends on where the imaged galaxies are: the
more distant they are, the greater the correlation of the images.
One therefore needs to know the statistical distribution of the
source galaxies, and ignorance of this can lead to relatively large
errors in recovered parameters.  In order to rectify this, most
lensing surveys obtain multi-colour photometry of the sources, from
which one can estimate their redshifts.  These `photometric
redshifts' are not as accurate as spectroscopic redshifts, but the
typical depth of survey required by lensing surveys makes
spectroscopy an impractical option for large numbers of sources.  3D
weak lensing uses the distance information of {\em individual}
sources, rather than just the {\em distribution} of distances. If
one has an estimate of the distance to each source galaxy, then one
can utilise this information and investigate lensing in three
dimensions. Essentially one has an estimate of the shear field at a
number of discrete locations in 3D.

There are several ways 3D information can be used: one is to
reconstruct the 3D gravitational potential or the overdensity field
from 3D lensing data.  We will look at this in Section~\ref{3Dmass}.
The second is to exploit the additional statistical power of 3D
information, firstly by dividing the sources into a number of shells
based on estimated redshifts. One then essentially performs a
standard lensing analysis on each shell, but exploits the extra
information from cross-correlations between shells. This sort of
analysis is commonly referred to as tomography, and we explore this
in Section~\ref{tomography}, and in Section~\ref{shearRatio}, where
one uses ratios of shears behind clusters of galaxies. Finally, one
can perform a fully-3D analysis of the estimated shear field. Each
approach has its merits. We cover 3D statistical analysis in Section
\ref{3Danalysis}. At the end of Section~\ref{intrinsic}, we
investigate how photometric redshifts can remove a potentially
important physical systematic: the intrinsic alignment of galaxies,
which could be wrongly interpreted as a shear signal. Finally, in
Section~\ref{shearintrinsic}, we consider a potentially very
important systematic error arising from a correlation between cosmic
shear and the intrinsic alignment of foreground galaxies, which
could arise if the latter responds to the local tidal gravitational
field which is partly responsible for the shear.

\subsection{\label{3Dmass}3D potential and mass reconstruction}

As we have already seen, it is possible to reconstruct the surface
density of a lens system by analysing the shear pattern of galaxies
in the background.  An interesting question is then whether the
reconstruction can be done in three dimensions, when distance
information is available for the sources.   It is probably
self-evident that mass distributions can be {\em constrained} by the
shear pattern, but the more interesting possibility is that one may
be able to {\em determine} the 3D mass density in an essentially
non-parametric way from the shear data.

The idea \cite{taylor2001} is that the shear pattern is derivable
from the lensing potential $\phi({\bf r})$, which is dependent on
the gravitational potential $\Phi({\bf r})$ through the integral
equation
\begin{equation}
\phi({\bf r}) = {2\over c^2}\int_0^r \,dr' \left({1\over r'}-{1\over
r}\right) \Phi({\bf r'})
\label{phi}
\end{equation}
where the integral is understood to be along a radial path (the Born
approximation), and a flat Universe is assumed in equation
(\ref{phi}).  In this Section we work with spherical coordinates
(we do not use the flat-sky approximation except where explicitely
noticed) and we note ${\bf r}=r \hat{\bf n}$ the position on the line of sight
in the direction $\hat{\bf n}$, at the comoving radial distance $r$ (which
we also noted $\chirad$ in Section~\ref{ch:2}).
The gravitational potential is related to the density
field via Poisson's equation (\ref{Poisson}).  There are two problems
to solve here; one is to construct $\phi$ from the lensing data, the
second is to invert equation (\ref{phi}).  The second problem is
straightforward: the solution is
\begin{equation}
\Phi({\bf r}) = {c^2\over 2}{\partial\over \partial
r}\left[r^2{\partial\over \partial r} \phi({\bf r})\right].
\end{equation}
From this and Poisson's equation $\nabla^2\Phi = (3/2)H_0^2\Om
\delta /a(t)$, we can reconstruct the mass overdensity field
\begin{equation}
\delta({\bf r}) = {a(t)c^2\over 3 H_0^2 \Om}
\nabla^2\left\{{\partial\over
\partial r}\left[r^2{\partial\over \partial r} \phi({\bf r})\right]\right\}.
\label{deltarphir}
\end{equation}
The construction of $\phi$ is more tricky, as it is not directly
observable, but must be estimated from the shear field. This
reconstruction of the lensing potential suffers from a similar
ambiguity to the mass-sheet degeneracy for simple lenses. To see
how, we first note that the complex shear field $\gamma$ is the
second derivative of the lensing potential (Eq.(\ref{defkappagam})):
\begin{equation}
\gamma({\r}) = \left[{1\over 2}\left({\partial^2\over \partial x^2}-
{\partial^2\over \partial y^2}\right) + i {\partial^2\over
\partial x \partial y}\right]\phi({\bf r}).
\end{equation}
As a consequence, since the lensing potential is real, its estimate
is ambiguous up to the addition of any field $f({\bf r})$ for which
\begin{equation}
{\partial^2 f(\r)\over \partial x^2}- {\partial^2 f(\r)\over
\partial y^2} = {\partial^2 f(\r)\over
\partial x \partial y} = 0.
\end{equation}
Since $\phi$ must be real, the general solution to this is
\begin{equation}
f(\r) = F(r) + G(r)x + H(r)y + P(r)(x^2 + y^2)
\end{equation}
where $F$, $G$, $H$ and $P$ are arbitrary functions of $r\equiv
|\r|$.  Assuming these functions vary smoothly with $r$, only the
last of these survives at a significant level to the mass density
(since the 3D Laplacian $\nabla^2$ in Eq.(\ref{deltarphir}) is 
dominated by the 2D Laplacian $\pl_x^2+\pl_y^2$ as for Eq.(\ref{kappa})),
and it corresponds to a sheet of overdensity
\begin{equation}
\delta = {4 a(t) c^2\over 3H_0^2 \Omega_m r^2} {\partial\over
\partial r} \left[r^2
{\partial\over
\partial r}P(r)\right].
\end{equation}
There are a couple of ways to deal with this problem.  For a
reasonably large survey, one can assume that the potential and its
derivatives are zero on average, at each $r$, or that the
overdensity has average value zero.  For further details, see
\cite{BT03}. Note that the relationship between the overdensity
field and the lensing potential is a linear one, so if one chooses a
discrete binning of the quantities, one can use standard linear
algebra methods to attempt an inversion, subject to some constraints
such as minimising the expected reconstruction errors.  With prior
knowledge of the signal properties, this is the Wiener filter.  See
\cite{HK03} for further details of this approach.

\subsection{\label{tomography}Tomography}

In the case where one has distance information for individual
sources, it makes sense to employ the information for statistical
studies.  A natural course of action is to divide the survey into
slices at different distances, and perform a study of the shear
pattern on each slice.  In order to use the information effectively,
it is necessary to look at cross-correlations of the shear fields in
the slices, as well as correlations within each slice \cite{hu1999}.
This procedure is usually referred to as tomography, although the
term does not seem entirely appropriate.

We start by considering the average shear in a shell, which is
characterised by a probability distribution for the source redshifts
$z=z(r)$,  $p(z)$.  We also take the opportunity to introduce the
more complicated {\em edth} derivative ($\edth$) on the curved sky,
which is required if the survey does not subtend a small angle on
the sky \cite{castro2005}. We shall not use make use of this much in
this review, but it is included for completeness.  The shear is the
second edth derivative of the lensing potential ,
\begin{equation}
\gamma({\bf r}) = {1\over 2}\edth\edth \phi({\bf r}) \simeq {1\over
2}(\partial_x+i\partial_y)^2\phi({\bf r})
\end{equation}
where the last equality holds in the flat-sky limit. If we average
the shear in a shell, giving equal weight to each galaxy, then the
average shear can be written in terms of an effective lensing
potential
\begin{equation}
\phi_{\rm eff}({\vtheta}) = \int_0^\infty dz\, p(z) \phi({\bf r})
\end{equation}
where the integral is at fixed ${\vtheta}$, and $p(z)$ is zero
outside the slice (we ignore errors in distance estimates such as
photometric redshifts; these could be incorporated with a suitable
modification to $p(z)$).  In terms of the gravitational potential,
the effective lensing potential is
\begin{equation}
\phi_{\rm eff}({\vtheta}) = {2\over c^2}\int_0^\infty dr
\,\Phi({\bf r}) g(r)
\end{equation}
where reversal of the order of integration gives the lensing
efficiency to be
\begin{equation}
g(r) = \int_{z(r)}^\infty dz'\, p(z') \left({1\over r}-{1\over
r'}\right),
\end{equation}
where $z'=z'(r')$ and we assume flat space.  If we perform a
spherical harmonic transform of the effective potentials for slices
$i$ and $j$, then the cross power spectrum can be related to the
power spectrum of the gravitational potential $P_\Phi(k)$ via a
version of Limber's equation:
\begin{equation}
\langle \phi^{(i)}_{\ell m}\phi^{*(j)}_{\ell' m'}\rangle =
C^{\phi\phi}_{\ell,ij}\, \delta_{\ell'\ell}\delta_{m'm}
\end{equation}
where
\begin{equation}
C^{\phi\phi}_{\ell,ij}=\left({2\over c^2}\right)^2 \int_0^\infty
dr\,\, {g^{(i)}(r)g^{(j)}(r)\over r^2}\, P_\Phi(\ell/r;r)
\end{equation}
is the cross power spectrum of the lensing potentials. The last
argument in $P_\Phi$ allows for evolution of the power spectrum with
time, or equivalently distance. The power spectra of the convergence
and shear are related to $C^{\phi\phi}_{\ell,ij}$ by \cite{hu2000}
\begin{eqnarray}
C^{\kappa\kappa}_{\ell,ij}&=&{\ell^2(\ell+1)^2\over
4}\,C^{\phi\phi}_{\ell,ij}\\\nonumber
C^{\gamma\gamma}_{\ell,ij}&=&{1\over 4}{(\ell+2)!\over
(\ell-2)!}\,C^{\phi\phi}_{\ell,ij}.
\end{eqnarray}
The sensitivity of the cross power spectra to cosmological
parameters is through various effects, as in 2D lensing: the shape
of the linear gravitational potential power spectrum is dependent on
some parameters, as is its nonlinear evolution; in addition the
$z(r)$ relation probes cosmology.  The reader is referred to
standard cosmological texts for more details of the dependence of
the distance-redshift relation on cosmological parameters.

\begin{figure}
\begin{center}
\epsfxsize=10.12 cm \epsfysize=8. cm {\epsfbox[22 284 590 684]{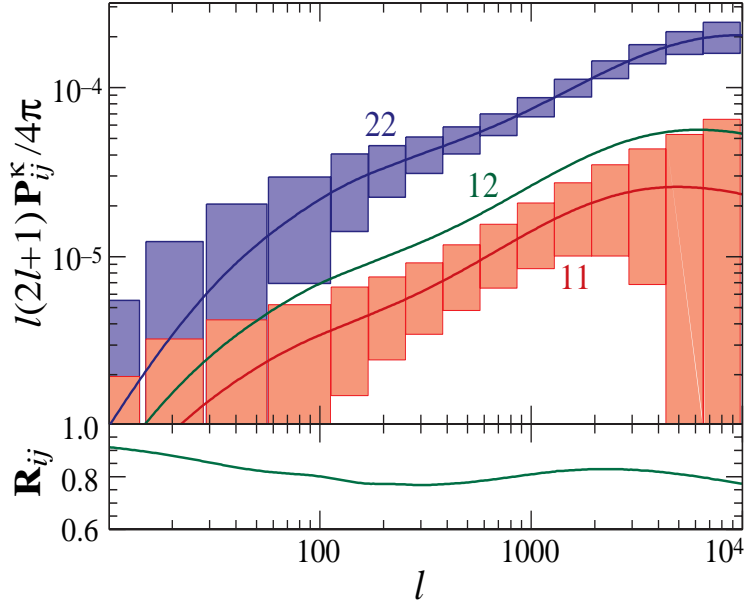}}
\end{center}
\caption{The power spectra of two slices, their cross power
spectrum, and their correlation coefficient. The galaxy population 
is split into two bins across a median redshift $z_{\rm median}=1$.
From Hu (1999).}\label{Hutomography}
\end{figure}
\begin{figure}
\begin{center}
\epsfxsize=10.12 cm \epsfysize=8. cm {\epsfbox[22 402 590 713]{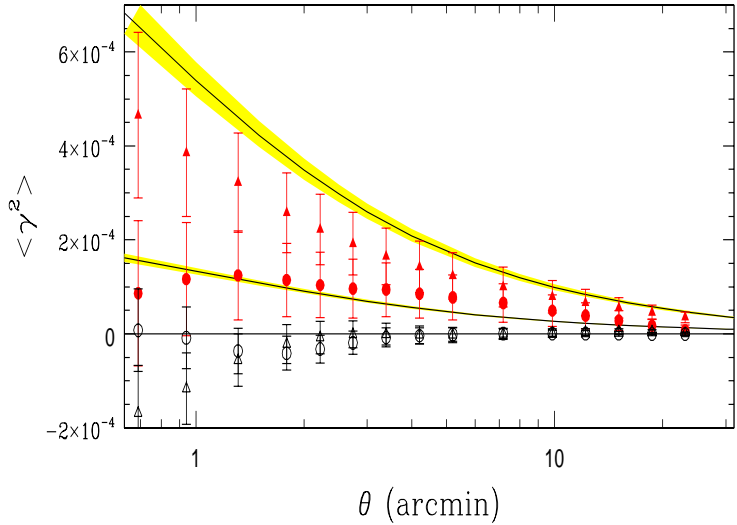}}
\end{center}
\caption{Tomographic shear measurement on the CFHTLS-deep (from
Semboloni et al 2006). Yellow lines show the $\pm 1 \sigma$ contours
around high and low redshift lensed galaxies. The filled triangles
and circles show the E-mode and the empty symbols the
B-mode.}\label{CFHTLStomography}
\end{figure}

Ref.\cite{hu1999} illustrates the power and limitation of tomography,
with two shells (Fig. \ref{Hutomography}).  As expected, the deeper
shell (2) has a larger lensing power spectrum than the nearby shell
(1), but it is no surprise to find that the power spectra from
shells are correlated, since the light from both passes through some
common material.  Thus one does gain from tomography, but, depending
on what one wants to measure, the gains may or may not be very much.
For example, tomography adds rather little to the accuracy of the
amplitude of the power spectrum, but far more to studies of dark
energy properties.  One also needs to worry about systematic
effects, as leakage of galaxies from one shell to another, through
noisy or biased photometric redshifts, can degrade the accuracy of
parameter estimation \cite{huterer2006,ma2006}. Figure
\ref{CFHTLStomography} shows the first tentative of tomographic
measurement using photometric redshift on the CFHTLS deep data.

\subsection{\label{shearRatio}The Shear Ratio test}

The shear contributed by the general large-scale structure is
typically about 1\%, but the shear behind a cluster of galaxies can
far exceed this.  As always, the shear of a background source is
dependent on its redshift, and on cosmology, but also on the mass
distribution in the cluster.  This can be difficult to model, so it
is attractive to consider methods which are decoupled from the
details of the mass distribution of the cluster.  Various methods
have been proposed \cite{JT03,BJ04,ZHS05}. The method currently
receiving the most attention is simply to take ratios of average
tangential shear in different redshift slices for sources behind the
cluster.

The amplitude of the induced tangential shear is dependent on the
source redshift $z_s$, and on cosmology via the angular diameter
distance-redshift relation $\De[\chirad(z_s)]$ by \cite{tayloretal2006}:
\begin{equation}
\gamma_t (z) = \gamma_t(z=\infty){\De[\chirad(z_s)-\chirad(z_l)] \over 
\De[\chirad(z_s)]},
\end{equation}
where $\gamma_{t,\infty}$ is the shear which a galaxy at infinite
distance would experience, and which characterises the strength of
the distortions induced by the cluster, at redshift $z_l$.
Evidently, we can neatly eliminate the cluster details by taking
ratios of tangential shears, for pairs of shells in source redshift:
\begin{equation}
R_{ij} \equiv {\gamma_{t,i}\over \gamma_{t,j}}
={\De[\chirad(z_j)]\,\De[\chirad(z_i)-\chirad(z_l)]\over
\De[\chirad(z_i)]\,\De[\chirad(z_j)-\chirad(z_l)]}.
\label{Rij}
\end{equation}
In reality, the light from the more distant shell passes through an
extra pathlength of clumpy matter, so suffers an additional source
of shear.  This can be treated as a noise term
\cite{tayloretal2006}. This approach is attractive in that it probes
cosmology through the distance-redshift relation alone, being (at
least to good approximation) independent of the growth rate of the
fluctuations. Its dependence on cosmological parameters is therefore
rather simpler, as many parameters (such as the amplitude of matter
fluctuations) do not affect the ratio except through minor
side-effects.  More significantly, it can be used in conjunction
with lensing methods which probe both the distance-redshift relation
and the growth-rate of structure. Such a dual approach can in
principle distinguish between quintessence-type dark energy models
and modifications of Einstein gravity.  This possibility arises
because the effect on global properties (e.g. $z(\chirad)$) is different
from the effect on perturbed quantities (e.g. the growth rate of the
power spectrum) in the two cases.  The method has a signal-to-noise
which is limited by the finite number of clusters which are massive
enough to have measurable tangential shear.  In an all-sky survey,
the bulk of the signal would come from the $10^5-10^6$ clusters
above a mass limit of $10^{14}M_\odot$.

\subsection{\label{3Danalysis}Full 3D analysis of the shear field}

An alternative approach to take is to recognise that, with
photometric redshift estimates for individual sources, the data one
is working with is a very noisy 3D shear field, which is sampled at
a number of discrete locations, and for whom the locations are
somewhat imprecisely known.  It makes some sense, therefore, to deal
with the data one has, and to compare the statistics of the discrete
3D field with theoretical predictions.  This was the approach of
\cite{H03,castro2005,Hetal06}. It should yield smaller statistical
errors than tomography, as it avoids the binning process which loses
information.

In common with many other methods, one has to make a decision
whether to analyse the data in configuration space or in the
spectral domain.  The former, usually studied via correlation
functions, is advantageous for complex survey geometries, where the
convolution with a complex window function implicit in spectral
methods is avoided.  However, the more readily computed correlation
properties of a spectral analysis are a definite advantage for
Bayesian parameter estimation, and we follow that approach here.

The natural expansion of a 3D scalar field which
is derived from a potential is in terms of products of spherical
harmonics and spherical Bessel functions, $j_\ell(kr)
Y_\ell^m(\theta,\varphi)$, using the spherical coordinates 
$(r,\theta,\varphi)$.  Such products, characterised by 3 spectral
parameters $(k,\ell,m)$, are eigenfunctions of the Laplace operator,
thus making it very easy to relate the expansion coefficients of the
density field to that of the potential (essentially via $-k^2$ from
the $\nabla^2$ operator). Similarly, the 3D expansion of the lensing
potential,
\begin{equation}
\phi_{\ell m}(k) \equiv \sqrt{2\over \pi} \int d^3{\bf r}\,
\phi({\bf r}) k j_\ell(kr) Y_\ell^m(\theta,\varphi),
\end{equation}
where the prefactor and the factor of $k$ are introduced for
convenience. The expansion of the complex shear field is most
naturally made in terms of spin-weight 2 spherical harmonics
$\phantom{.}_2Y_\ell^m$ and spherical Bessel functions, since
$\gamma={1\over 2}\edth\edth\phi$, and $\edth\edth Y_\ell^m \propto
\,\phantom{.}_2Y_\ell^m$:
\begin{equation}
\gamma({\bf r})=\sqrt{2\over \pi}\sum_{\ell m}\int dk\, \gamma_{\ell
m} \,k\,j_\ell(kr)\, \phantom{.}_2Y_\ell^m(\theta,\varphi).
\end{equation}
The choice of the expansion becomes clear when we see that the
coefficients of the shear field are related very simply to those of
the lensing potential:
\begin{equation}
\gamma_{\ell m}(k) = {1\over 2}\sqrt{(\ell+2)!\over (\ell-2)!} \,\,
\phi_{\ell m}(k).
\end{equation}
The relation of the $\phi_{\ell m}(k)$ coefficients to the expansion
of the density field is readily computed, but more complicated as
the lensing potential is a weighted integral of the gravitational
potential.  The details will not be given here, but relevant effects
such as photometric redshift errors, nonlinear evolution of the
power spectrum, and the discreteness of the sampling are easily
included. The reader is referred to the original papers for details
\cite{H03,castro2005,Hetal2006}.

In this way the correlation properties of the $\gamma_{\ell m}(k)$
coefficients can be related to an integral over the power spectrum,
involving the $z(r)$ relation, so cosmological parameters can be
estimated via standard Bayesian methods from the coefficients.
Clearly, this method probes the dark energy via both the growth rate
and the $z(r)$ relation.  The method has recently been applied for
the first time, to the COMBO-17 survey\cite{kitching2006}.

\subsection{\label{forecasts}Parameter forecasts from 3D lensing methods}

In this Section we summarise some of the forecasts for cosmological
parameter estimation from 3D weak lensing.  We concentrate on the
statistical errors which should be achievable with the shear ratio
test and with the 3D power spectrum techniques.  Tomography should
be similar to the latter.  We show results from 3D weak lensing
alone, as well as in combination with other experiments.  These
include CMB, supernova and baryon oscillation studies.  The methods
generally differ in the parameters which they constrain well, but
also in terms of the degeneracies inherent in the techniques.  Using
more than one technique can be very effective at lifting the
degeneracies, and very accurate determinations of cosmological
parameters, in particular dark energy properties, may be achievable
with 3D cosmic shear surveys covering thousands of square degrees of
sky to median source redshifts of order unity.

\begin{figure}
\includegraphics[width=110mm, angle=0]{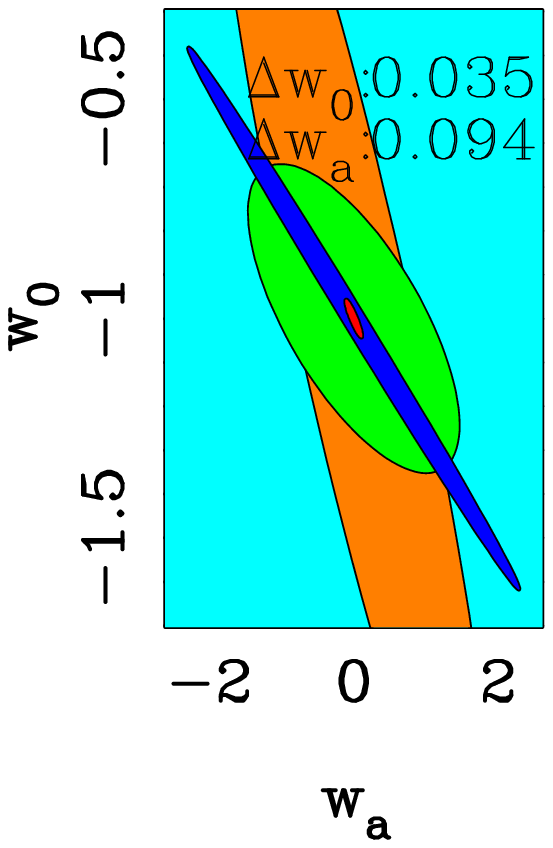}
\caption{The accuracy expected from the combination of experiments
dedicated to studying dark energy properties.  The equation of state
of dark energy is assumed to vary with scale factor $a$ as $w(a)=w_0
+ w_a(1-a)$, and the figures show the 1-sigma, 2-parameter regions
for the experiments individually and in combination.  The supernova
study fills the plot, the thin diagonal band is Planck, the
near-vertical band is BAO, and the ellipse is the 3D lensing power
spectrum method.  The small ellipse is the expected accuracy from
the combined experiments. From Heavens et al. (2006).}
\label{4Exp}
\end{figure}

\begin{figure}
\includegraphics[width=110mm, angle=0]{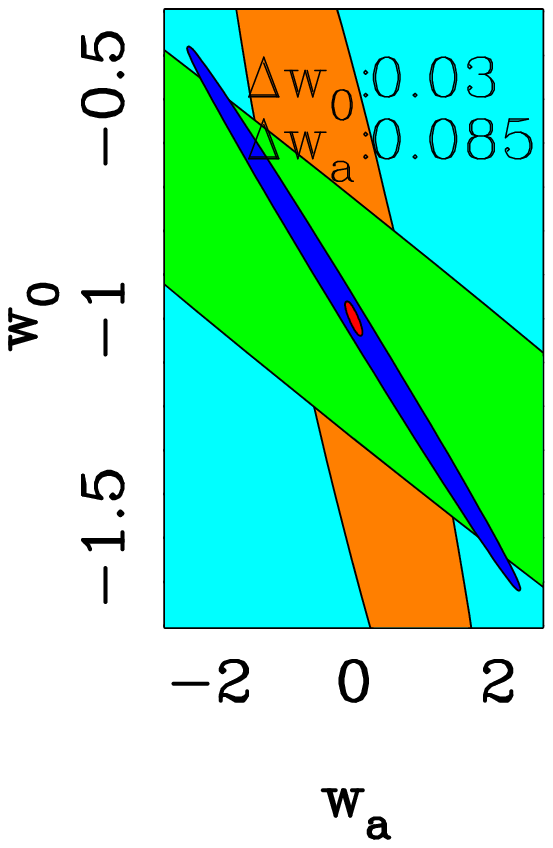}
\caption{As in Fig.~\ref{4Exp}, but with the shear ratio test as the
lensing experiment.  Supernovae fill the plot, Planck is the thin
diagonal band, BAO the near-vertical band, and the shear ratio is
the remaining 45 degree band.  The combination of all experiments is
in the centre.   From Taylor et al. (2006)}
\label{figratio}
\end{figure}

The figures~\ref{4Exp} and \ref{figratio} show the accuracy which 
might be achieved with a number
of surveys designed to measure cosmological parameters.  We
concentrate here on the capabilities of each method, and the methods
in combination, to constrain the dark energy equation of state, and
its evolution, parametrised by \cite{cheval01}
\begin{equation}
w(a) = {p\over \rho c^2} = w_0 + w_a(1-a)
\end{equation}
where the behaviour as a function of scale factor $a$ is, in the
absence of a compelling theory, assumed to have this simple form.
The constant value $w=-1$ would arise if the dark energy behaviour 
was actually a cosmological constant.

The assumed experiments are: a 5-band 3D weak lensing survey,
analysed either with the shear ratio test, or with the spectral
method, covering 10,000 square degrees to a median redshift of 0.7,
similar to the capabilities of a groundbased 4m-class survey with a
several square degree field; the Planck CMB experiment (14-month
mission); a spectroscopic survey to measure baryon oscillations
(BAO) in the galaxy matter power spectrum, assuming constant bias,
and covering 2000 square degrees to a median depth of unity, and a
smaller $z=3$ survey of 300 square degrees, similar to WFMOS
capabilities on Subaru; a survey of 2000 Type Ia supernovae to
$z=1.5$, similar to SNAP's design capabilities.

We see that the experiments in combination are much more powerful
than individually, as some of the degeneracies are lifted.  Note
that the combined experiments appear to have rather smaller error
bars than is suggested by the single-experiment constraints.  This
is because the combined ellipse is the projection of the product of
several multi-dimensional likelihood surfaces, which intersect in a
small volume.  (The projection of the intersection of two surfaces
is not the same as the intersection of the projection of two
surfaces). The figures show that errors of a few percent on $w_0$
are potentially achievable, or, with this parametrisation, an error
of $w$ at a `pivot' redshift of $z\simeq 0.4$ of under 0.02.  This
error is essentially the minor axis of the error ellipses.

\subsection{\label{intrinsic}Intrinsic alignments}

The main signature of weak lensing is a small alignment of the
images, at the level of a correlation of ellipticities of $\sim
10^{-4}$.  One might be concerned that physical processes might also
induce an alignment of the galaxies themselves.  The possible effect
is immediately apparent if one considers that the shear is often
estimated from the ellipticity of a galaxy, which includes the
intrinsic ellipticity of the source $e_s$ :
\begin{equation}
e\simeq e_S + 2\gamma.
\end{equation}
A useful statistic to consider is the shear correlation function,
which would normally be estimated from the ellipticity correlation
function:
\begin{equation}
\langle e e^* \rangle = 4\langle \gamma \gamma^* \rangle + \langle
e_s e_s^* \rangle + 4 \langle \gamma e^*\rangle.
\end{equation}
This equation is schematic, referring either to galaxies separated
by some angle on the sky, or by a 3D separation in the case of a 3D
analysis.  The first term is the cosmic signal one wishes to use;
the second term is the {\em intrinsic alignment} signal, and the
third is the {\em shear-intrinsic alignment} signal, which we will
consider later.  Until recently, both these additional terms were
assumed to be zero.  The hope was that even galaxies close together
on the line of sight would typically be at such large physical
separations that physical processes which could correlate the
orientations would be absent. However, the lensing signal is very
small, so the assumption that intrinsic alignment effects are
sufficiently small needs to be tested.  For the intrinsic alignment
signal, this was first done in a series of papers by a number of
groups in 2000-1 \cite{Hetal2000,croft2000,critten2001,catelan2001},
and the answer is that the effect may not be negligible, and is
expected to be strongly dependent on the depth of the survey. This
is easy to see, since at fixed angular separation, galaxies in a
shallow survey will be physically closer together in space, and
hence more likely to experience tidal interactions which might align
the galaxies.  In addition to this, the shallower the survey, the
smaller the lensing signal.  In a pioneering study, the alignments
of nearby galaxies in the SuperCOSMOS survey were investigated
\cite{brown}. This survey is so shallow (median redshift $\sim 0.1$)
that the expected lensing signal is tiny.  A non-zero alignment was
found, which agrees with at least some of the theoretical estimates
of the effect.  The main exception is the numerical study of
\cite{jing2001}, which predicts a contamination so high that it
could dominate even deep surveys. For deep surveys, the consensus is
that the effect is expected to be rather small, but if one wants to
use weak lensing as a probe of subtle effects such as the effects of
altering the equation of state of dark energy, then one cannot
ignore it. There are essentially two options - either one tries to
calculate the intrinsic alignment signal and subtract it, or one
tries to remove it altogether.   The former approach is not
practical, as, although there is some agreement as to the general
level of the contamination, the details are not accurately enough
known.  The latter approach is becoming possible, as lensing surveys
are now obtaining estimates of the distance to each galaxy, via
photometric redshifts (spectroscopic redshifts are difficult to
obtain, because one needs a rather deep sample, with median redshift
at least 0.6 or so, and large numbers, to reduce shot noise due to
the random orientations of ellipticities).   With photometric
redshifts, one can downweight or completely remove physically close
galaxies from the pair statistics (such as the shear correlation
function) \cite{HH03,KS02}. Thus one removes a systematic error in
favour of a slightly increased statistical error.  The analysis in
\cite{heymans2004} explicitly removed close pairs and shows that it can
be done very successfully.

\subsection{\label{shearintrinsic}Shear-Intrinsic alignment
correlation}

The cross term $\langle \gamma e^*\rangle$ was neglected entirely
until it was pointed out\cite{HS04} that it was not necessarily
zero. The idea here is that the local tidal gravitational field
contributes to the shear of background images, and if it also
influenced the orientation of a galaxy locally, then it could induce
correlations between foreground galaxies and background galaxies,
even though they may be physically separated by gigaparsecs.  This
term is more problematic for cosmic shear studies, because it is not
amenable to the simple solutions which work well for the intrinsic
alignment signal.  It is conceivable that this effect is the
limiting systematic effect in cosmic shear studies, as it seems
necessary actually to model it and remove it. Studies of the SDSS
\cite{Mandel05} measured a significant signal in a related
statistic, for very luminous galaxies, and a study of N-body
simulations supported the view that the effect was likely to be
non-negligible, at the level of up to 10\% of the cosmic shear
signal. On a more positive note, it seems
\cite{HS04,king2005,Hetal06b} that the term scales with source and
lens angular diameter distances in proportion to the lensing
efficiency $\De(\chirad_l)\,\De(\chirad_s-\chirad_l)/\De(\chirad_s)$. 
This is reasonable, and
also very useful, as it makes the parametrisation of the
shear-intrinsic alignment much more straightforward.  One can either
use templates \cite{king2005} or parametrise the contamination as a
single function of separation, and marginalise over these nuisance
parameters in the estimation of cosmological parameters.



\section{Non-Gaussianities}
\label{Non-Gaussianities}

The two-point statistics discussed in Section~\ref{ch2:2Pointstatistics}
can be used to constrain cosmological parameters. However, since they
can be expressed in terms of the convergence power $P_\kappa(\ell)$
they mainly depend on the same combination of parameters. Thus, from
Eq.(\ref{kappa}) we can expect $\lag\bar\kappa^2\rag \sim \sigma_8^2\Om^2$
if we neglect the dependence on cosmology of comoving distances,
where $\sigma_8$ is the normalization of the linear power-spectrum.
A more careful analysis \cite{Bernardeau_et_al1997} actually gives
the scaling of Eq.(\ref{scaling}). In order to lift this degeneracy between 
the parameters $\Om$ and $\sigma_8$ one can combine weak lensing
observations with other cosmological probes such as the CMB, as we shall
discuss in Section~\ref{External-data-sets}, or use 3D information as seen
in Section~\ref{ch4:3Dweaklensing} (e.g. Eq.(\ref{Rij})).
An alternative procedure is to consider higher-order moments of weak lensing
observables.
Indeed, even if the initial conditions are Gaussian, since the dynamics is
non-linear non-Gaussianities develop and in the non-linear regime the
density field becomes strongly non-Gaussian (this is an unstable
self-gravitating expanding system). This can be seen from the constraints
$\lag\delta\rag=0$ and $\delta \geq -1$ (because the matter density $\rho$
is positive) which imply that in the highly non-linear regime
($\lag\delta^2\rag\gg 1$) the probability distribution of the density
contrast $\delta$ must be far from Gaussian. Since weak gravitational lensing
effects arise from the matter distribution (see Eq.(\ref{Psi})) high-order
correlation functions of both the 3D density field and weak-lensing
observables are non-zero and could be used to extract additional information.

\subsection{Bispectrum and three-point functions}

\begin{figure*}
\begin{center}
\leavevmode\epsfxsize=14.cm \epsfbox{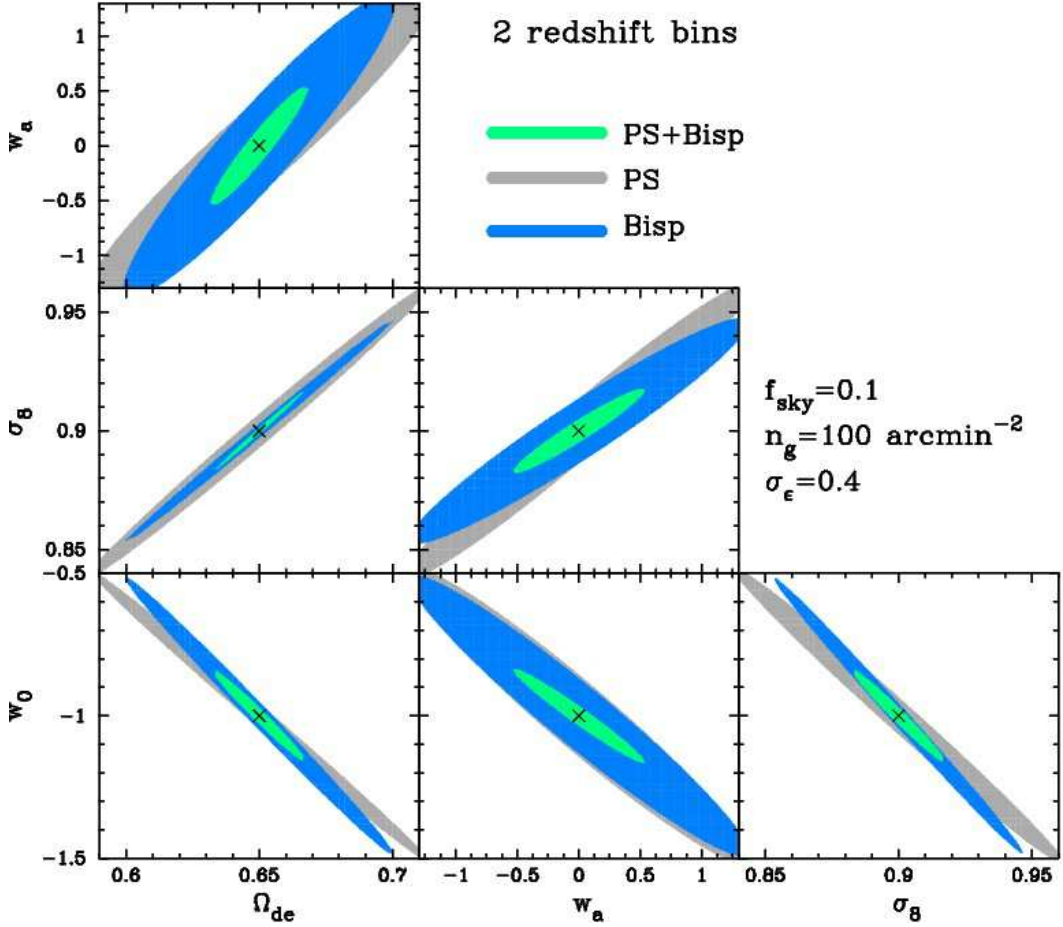}
\end{center}
\caption{Projected $68\%$ confidence level constraints in the
parameter space of $\Omega_{\rm de}$, $w_0$, $w_a$ and $\sigma_8$
from the lensing power spectrum and the bispectrum in two redshift
bins, as indicated. The coefficients $w_0$ and $w_a$ parameterize
the equation of state of the dark energy component. The results
shown are obtained assuming priors on $n$, $\Omega_{\rm b}h^2$ and
$h$ expected from the Planck mission. The sky coverage and number
density are taken to be $f_{\rm sky}=0.1$ and $n_g=100 $
arcmin$^{-2}$, and angular modes $50\le l\le 3000$ are used. From
Takada \& Jain (2004).}
\label{fig:chins2}
\end{figure*}

The three-point correlation function is the lowest-order statistics which
can be used to detect non-Gaussianity. In Fourier space it is called the
bispectrum which is defined as:
\beq
\lag \delta(\bk_1)\delta(\bk_2)\delta(\bk_3) \rag = (2\pi)^3 
\delta_D(\bk_1+\bk_2+\bk_3) B(k_1,k_2,k_3)
\label{bispecdelta}
\eeq
for the 3D matter density contrast,
where the Dirac factor results from statistical homogeneity. Isotropy
also implies that $B(k_1,k_2,k_3)$ only depends on the length
of the three wavenumbers $\bk_1,\bk_2,\bk_3$ or alternatively on two
lengths $k_1,k_2$ and the angle $\alpha_{12}$ between both vectors.
A key feature of the bispectrum (\ref{bispecdelta}) is that in the large-scale
limit its dependence on the normalization of the power-spectrum can be
factorized out \cite{Bernardeau_et_al1997}. Indeed, at large scales where
the density contrast is much smaller than unity and quasi-linear perturbation
theory is valid one can expand the density contrast as a perturbative series
of the form:
\beq
\delta(\bk,z)= \delta^{(1)}(\bk,z) + \delta^{(2)}(\bk,z) + ...
\label{deltaexp}
\eeq
where $\delta^{(q)}$ is of order $q$ over the initial density field
($\delta^{(1)}$ is simply the linear density contrast $\delta_L$). Then,
substituting into the three-point function we obtain:
\beqa
\lefteqn{ \lag \delta(\bk_1)\delta(\bk_2)\delta(\bk_3) \rag =
\lag \delta^{(1)}(\bk_1)\delta^{(1)}(\bk_2)\delta^{(1)}(\bk_3) \rag +
\lag \delta^{(2)}(\bk_1)\delta^{(1)}(\bk_2)\delta^{(1)}(\bk_3) \rag }
\nonumber \\
&& + \lag \delta^{(1)}(\bk_1)\delta^{(2)}(\bk_2)\delta^{(1)}(\bk_3)
+ \lag \delta^{(1)}(\bk_1)\delta^{(1)}(\bk_2)\delta^{(2)}(\bk_3)
+ ...
\label{bispecdeltaexp}
\eeqa
where the dots stand for terms of order $(\delta^{(1)})^5$ and beyond.
For Gaussian initial conditions the first term vanishes whereas the three
other terms are of order $(\delta^{(1)})^4$ so that the quantity:
\beq
Q(k_1,k_2,k_3) = \frac{B(k_1,k_2,k_3)}
{P(k_2)P(k_3)+P(k_1)P(k_3)+P(k_1)P(k_2)}
\label{Qdelta}
\eeq
is independent of the normalization of the linear density power-spectrum
$P_L(k)$ at large scales.
In this manner one can separate the dependence on $\sigma_8$ from the
dependence on other cosmological parameters.
Using the small-angle
approximation the $\vell$-space three-point correlation of the convergence
reads \cite{Bernardeau_et_al2003}:
\beq
\lag \kappa(\vell_1)\kappa(\vell_2)\kappa(\vell_3) \rag = (2\pi)^2
\delta_D(\vell_1+\vell_2+\vell_3) B_{\kappa}(\ell_1,\ell_2,\ell_3) ,
\label{bispeckappa}
\eeq
with:
\beq
B_{\kappa}(\ell_1,\ell_2,\ell_3) = \int \d\chirad \, \frac{\wh^3}{\De^4} \,
B\left(\frac{\ell_1}{\De},\frac{\ell_2}{\De},\frac{\ell_3}{\De}\right).
\label{bispecdeltakappa}
\eeq
Then, as in Eq.(\ref{Qdelta}) one can consider ratios such as
$B_{\kappa}(\ell_1,\ell_2,\ell_3)/(P_{\kappa}(l_2)P_{\kappa}(l_3)+..)$
to lift the degeneracy between the parameters $\Om$ and $\sigma_8$
\cite{Bernardeau_et_al1997}. Using tomography (i.e. redshift binning of the
sources) also helps to constrain cosmological parameters such as the
equation of state of the dark energy component, as studied in 
Ref.~\cite{Takada_Jain2004}.
We display their results in Fig.~\ref{fig:chins2} which shows that
bispectrum tomography can improve parameter constraints
significantly, typically by a factor of three, compared to just power
spectrum tomography.

In practice, most of the angular range probed by weak lensing surveys is
actually in the transition domain from the linear to highly non-linear regimes
(from $10'$ down to $1'$). Therefore, it is important to have a reliable
prediction for these mildly and highly non-linear scales, once the cosmology
and the initial conditions are specified. Since there is no rigorous
analytical framework to fully describe this regime numerical simulations play
a key role to obtain the non-linear evolution of the matter power spectrum
and of higher-order statistics \cite{Peacock_Dodds1996,Smith_et_al2003}.
Based on these simulation results and analytical insight it is
possible to build analytical models which can describe the low order moments
of weak lensing observables such as the bispectrum
\cite{vw2001}.
Using a halo model as described in Appendix~\ref{Halo models},
Refs.~\cite{Takada_Jain2003a,Takada_Jain2003b}
investigated the real-space three-point correlation
of the convergence
$\lag\kappa(\vtheta_1)\kappa(\vtheta_2)
\kappa(\vtheta_3)\rag $. They studied its dependence on the triangle
geometry $(\vtheta_1,\vtheta_2,\vtheta_3)$
and on the parameters of the halo model \cite{Takada_Jain2003a} and compared
these predictions with numerical simulations \cite{Takada_Jain2003b}.

As seen earlier it is more convenient for observational purposes to
consider the shear rather than the convergence since it is the
former which is directly measured (in fact what is actually measured
is the reduced shear $\gamma/(1-\kappa)$ which can be approximated
by $\gamma$ in the weak-lensing regime, \cite{schneiderseitz1995}).
However, since $\vgamma$ is a 2-component field
there are many ways to combine shear triplets. Here we defined the
shear spin-2 ``vector'' as $\vgamma=\gamma_1 \vex+\gamma_2 \vey$ where
$\vex,\vey$ are the 2D basis vectors using the flat sky
approximation which is valid for small angles (let us recall that 
$\vgamma$ is not truly a vector since its components change as 
$\cos(2\psi)$ and $\sin(2\psi)$ under a rotation of $\psi$ of 
coordinate axis, as seen from Eq.(\ref{gammal})). Besides, one must
take care not to define statistics which depend on the choice of the
coordinate system. A possible approach is to consider scalar
quantities such as the aperture mass $\Map$ which can be expressed
both in terms of the convergence or shear fields. However, this may
not be optimal from a signal-to-noise perspective since the
integration over the window radius $\theta_s$ may dilute the
cosmological signal as contributions from triangle configurations
where the shear three-point function is positive or negative can
partly cancel out. Moreover, the additional information contained in
the detailed angular behavior of the shear three-point correlation
can be useful to constrain cosmology and large-scale structures.
Therefore, it is interesting to build estimators designed for the
high-order correlations of the shear field.

\begin{figure}
\begin{tabular}{c}
{\epsfxsize=7cm\epsfysize=6cm\epsffile{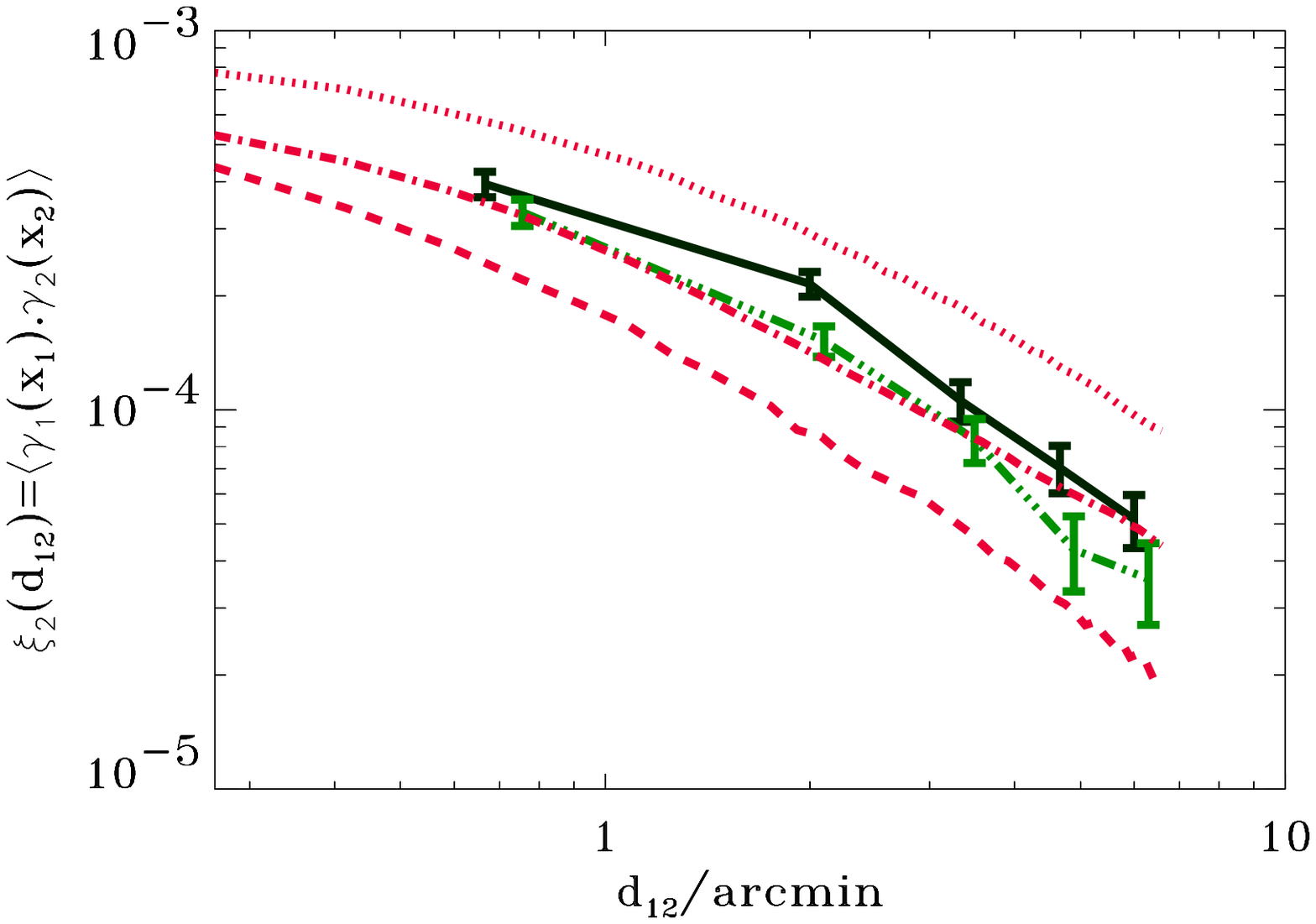}}
{\epsfxsize=7cm\epsfysize=6cm\epsffile{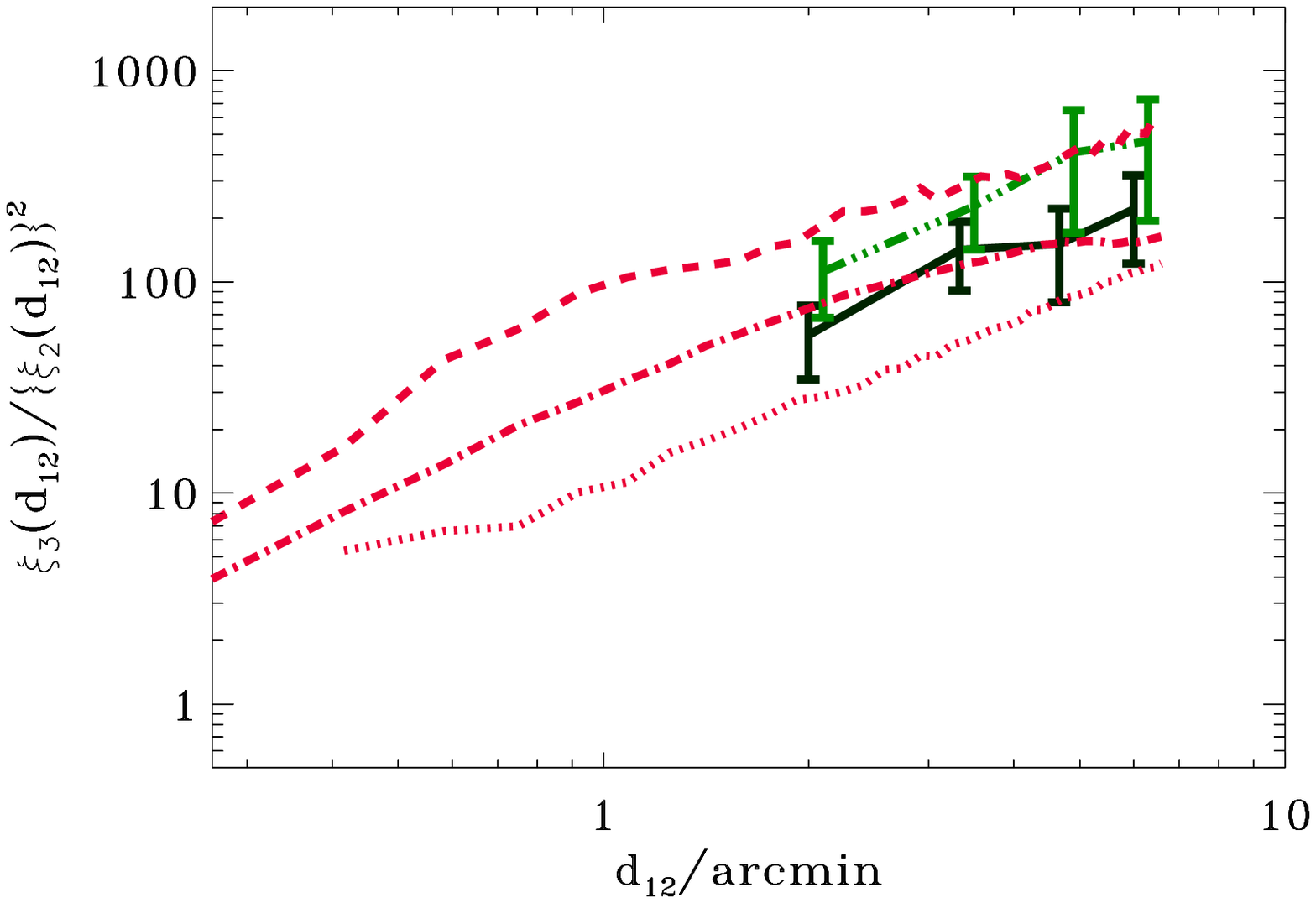}}
\end{tabular}
\caption{Results for the VIRMOS-DESCART survey for the two point correlation
function (left) and the reduced three point function (right).  The solid line
 with error bars shows the raw results, when both the $E$ and $B$ contributions
to the two-point correlation functions are included.  The dot-dashed
line with error bars corresponds to measurements where the
contribution of the $B$ mode has been subtracted out from the
two-point correlation function (but not from $\xi_3$ there is no
known way to do it). These measurements are compared to results
obtained in $\tau $CDM, OCDM and $\Lambda$CDM simulations (dashed,
dotted and dot-dashed lines respectively). From Bernardeau et al. (2002).}
\label{xidetec}
\end{figure}

One strategy investigated in Ref.\cite{Bernardeau_et_al2003} is to
study the mean shear pattern $\vgamma(\vtheta)$ around a pair of points
$\vtheta_1,\vtheta_2$ through the quantity $\lag\vgamma_3\vtheta)\rag
=\lag[\vgamma(\vtheta_1).\vgamma(\vtheta_2)]\vgamma(\vtheta)\rag$.
This study, based on analytical results (using the behavior of the density
three-point correlation in the quasi-linear regime and its simplest extension
to smaller scales) and numerical simulations,
shows that this mean shear is almost uniform, and perpendicular
to $\vtheta_{12}$, over an elliptic area that covers the segment 
$\vtheta_{12}$ which joins both points. This suggests to measure the average of
$\lag\vgamma_3(\vtheta)\rag$ over this ellipse so as to avoid cancellations.
In this manner \cite{BMVW} managed to obtain from the
VIRMOS-DESCART Lensing Survey the first detection of non-Gaussianities
in a weak lensing survey. We display their results in Fig.~\ref{xidetec}
which shows that the amplitude and shape of the signal agree with theoretical
predictions from numerical simulations. Although the measures are still too
noisy to provide useful constraints on cosmology they show such
weak-lensing observations to be a very promising tool.
On the other hand, \cite{Bernardeau2005} also obtained explicit analytical
expressions for the shear three-point correlations from the one-halo term
which appears within halo models (when all points are assumed to lie within
the same dark matter halo, see Appendix~\ref{Halo models}) and recovered the
pattern shown by numerical simulations. These results may serve as a guideline
to build optimized estimators for the shear three-point correlations.

A more systematic approach presented in \cite{Schneider_Lombardi2003}
is to look for natural components which transform in a simple way through
rotations. Thus, to handle the three point function
$\lag\gamma(\vtheta_1)\gamma(\vtheta_2)
\gamma(\vtheta_3)\rag$ one first
defines the ``center'' $\bc$ of the three directions
$\vtheta_1,\vtheta_2,\vtheta_3$ on the sky
($\bc$ may be taken for instance as the centroid, the circumcenter or the
orthocenter of the triangle). Next, at each point one defines $\gamma_+$
as the component of the shear along the direction that separates $\bc$ and
$\vtheta_i$ and $\gamma_{\times}$ as the component along this
direction rotated by $45^{\circ}$. From these tangential and cross components
one introduces the complex shear $\gamma^{(\bc)}=\gamma_++i\gamma_{\times}$
and the ``natural components'' are defined as the four complex combinations:
\beqa
\lefteqn{\Gamma^{(0)}= \lag\gamma^{(\bc)}(\vtheta_1)\gamma^{(\bc)}(\vtheta_2)\gamma^{(\bc)}(\vtheta_3)\rag , \;\;\;\;\;
\Gamma^{(1)}= \lag\gamma^{(\bc)*}\gamma^{(\bc)}\gamma^{(\bc)}\rag , }
\nonumber \\
&& \Gamma^{(2)}= \lag\gamma^{(\bc)}\gamma^{(\bc)*}\gamma^{(\bc)}\rag , \;\;\;\;\;
\Gamma^{(3)}= \lag\gamma^{(\bc)}\gamma^{(\bc)}\gamma^{(\bc)*}\rag .
\label{Gammadef}
\eeqa
Clearly the $\Gamma^{(i)}$ only depend on the geometry of the triangle but
to avoid ambiguities and miscalculations one must ensure that the points are
always labeled in the same direction (e.g. counterclockwise,
\cite{Schneider2005}).
Each of these $\Gamma^{(i)}$ is invariant only under special rotations
but the important feature is that, under a general rotation, the different
$\Gamma^{(i)}$ do not mix but are simply multiplied by a phase factor.
Note however that the four $\Gamma^{(i)}$ are not independent as they arise
from the same matter distribution and three-point statistics are fully
described by the projected matter bispectrum \cite{Schneider_et_al2005}.
On the other hand, such interrelations provide a redundancy which might
be used to detect noise sources or B modes.

\begin{figure}
\begin{center}
\leavevmode\epsfxsize=14cm \epsfbox{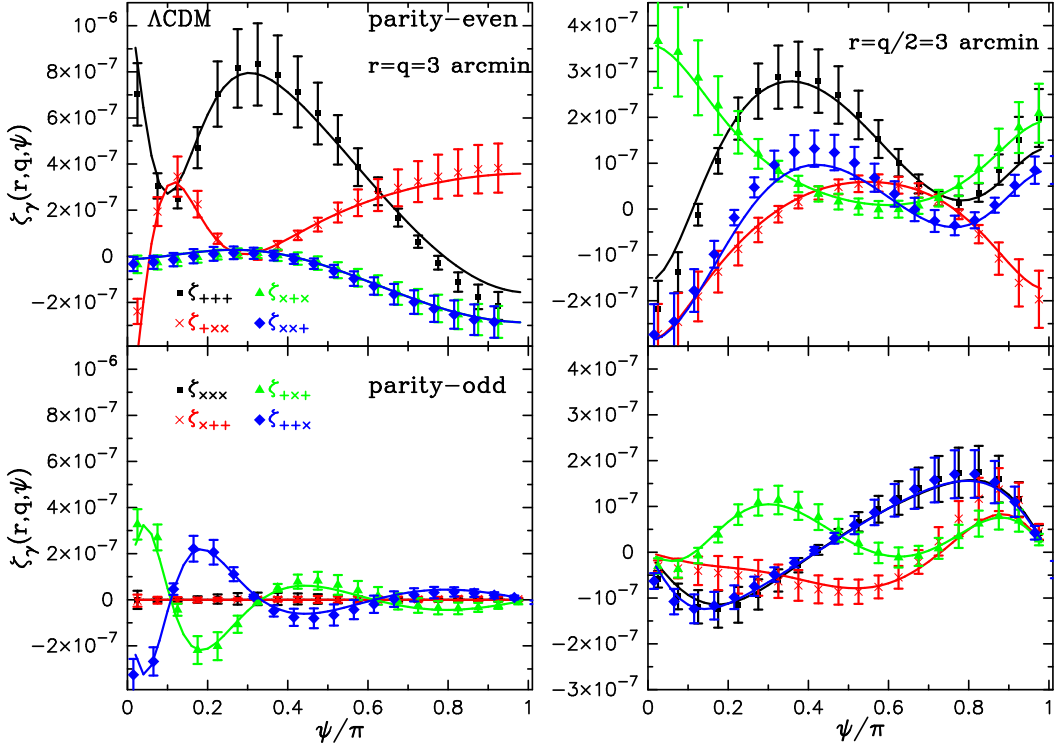}
\end{center}
\caption{The eight shear 3-point correlation functions for the
$\Lambda$CDM model against triangle configurations ($\psi$ is the
angle between the two sides of length $q$ and $r$).  The upper and
lower plots show the results for the parity-even and -odd functions,
respectively. Note that range on the y-axis for the right panel is
about two times smaller than in the left panel.  The solid curves
show the halo model predictions for the eight shear 3-point
correlation functions, while the symbols are the simulation results
as indicated. From Takada \& Jain (2003).}
\label{fig:s3ptlcdm}
\end{figure}

An alternative method to study the shear three-point function is to
divide the possible combinations into even and odd quantities
through parity transformations \cite{Zaldarriaga_Scoccimarro2003}.
For instance, one chooses for the center of the triangle the
barycenter $\bc=(\vtheta_1+\vtheta_2+{\vtheta}_3)/3$ and defines again 
the tangential and cross
components of the shear, $\gamma_+$ and $\gamma_{\times}$, from the
direction that separates $\bc$ and $\vtheta_i$ and from
this direction rotated by $45^{\circ}$. Clearly this rotation
changes direction under a parity transformation so that
$\gamma_{\times}\rightarrow-\gamma_{\times}$ (since $\gamma$ is a
spin-2 field, see Eq.(\ref{gammal}), and the relative rotation
is $2\times 45^{\circ}=90^{\circ}$) whereas
$\gamma_+\rightarrow\gamma_+$. Therefore, we obtain four parity-even
three-point correlations: $\lag\gamma_+\gamma_+\gamma_+\rag$,
$\lag\gamma_+\gamma_{\times}\gamma_{\times}\rag$,
$\lag\gamma_{\times}\gamma_+\gamma_{\times}\rag$,
$\lag\gamma_{\times}\gamma_{\times}\gamma_+\rag$, and four
parity-odd three-point correlations:
$\lag\gamma_{\times}\gamma_{\times}\gamma_{\times}\rag$,
$\lag\gamma_{\times}\gamma_+\gamma_+\rag$,
$\lag\gamma_+\gamma_{\times}\gamma_+\rag$,
$\lag\gamma_+\gamma_+\gamma_{\times}\rag$. As a consequence, for
some symmetric configurations some odd functions must vanish
\cite{Takada_Jain2002}. In particular, for equilateral triangles all
odd functions vanish. This property assumes that the shear results
only from weak-lensing (which only produces $E$ modes), whereas
source galaxy clustering, intrinsic alignments and observational
noise can produce both $E$ and $B$ modes 
(see Sections~\ref{ch2:2Pointstatistics} and \ref{ch4:3Dweaklensing}). 
The advantage of this
procedure is that by focusing on even functions one avoids to dilute
the signal by combining the estimators with parts which contain no
weak-lensing information (odd functions for symmetrical triangle
geometries). Besides, the parity-odd functions can be used to
monitor the noise or to estimate the contribution associated with
higher-order effects beyond the Born approximation, source
clustering or intrinsic alignments.
Ref.~\cite{Zaldarriaga_Scoccimarro2003} used a halo model (see
Appendix~\ref{Halo models}) to investigate the behavior of these
three-point correlations as a function of the triangle geometry.
Ref.~\cite{Takada_Jain2003b} found that the halo model agrees well
with numerical simulations at scales $>1'$ and could be used to
obtain predictions for shear statistics in order to lift
degeneracies in cosmological parameters. We display their results in
Fig.~\ref{fig:s3ptlcdm} which also shows that odd functions are
smaller than even ones and vanish for symmetric geometries. They
also note that future weak-lensing observations may be able to
constrain the parameters of the halo model such as the mean halo
density profile and halo mass function. On the other hand, using
ray-tracing simulations Ref.~\cite{Takada_Jain2002} also evaluated
the signal-to-noise taking into account the noise associated with
galaxy intrinsic ellipticities. They found that a deep lensing
survey of area $10$ deg$^2$ should be sufficient to detect a
non-zero signal but an accurate measure would require an area
exceeding $100$ deg$^2$.

\subsection{Cumulants and probability distributions}
\label{Cumulants}

\begin{figure}
\begin{tabular}{c}
{\epsfxsize=7cm\epsfysize=7cm\epsfbox[36 433 310 705]{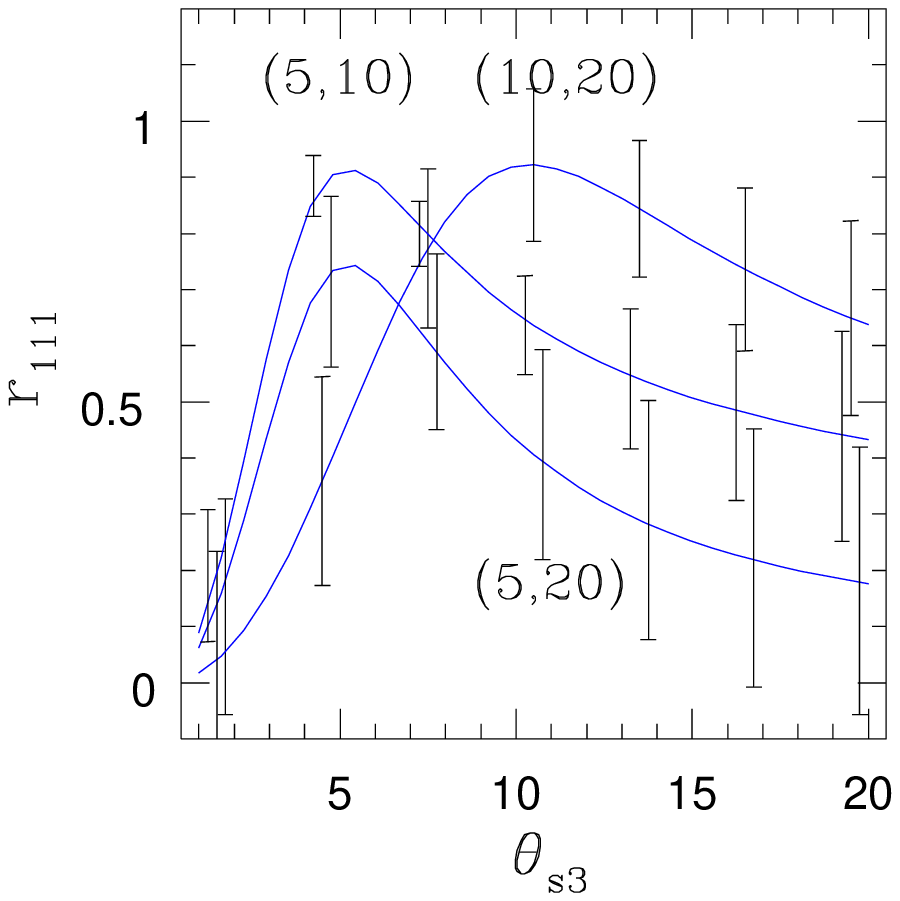}}
{\epsfxsize=7cm\epsfysize=7cm\epsfbox[52 433 324 715]{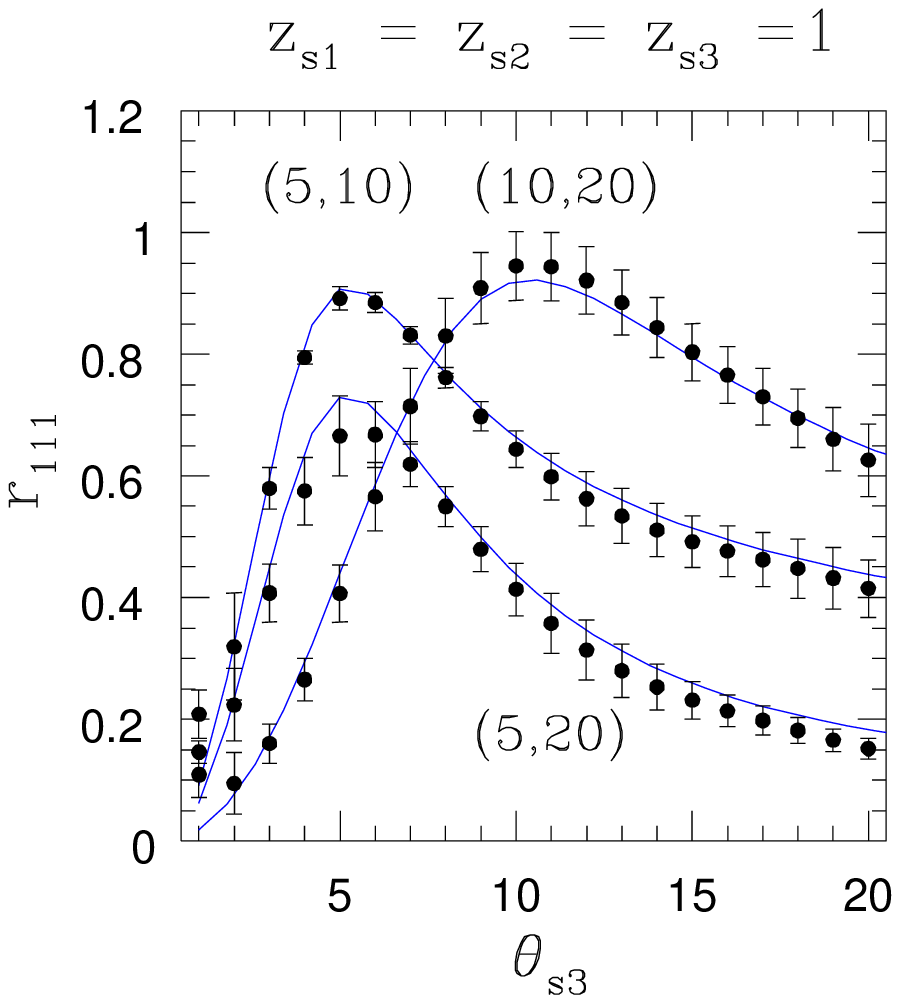}}
\end{tabular}
\caption{ {\it Left panel:} the correlator $r_{111}$ of the aperture-mass
for the full
wide SNAP survey is plotted as a function of smoothing angle
$\theta_{s3}$ for a fixed pair of $(\theta_{s1},\theta_{s2})$. The
pair $(\theta_{s1},\theta_{s2})$ for each curve is indicated in the
plot. Error bars denote the $1-\sigma$ scatter around the mean,
associated with galaxy intrinsic ellipticities and cosmic variance.
{\it Right panel:} the three-point correlation coefficient $r_{111}$ 
of the aperture-mass
as a function of smoothing angle $\theta_{s3}$ for a fixed pair of
$(\theta_{s1},\theta_{s2})$. The three source redshifts are equal:
$z_{s1}=z_{s2}=z_{s3}=1$. The solid curve is the analytical model
(\ref{stellar}) while solid points with error-bars are measurements
from simulation data. From Munshi \& Valageas (2005b).}
\label{fig:r111}
\end{figure}

A simpler quantity than the three-point functions discussed above is
provided by the third-order cumulant $\lag {\bar X}^3\rag$ of smoothed
weak-lensing observables (\ref{Xx}) such as the smoothed convergence
$\bar\kappa$ or the aperture-mass $\Map$. In the quasi-linear regime
where a perturbative approach is valid (with Gaussian initial
conditions) one can see from
Eqs.(\ref{bispecdeltaexp})-(\ref{bispecdeltakappa}) that the
skewness $S_3^{(\kappa)}=\lag\barkappa^3\rag/\lag\barkappa^2\rag^2$ of
the smoothed convergence is independent of the matter density power
spectrum normalization $\sigma_8$ (the same property is clearly
valid for other observables like $\Map$ which are linear over the
matter density field). Therefore, by measuring the second- and 
third-order moments of the convergence or of the aperture mass at 
large angular scales one can obtain a constraint on $\Om$
\cite{Bernardeau_et_al1997}. Indeed, from Eq.(\ref{kappa}) we see
that $\kappa \sim \Om$ (neglecting the dependence of cosmological
distances $\De$ on $\Om$) hence we can expect a strong dependence on
$\Om$ of the skewness as $S_3^{(\kappa)}\sim\Om^{-1}$. A numerical
study shows indeed that for sources at redshift $z_s \simeq 1$ the
skewness scales roughly as $S_3^{(\kappa)}\sim\Om^{-0.8}$
\cite{Bernardeau_et_al1997}. Alternatively, from the skewness of
weak lensing observables one can derive the skewness of the matter
density field in the linear regime and check that the scenario of
the growth of large-scale structures through gravitational
instability from initial Gaussian conditions is valid. In order to
increase the information content which can be extracted from
low-order cumulants one can consider generalized moments such as
$\lag\Map(\theta_{s1})\Map(\theta_{s2})\Map(\theta_{s3})\rag$ which
cross-correlate the aperture-mass $\Map(\theta_{si})$ associated
with three different filter radii $\theta_{s1},\theta_{s2}$ and
$\theta_{s3}$. Then, the amplitude of such cumulants can be used to
constrain cosmological parameters whereas the dependence on the
angular radius or the angular separation of the various filters
helps constraining the properties of the large-scale density field
\cite{munshivalageas2005b,Kilbinger_Schneider2005}. For instance,
one can introduce correlation coefficients $r_{pqs}$ such as: 
\beq
r_{pqs} = \frac{\lag \barX_1^p \barX_2^q \barX_3^s \rag_c} 
{\lag\barX_1^{p+q+s}\rag_c^{p/(p+q+s)} \lag\barX_2^{p+q+s}\rag_c^{q/(p+q+s)}
\lag\barX_3^{p+q+s}\rag_c^{s/(p+q+s)}} 
\label{rpqs} 
\eeq 
where $\barX_i$ is
the aperture-mass or the convergence smoothed over scale
$\theta_{si}$ (the source redshift distributions $n_i(z_s)$ may also
be different). These correlation coefficients describe the
information associated with three-point cumulants $\lag\barX_1^p \barX_2^q
\barX_3^s \rag_c$ which goes beyond the one-point cumulants $\lag
\barX^p\rag_c$. We show in Fig.~\ref{fig:r111} the predictions of an
analytical model based on a hierarchical {\it ansatz}
(Appendix~\ref{Hierarchical models}) for the aperture-mass
statistics, applied to the planned SNAP survey (left panel) and
compared with numerical simulations (right panel). The behavior of
these correlation coefficients can be used to discriminate between
models of the density field \cite{munshivalageas2005b} and to check
that the observed non-Gaussianities arise from non-linear
gravitational clustering.
On the other hand, the left panel in Fig.~\ref{fig:r111} also shows how 
the aperture-mass is correlated between different angular scales. 
This correlation decreases faster than for the smoothed convergence or 
the smoothed shear because the filter $W_{\Map}(\ell\theta_s)$ is more 
narrow than $W_{\kappa}$ in Fourier space, see Fig.~\ref{kernels}. 
This is actually useful if one intends to derive constraints on cosmology
from weak lensing surveys since it means that the errors associated with 
sufficiently different scales are uncorrelated. This also holds
for the two-point moments discussed in Section~\ref{ch2:2Pointstatistics}
and for higher-order moments. Of course, the power-spectrum $P_{\kappa}(\ell)$
and higher-order generalization such as the bispectrum 
$B_{\kappa}(\ell_1,\ell_2,\ell_3)$ are even less correlated and contain
all the relevant information.

In practice, because of the numerous holes within the survey area one first
computes shear three-point correlations by summing over galaxy triplets
and next writes $\lag\Map^3\rag$ as an integral over these three-point
correlations (using the fact that $\Map$ can be written in terms of the
shear $\gamma$), see for instance \cite{Jarvis_et_al2004}. Applying this method
to the VIRMOS-DESCART data Ref.~\cite{Pen_et_al2003b} were able to detect
$S_3^{(\Map)}$ and to infer an upper bound $\Om<0.5$ by comparison with
simulations.
Next, one could measure higher order moments of weak lensing
observables. Note that for the shear components odd order moments vanish by
symmetry so that one needs to consider the fourth-order moment to go beyond
the variance \cite{Takada_Jain2002}. However, higher order moments are
increasingly noisy \cite{Valageas_Munshi_Barber2005} so that it has not been
possible to go beyond the skewness yet.

\begin{figure}
\begin{tabular}{c}
{\epsfxsize=7.4cm\epsfysize=7.5cm\epsfbox[68 140 688 715]{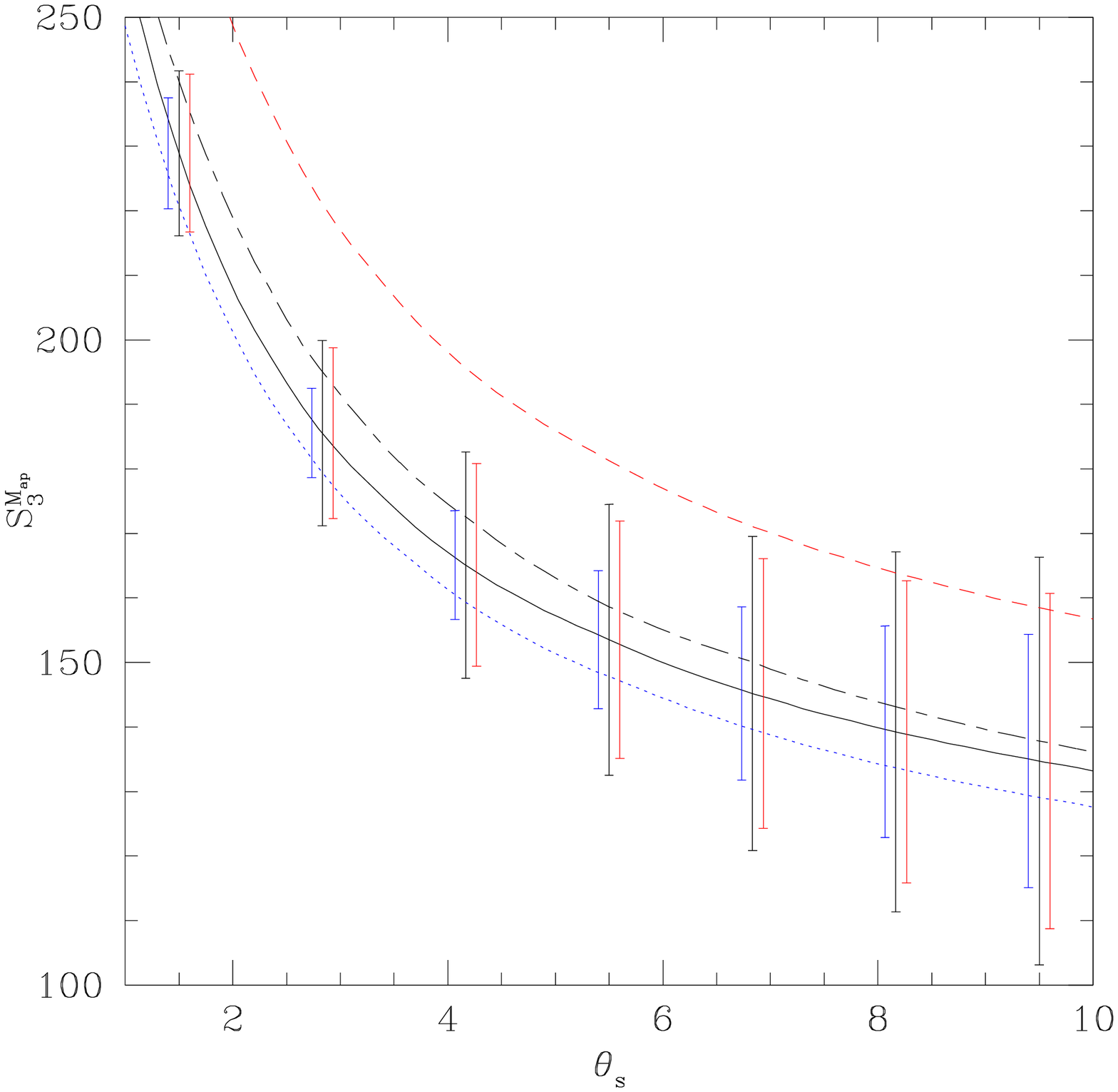}}
{\epsfxsize=7.4cm\epsfysize=7.5cm\epsfbox[118 120 688 715]{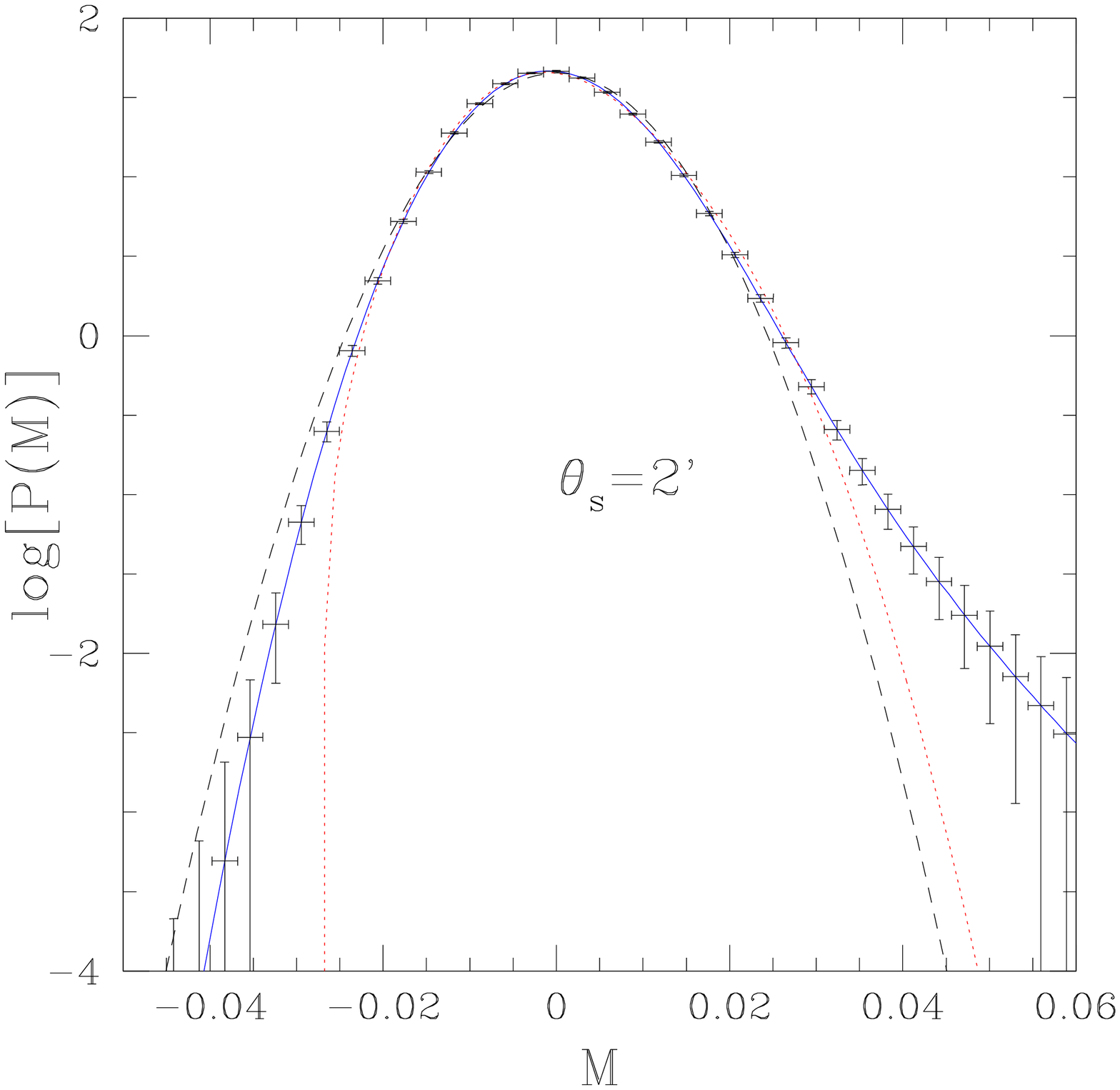}}
\end{tabular}
\caption{ {\it Left panel:} The skewness
$S_3^{\Map}=\lag\Map^3\rag/\lag\Map^2\rag^2$ of the aperture-mass
(solid curve), for the SNAP survey. The central error bars show the
$1-\sigma$ dispersion due to galaxy intrinsic ellipticities and
cosmic variance. The smaller error bars which are slightly shifted
to the left show the dispersion obtained by neglecting non-Gaussian
contributions to the dispersion whereas the smaller error bars which
are slightly shifted to the right show the dispersion obtained from
the estimator built from the cumulant rather than the moment. We
also show the effect of a $10\%$ increase of $\Om$ (lower dotted
curve), of a $10\%$ increase of $\sigma_8$ (central dot-dashed
curve) and of a $10\%$ decrease of the redshift $z_0$ (upper dashed
curve). {\it Right panel:} The pdf $\cP(M)$ for the estimator $M$
associated with the aperture-mass $\Map$. Note that the Gaussian
noise introduced by intrinsic ellipticities makes $\cP(M)$ closer to
the Gaussian than the actual pdf $\cP(\Map)$ which only takes into
account gravitational lensing. The solid line shows the theoretical
prediction, the dashed line is the Gaussian and the dotted line is
the Edgeworth expansion up to the first non-Gaussian term (the
skewness). The error bars show the $1-\sigma$ dispersion. From
Valageas, Munshi \& Barber (2005).}
 \label{figlPMapt2}
\end{figure}

In order to compare observations with theory one needs to use
numerical simulations or to build analytical models which can
describe the low order moments of weak lensing observables or their
full probability distribution, as described in
Appendix~\ref{Analytical modeling of gravitational clustering and
weak-lensing statistics}. This can be done through a hierarchical
{\it ansatz} where all higher-order density correlations are
expressed in terms of the two-point correlation
\cite{Barber_Munshi_Valageas2004,Valageas_Munshi_Barber2005}. Then,
the probability distribution of weak lensing observables can be
directly written in terms of the probability distribution of the
matter density (Appendix~\ref{Hierarchical models}). In some cases
the mere existence of this relationship allows one to discriminate
between analytical models for the density field which are very
similar \cite{Munshi_Valageas_Barber2004}. Alternatively, one can
use a halo model (Appendix~\ref{Halo models}) where the matter
distribution is described as a collection of halos
\cite{Cooray_Sheth2002} and the low order moments of weak lensing
observables can be derived by averaging over the statistics of these
halos \cite{Takada_Jain2003a,Takada_Jain2003b}. On the other hand,
one can use weak lensing to constrain halo properties and to detect
substructures \cite{Dolney_Jain_Takada_2004}. For the particular
case of the smoothed convergence the probability distribution 
function (PDF) $\cP(\barkappa)$ can be expressed in terms of the 
PDF $\cP(\delta)$ of the 3D matter density contrast within some simple
approximations, see Eq.(\ref{Pkappa}). This allows one to apply to
the convergence simple models which were originally devised for the
3D density field such as the lognormal model \cite{Taruya_et_al2002}
or more elaborate ones
\cite{Valageas2000,Wang_et_al2002,Barber_Munshi_Valageas2004}. For
more complex weak-lensing observables such as the aperture-mass
these approximations can no longer be used and one needs an explicit
model of the density correlations (see Appendix~\ref{Analytical
modeling of gravitational clustering and weak-lensing statistics})
to derive their cumulants or their PDF. We display in
Fig.~\ref{figlPMapt2} the results obtained from a hierarchical {\it
ansatz} for the aperture-mass, for the skewness (left panel) and the
PDF of an estimator of $\Map$ \cite{Valageas_Munshi_Barber2005}. In
such calculations one must take into account the noise associated
with the intrinsic ellipticities of galaxies and the cosmic
variance. However, the left panel of Fig.~\ref{figlPMapt2} shows
that despite these sources of noise future surveys such as SNAP
should be able to constrain cosmological parameters at a level of
$10\%$ from such low-order moments, whereas the right panel shows
that the tails of the PDF $\cP(\Map)$ should allow one to extract
information beyond low-order moments.

\subsection{Primordial non-Gaussianities}

So far we have discussed the non-Gaussianities associated with the
non-linearity of the gravitational dynamics, assuming Gaussian initial
conditions. However, results from future surveys can also
be very useful in constraining primordial non-Gaussianity predicted
by some early universe theories \cite{Takada_Jain2004}.
On the other hand, generalised theories of gravity can
have very different predictions regarding gravity-induced 
non-Gaussianities as compared to General Relativity, which can also 
be probed using future data. A joint
analysis of power spectrum and bispectrum from weak lensing
surveys will provide a very powerful way to constrain, not only
cosmological parameters, but early universe theories and alternative
theories of gravitation, see also \cite{Schmid05}.
Thus, a recent work \cite{Bernardeau2005} has studied
the possibility of constraining higher-dimensional gravity
from cosmic shear three-point correlation function.


\section{\label{ch:6} Data Reduction from Weak Lensing Surveys}

\subsection{Shape measurement}
\label{Shape-measurement}

Weak lensing by large scale structures induces a coherent alignment
of galaxy shapes across large angular distances. A crucial step for
its measurement is the accurate estimation of galaxy shapes, free of
biases and systematics. A common approximation is to describe
galaxies as simple elliptical objects, for which the quadrupole of the
light distribution is a fair estimate of the shear $\gamma$. If we
call $f(\vtheta)$ the 2-dimensional light distribution of the galaxy
image, then the quadrupole moment $Q_{ij}$ is defined in its
simplest form as:
\begin{equation}
Q_{ij}={\int~{\rm d}\theta_i {\rm d} \theta_j~f(\vtheta)(\theta_i-\hat
\theta_i)(\theta_j-\hat \theta_j)\over \int~{\rm d}\theta_i {\rm d} \theta_j~f(\vtheta)},
\label{Qij}
\end{equation}
where $\hat \theta_i$ is the centroid of the light distribution. The
ellipticity $e_i=(e_1,e_2)$ of the galaxy is given by (using the same definition
as in Section~\ref{Estimators_and_their_covariance} with
Eqs.(\ref{eS_e})-(\ref{e_2gamma})):
\begin{equation}
(e_1,e_2)=\left( \frac{Q_{11}-Q_{22}}{Q_{11}+Q_{22}} ;
\frac{2~Q_{12}}{Q_{11}+Q_{22}} \right).
\label{def_elli}
\end{equation}
When galaxies are not lensed (in the source plane), their average
ellipticity is zero $\langle e_i \rangle=0$ from the assumption of
isotropy. Lensed galaxies exhibit an average non-zero distortion
$\delta_i=\langle e_i\rangle$ over the coherence scale of
the lensing potential $\phi$ because the shear $\gamma_i$ stretches
galaxy shapes locally in the same direction. The distortion is given
by Eq.(\ref{def_elli}), where $Q$ is the quadrupole moment of the
galaxy image $Q^I$. The relation between the source quadrupole $Q^S$
and $Q^I$ depends on the magnification matrix $\cA$
(see Section~\ref{ch:2}):
\begin{equation}
Q^I={\cal A}^{-1}Q^S{\cal A}^{-1},\label{Qdashed}
\end{equation}
where
\begin{equation}
{\cal A}=\left(
\begin{array}{cc}
1-\kappa-\gamma_1 & -\gamma_2 \\
-\gamma_2 & 1-\kappa+\gamma_1 \label{ampmatrix}
\end{array}
\right).
\end{equation}
\begin{figure}
\begin{center}
\epsfig{file=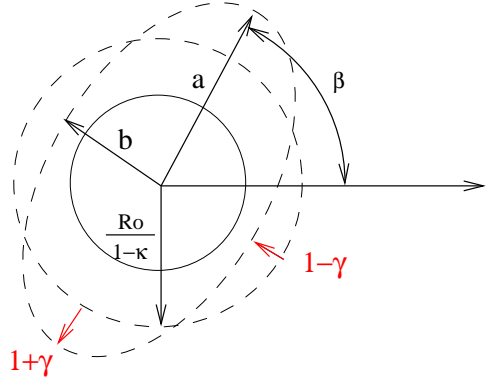,height=5cm}
\end{center}
\caption[]{Schematic representation of a lensed galaxy. In the
source plane, the galaxy is circular and has radius $R_0$.
Convergence $\kappa$ stretches its average radius to
$R_0/(1-\kappa)$, and the shear $\gamma$ distort the galaxy along
some angle $\beta$. $a$ and $b$ are the semi-major and semi-minor
axis respectively.} \label{lensedgal.ps}
\end{figure}
The matrix $\cA$ of Eq.(\ref{ampmatrix}) describes the mapping between
the source and image planes. The source galaxy surface brightness
$f^S(\vtheta)$ is stretched along the eigen axis of $\cal A$, and the
observed galaxy light distribution becomes:
\begin{equation}
f(\vtheta)=\left[1+\left({\cal
A}-I\right)_{ij}\theta_j\partial_i\right]f^S(\vtheta)
\label{flens}
\end{equation}
Figure~\ref{lensedgal.ps} shows what is happening to the shape of a
galaxy in the weak lensing regime. The quantities of interest can be
listed as follows (see also Section~\ref{ch:2}):

\begin{itemize}
\item[$\bullet$] Convergence $\kappa$
\item[$\bullet$] Distortion $|\delta |$=${a^2-b^2\over a^2+b^2}$
\item[$\bullet$] Shear $|\gamma |$=${a-b\over a+b}$
\item[$\bullet$] Magnification $\mu=\left[{\rm det}({\cal A})\right]^{-1}=\left[(1-\kappa)^2-\gamma^2\right]^{-1}$
\end{itemize}
The convergence and magnification cannot be easily measured because
we do not know the size of galaxies in the source plane. Therefore,
we focus on the shear (and the related aperture-mass $\Map$,
see Eq.(\ref{Map})) as estimators of weak lensing
effects. However we would like to point out that magnification
has been measured successfully on SLOAN from the excess of distant
quasars around foreground galaxies \cite{scrantonetal2005} and this
technique could potentially be a powerful probe of cosmology in the
future. The distortion offers the advantage that we know
statistically its value in the source plane, i.e. $\langle
\delta_i\rangle=0$, meaning that source galaxies are randomly
aligned (but see Sections~\ref{intrinsic} and \ref{shearintrinsic}).

\subsection{Point Spread Function correction}

One cannot easily measure the distortion $\delta_i$ for two reasons:
the optics of the telescope induce geometrical distortions that can
be misinterpreted as a weak lensing signal if not carefully
corrected: the stars, which provide a picture of the Point Spread
Function (PSF) of the telescope's aperture, have complicated,
elongated shapes. The second problem is that the weak lensing signal
is better measured on distant galaxies. Indeed, since the line of sight
is more extended weak lensing effects have a greater magnitude, see
Eq.(\ref{Psi}). However, these distant galaxies are small and faint,
and the sky noise for these objects is large. Consequently, a naive
estimation of the quadrupole $Q_{ij}$ from Eq.(\ref{Qij}) is essentially
a measure of sky noise.

Refs.\cite{KSB95,LK97} showed that one can overcome both problems by 1)
reducing the noise at the edge of the galaxy using a gaussian filter,
2) expressing the effect of the PSF convolution analytically as a
perturbation expansion, and use the first-order term to correct the
galaxy shapes. The KSB method \cite{KSB95} can be summarized as
follows: the observed ellipticity $e^{obs}_i$ is the sum of three
terms, the first is the intrinsic ellipticity $e^{int}_i$ of the
galaxy before lensing and before convolution with the PSF.
$e^{int}_i$ is unobservable, but the isotropy of space implies
$\langle e^{int}_i \rangle=0$. The second term describes the galaxy
shape response to an anisotropic PSF (which depends on a measurable
quantity called the smear polarizability tensor $P^{sm}_{ij}$), and
the last term describes the response to the isotropic PSF (which
depends on the ``preseeing" shear polarizability $P^\gamma$, also
measurable). Therefore the observed ellipticity becomes
\begin{equation}
e_i^{obs}=e_i^{int}+P^{sm}_{ij}p^{*}_j+P^\gamma \gamma_i,
\end{equation}
where $\gamma_i$ is the shear we want to measure and $p^*$ is the
stellar ellipticity estimated at the galaxy position. $P^{sm}$,
$p^*$ and $P^\gamma$ can all be estimated from the image (see
\cite{KSB95} for the details). Some refinements should be included
when shapes are measured on space data
\cite{Hetal98,Refregier_Rhodes_Groth2002}, but
they do not change the philosophy of the method, which remains a
first-order perturbative approach. The estimation of $p^*$ at the
galaxy position is done by first measuring $p^*$ on the stars and
then interpolating its value assuming a second-order polynomial
variation across the CCDs. This is a crucial step since an
inaccurate model could lead to significant $B$ modes in the signal
\cite{vanWaerbeke_et_al2002}, fortunately various models have been proposed to
account for non-polynomial variations \cite{H2004}. Ref.~\cite{JJB06}
has shown that we can perform a singular value decomposition method
of the PSF variation between individual exposures in order to
improve the correction. This would be a particularly useful approach
for lensing surveys planning to observe the same part of the sky
hundreds of times (LSST, ALPACA).

The ultimate accuracy of galaxy shape correction is still a wide
open question. Whether or not there is a fundamental limit in the
measurement of galaxy shapes is a particularly critical issue for
the design of future gravitational lensing surveys
(see Section~\ref{Statistical-and-Systematic-Errors}).
KSB has been historically the first shape measurement method
working and for that reason it has been tested intensively, but its
accuracy is not expected to be better than 5-10\% \cite{erben01}.
Clearly this is not enough for precision cosmology which seeks
sub-percent precision on shape measurement.

Many post-KSB techniques have been developed over the past five
years in order to improve upon the original KSB approach, most of
them are being intensively tested only now \cite{STEP1,STEP2}. A
quick summary of these methods follows. The reader can obtain the
details by looking at the original papers or the description given
in \cite{STEP1}. Ref.~\cite{KK99}
proposed to model the
PSF and the galaxies by a sum of Gaussians of different widths. The
pre-seeing galaxy shape is recovered by a $\chi^2$ minimisation
between the galaxy model and the measured profile. \cite{K2000} extended
the original KSB by properly modelling the PSF with a realistic
kernel, dropping the assumption of a Gaussian profile. The PSF is
then circularized prior to measuring the galaxy shapes (the
circularization technique is also used by \cite{smithetal2001}). A
radically different approach consists in projecting the shapes
(galaxies and stars) on a basis of orthogonal functions
\cite{BJ2002,AR2003,RB2003}. The effect of convolution and shear can
be expressed analytically on the basis functions and the solution of
the preseeing galaxy shape can be found by a straightforward matrix
inversion. This technique turns out to be important not only for
shape measurement, but also for simulating realistic galaxy
profiles. It also offers in principle a total control of the
different processes changing the shape of a galaxy (shear,
amplification, PSF, etc...). Mathematically, a galaxy with profile
$f(\vtheta)$ can be decomposed over a set of basis functions
$B_{(n_1,n_2)}(\vtheta;\beta)$ as
\begin{equation}
f(\vtheta)=\sum_{(n_1,n_2)} f_{(n_1,n_2)}B_{(n_1,n_2)}(\vtheta; \beta),
\end{equation}
where $\beta$ is a scaling parameter adjusted to the size of the
galaxy we want to analyse, $f_{(n_1,n_2)}=f_{\bf n}$ are called the
shapelets coefficients, and $B_{(n_1,n_2)}$ could be any family of
polynomial functions fulfilling our favorite recurrence relation
(orthogonality, orthonormality, etc...). Although the description is
given here for a cartesian coordinate system, the same formalism can
be developed for any coordinate system better suited for galaxy
shape analysis (see \cite{MR2005} for a derivation using polar
coordinates). The convolution of a lensed galaxy $f(\vtheta)$ with a
stellar profile $g(\vtheta)$ produces a galaxy with profile $h(\vtheta)$
such that its shapelets coefficients are:
\begin{equation}
h_{\bf n}=\sum_{{\bf m},{\bf l}} C_{\bf nml}f_{\bf m}g_{\bf l}
=\sum_{\bf m} P_{\bf nm}f_{\bf m},
\end{equation}
where $C$ and $P=Cg$ are matrices that can be measured on the data.
The preseeing galaxy profile can be formally obtained from the
matrix inversion $f=P^{-1} h$. \cite{BJ2002} developed a similar
method but they also perform a circularization of the PSF like in
\cite{smithetal2001} prior to the measurement of the pre-seeing
galaxy shape. \cite{KK06} has implemented a version of the shapelet
technique based on ideas developed in \cite{KK99} where galaxies and
PSF are decomposed on a fixed set of simple profiles. Currently,
there are six different shape measurement techniques, and several
implementations of KSB, corresponding to more than a dozen of
different pipelines. This clearly demonstrates the richness of the
topic, but one should ensure that they all lead to the same shear
measurement. A comparison of the performances of all these
techniques is shown in \cite{HS03}, but the authors focussed on
analytical galaxy profiles instead of real, noisy, profiles. They
found that KSB was performing the best, which is surprising given
the number of approximations involved. The Shear TEsting Program
STEP
\cite{STEP1,STEP2}\footnote{http://www.physics.ubc.ca/$\sim$heymans}
has been setup in order to systematically perform intensive tests of
shape measurement methods. Its goal is to test the different
pipelines on simulated and real data sets and to find the ultimate
limit of shape measurement from space and ground based images.
\begin{figure}
\begin{center}
\epsfig{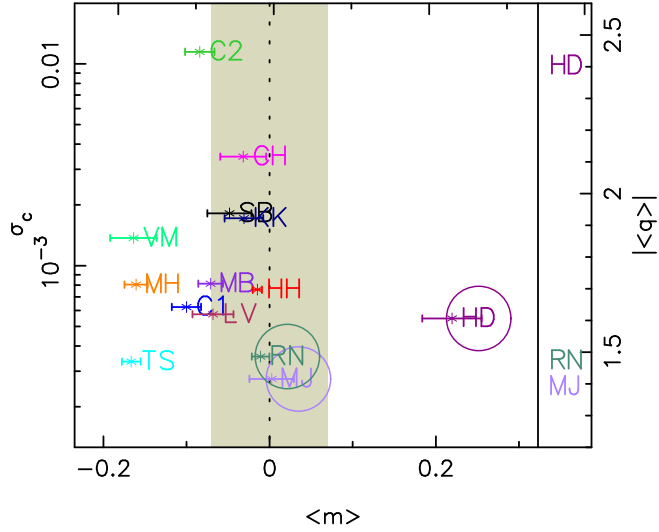}
\end{center}
\caption[]{From Heymans et al. (2006), plot showing the calibration
bias $m$, PSF residual and non-linearity $\langle q \rangle$ for
different pipelines (see Heymans et al. 2006 for a complete list of
references and detailed description of the methods).}
\label{steplatest}
\end{figure}

\cite{STEP1} measured the difference between the measured shear
$\gamma^{\rm meas}_i$ and the true shear $\gamma^{\rm true}_i$ on a
large number of simulated images with very different PSF simulating
various optical defects of the telescope. The chosen parametrization
was:

\begin{equation}
\gamma^{\rm meas}_1 - \gamma_1^{\rm true} = q(\gamma_1^{\rm true})^2
+ m\gamma_1^{\rm true} + c_1,
\end{equation}
where $q$, $m$ and $c_1$ are measured from different realizations of
image quality (PSF anisotropy and seeing size). Figure~\ref{steplatest}
shows the r.m.s. of these parameters. It shows the
main result from STEP, indicating that shape measurement accuracy is
within the few percent range, at least one order of magnitude above
the required accuracy for future experiments such as SNAP and LSST.
This figure also shows that KSB is performing remarkably well, in
agreement with \cite{HS03}, and the newer methods are potentially
even more powerful.

Among all the methods, the very attractive feature of the shapelets
lies in the fact that each transformation experienced by a galaxy
(PSF, shear, convergence, etc...) can be expressed as a simple set
of operators acting linearly on the set of basis functions
describing the galaxy. A general transformation of the galaxy shape
is therefore a linear process which can formally be solved in one
pass to provide the pre-seeing shape. The shapelets, or similar
approaches \cite{BJ2002}, provide in principle a perfect description
of the galaxy shapes. Is it the ultimate method with the best
possible accuracy? In the weak lensing regime for instance, the
quadrupole of the light distribution fully describes the shear
$\gamma$, a two-components spin-weight 2 object, and it is
unnecessary to measure the details of the galaxy shape. In that
case, the shapelets might appear like overkill, and limiting the
number of basis functions is appropriate \cite{KK06}. The solution
of the shape measurement problem is a trade-off between an accurate
description of the galaxy morphology (i.e. galaxy structures) and an
unbiased measure of the second order moments: we want to describe
the galaxy shape with enough, but not too many, details. This
optimal trade-off depends on the weak lensing information we want to
extract from the galaxy distortion. For instance, in the weak
lensing regime, it is assumed that the shear does not vary across
the galaxy, and the quadrupole of the light distribution is then a
complete description of the lensing effect. This is equivalent to
saying that the centroid of a lensed galaxy is the lensed centroid
of the source galaxy. Mathematically, this means that
\begin{equation}
\left({\vtheta}^S-{\vtheta}^{SC}\right)\simeq
{\cal A}\left({\vtheta}^I-{\vtheta}^{IC}\right),
\label{weaklensapprox}
\end{equation}
where ${\vtheta}^S$ and ${\vtheta}^I$ are the angular position
of a galaxy source and image respectively, and ${\vtheta}^{SC}$
and ${\vtheta}^{IC}$ are the centroid position of the source and
image respectively. Eq.(\ref{Qdashed}) is a direct consequence of this
approximation. It obviously breaks down near the critical line where
the determinant of the amplification matrix $\cal A$ is zero and the
magnification is infinite. The source galaxy is then strongly
distorted and gravitational arcs are observed \cite{FM94}. In the
intermediate regime where source galaxies are mildly distorted and
have an arc-like shape, weak lensing is just an approximation, and
the quadrupole of the light distribution is not enough to quantify
the lensing effect. A higher-order description of the galaxy shape
becomes necessary \cite{GN02}. Refs.\cite{GB05,Bacon2006,MRRBB2006}
developed the theory of {\it flexion} which is a description of the
next order of shear measurements, the octopole. In that case,
Eq.(\ref{weaklensapprox}) is not valid and should be replaced by:

\begin{equation}
\theta_i'={\cal A}_{ij}\theta_j+{1\over 2} D_{ijk}\theta_i\theta_j,
\end{equation}
where $D_{ijk}=\partial_k {\cal A}_{ij}$ is given by:

\begin{equation}
D_{ij1}=\left(
\begin{array}{cc}
-2\gamma_{1,1}-\gamma_{2,2} & \quad-\gamma_{2,1} \\
-\gamma_{2,1} & -\gamma_{2,2} \\
\end{array}
\right) \ \ \ ; \ \ \
D_{ij2}=\left(
\begin{array}{cc}
-\gamma_{2,1} & -\gamma_{2,1} \\
-\gamma_{2,1} & 2\gamma_{1,2}-\gamma_{2,1} \\
\end{array}
\right).
\end{equation}
Higher orders of galaxy shapes are a probe of higher order
derivatives of the gravitational lensing potential \cite{SEF}. This is
therefore particularly relevant for lensing by cluster of galaxies
and space quality images, because the latter is well suited for an
accurate measurement of galaxy shapes beyond the quadrupole. The new
relation between the source and image galaxy profile is given by

\begin{equation}
f(\vtheta)=\left[1+\left[({\cal A}-{\cal I})_{ij}\theta_j+
{1\over 2}D_{ijk}\theta_j \theta_k\right]\partial_i\right]f^S(\vtheta),
\end{equation}
which replaces Eq.(\ref{flens}). The flexion terms can be conveniently
expressed in terms of shapelet operators \cite{GB05}. The practical
utility of flexion has not been demonstrated yet, but progress is
being made to show whether or not it is measurable. With the
development of CCD detectors in space it is likely that flexion
could provide useful lensing information \cite{Bacon2006}.

Lots of progress has been made in the measurement of galaxy shapes
since the original KSB paper in 1995. The most recent shape
measurement method \cite{KK06} shows that the one percent accuracy
goal has not yet been reached, and most of the effort now consists
of showing that this goal can be met in order to perform high
precision cosmology with weak lensing; this is the primary goal of
the STEP project. With the development of these techniques and
improving image quality, it becomes possible to measure higher order
shear effects such as the octopole. Shapelets or a similar method
are particularly useful to extract high-order morphology
information, while KSB is enough for percent precision shear
measurement \cite{STEP1,STEP2}.

\begin{figure}
\begin{center}
\epsfig{file=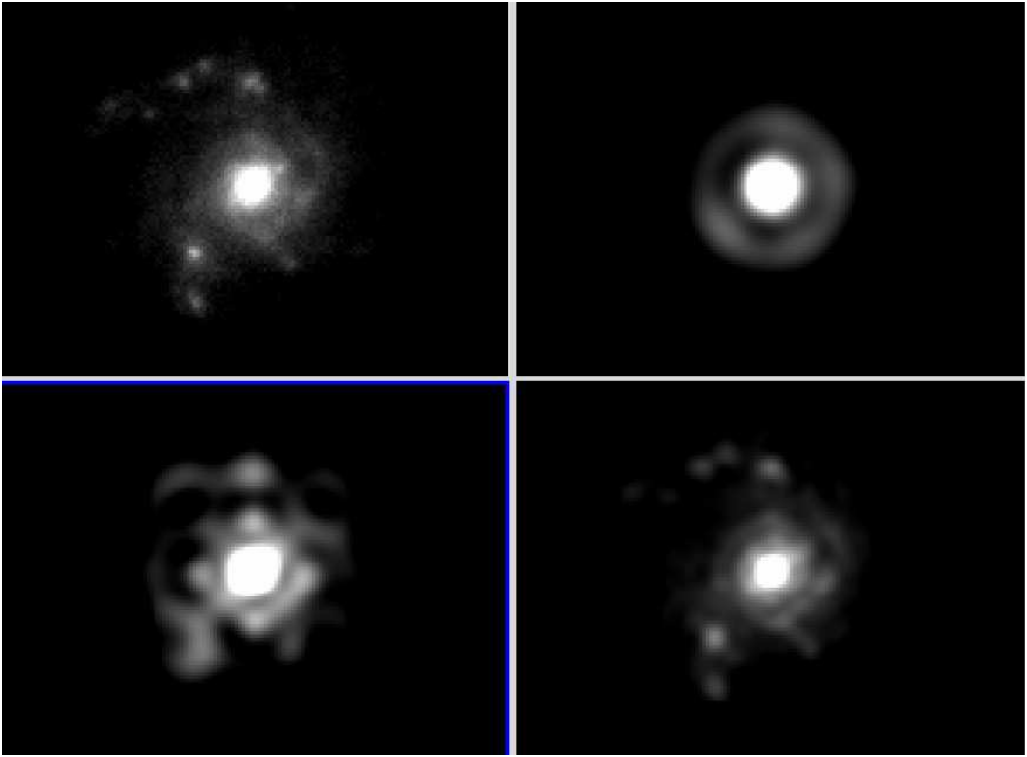,height=10cm}
\end{center}
\caption[]{Simulating galaxies with shapelets: top-left is a galaxy
from the Hubble Ultra Deep Field and the other panels show its
simulated image using the shapelets with $n_{\rm max}=5, 10, 30$.}
\label{simulgal}
\end{figure}

The shapelets can also be used to
generate realistic images of galaxies, which is important for
testing shape measurement methods. Here one uses a training set to
{\it teach} a galaxy image simulator how to generate a realistic
combination of shapelet coefficients in order to reproduce
statistically real data sets. One such galaxy simulation is shown in
Figure~\ref{simulgal}. Ref.\cite{MRRBB2006} for instance simulated
galaxies using the shapelets, whose distribution has been trained
using the Hubble Ultra Deep Field.

\subsection{Statistical and Systematic Errors}
\label{Statistical-and-Systematic-Errors}

In this Section, we consider the potential sources of error and
systematics in the shear measurement, not the error caused by
intrinsic alignment (see Section~\ref{intrinsic}) and selection biases
correlated with the shear orientation and amplitude \cite{HS03}.
The latter was shown to be negligible \cite{STEP1}.

\begin{figure}
\begin{center}
\epsfig{file=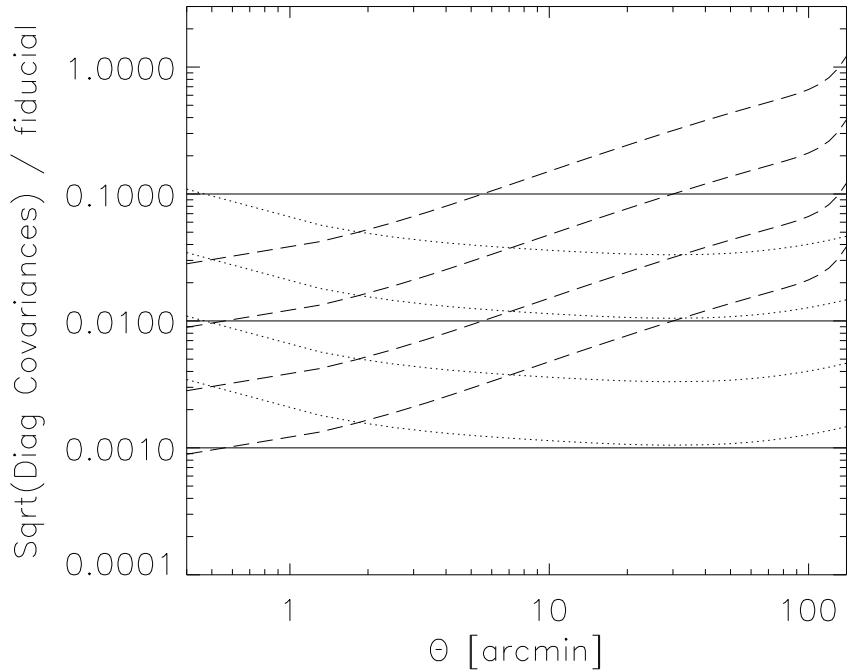,height=10cm}
\end{center}
\caption[]{Two sources of error: statistical noise and cosmic
variance. The curves show the diagonal part of the covariance matrix
for either source of noise, normalised by the fiducial model shear
variance. The thick solid lines show various levels of
multiplicative errors from 0.01 to 0.001.} \label{covariancematrix}
\end{figure}

A complete description of galaxy shape is certainly not necessary
for most weak lensing applications. According to \cite{STEP1}, the
main limitation in shear measurement comes from the calibration of
the shear amplitude: the PSF anisotropy is relatively easy to
correct, but the isotropic correction due to the seeing is still not
accurate to better than a few percent (see Figure~\ref{steplatest}).
An additive error should also be considered, as suggested by Figure
\ref{steplatest}. Ref.\cite{huterer2006} has shown that, in order not
to degrade significantly the cosmological parameters estimation from
future weak lensing experiments (SNAP and LSST), the additive error
has to be less then $10^{-4}$, which is one order of magnitude better
than what can be achieved today (see Figure~\ref{steplatest}). It is
still unclear how to estimate the impact of the additive error for
realistic surveys, since the redshift, color and morphology
dependence might be quite complicated and are not yet well
understood. The multiplicative (i.e. calibration) error of the order
of one percent does not seem to degrade dramatically the
cosmological parameters constraints \cite{huterer2006}, but this is
because the cosmological constraints come from the largest scales,
where the error budget is dominated by cosmic variance and not by
the multiplicative error \cite{vwetal06}. Figure~\ref{covariancematrix}
shows that a large multiplicative error
degrades significantly the weak lensing signal at angular scales less
than $10$ arcminutes, and has no effect at scales $20$ arcminutes
and above. Therefore a complete scientific use of a lensing survey
is also dependent on our ability to reduce the multiplicative error,
not only the additive error.
With the STEP effort \cite{STEP1,STEP2} and the large amount of
ground and space based data available from current weak lensing
surveys, the shape measurement issue will probably be solved within
the next 2 or 3 years. The major source of error for weak lensing
survey will then become our ability to estimate the photometric
redshifts of the source galaxies \cite{vwetal06}. This is a powerful
technique, but the redshift error is large and often there are
multiple redshift solutions due to spectral features not covered by
the set of filters. This is clearly shown in \cite{ilbert06}, where
photometric redshift degeneracies are left with the 5 filters ugriz
of the MEGACAM camera. Ref.\cite{huterer2006} has shown that the
requirement of photometric redshift precision is tight if we want to
achieve tomography: for instance the average redshift needs to be
accurate to better than 1\%. A large number of filters in the
optical and near infrared coverage would then be necessary.
Ref.\cite{kitching2006} finds that the requirements are less severe for 3D
weak lensing (see Section~\ref{ch4:3Dweaklensing}).

\section{Simulations}
\label{Simulations}

In order to obtain good signal-to-noise for estimates of
cosmological parameters, it is necessary to probe many different
scales, including small scales where linear or second-order
perturbation theory is not valid. As we have no exact analytical
description of matter clustering at small or intermediate scales,
numerical N-body simulation techniques are employed to
study gravitational clustering in an expanding background. Numerical
techniques typically use ray-tracing techniques through N-body
simulations to study weak lensing of background sources. Other
methods include line-of-sight integration of shear. Although often
only limited by computational power, numerical techniques too depend
on various approximations which can only be verified by consistency
checks against analytical results.

Simulating (strong) lensing by individual objects can provide
valuable information regarding background cosmology through the
statistics of arcs, see e.g. \cite{turner1984, narayan1991} for
early studies. Simulating weak lensing surveys on the other hand
probes inhomogeneous matter distribution by large scale structure in
the Universe. Early attempts to simulate weak gravitational lensing
by inhomogeneous dark matter distribution was initiated by various
authors in the early 1990s. These studies include 
\cite{weiss1988,jarros1990,lee1990,babul1991,Bartelmann_Schneider2001,Blandford_et_al1991}.
However the most detailed studies in this direction started emerging
by the late 1990s. \cite{wambsganss1998}
following up previous work done by
\cite{wambsganss1995,wambsganss1997} presented a detailed analysis of
lensing by large scale structure. On a slightly larger angular scale
this study was extended and complemented later by the work done by
\cite{Jain_Seljak_White2000}  who constructed shear and convergence 
maps using
ray-tracing simulations. With ever increasing computational power
the  typical size of the sky which can be simulated using ray tracing
experiments has increased over the years and recent studies can now
focus even on scales comparable to tens of degrees while still
resolving smaller angular scales well. These numerical developments
have also stimulated improved analytical modelling of weak lensing
using perturbative techniques at larger angular scales and
halo-based models or the hierarchical {\it ansatz} at small angular scales.

\subsection{Ray tracing}

In case of experiments involving ray-tracing simulations, one combines
several large-scale boxes obtained from cosmological N-body simulations
to build a large simulated volume from the observer up to the source
plane at redshift $z_s$. Then, the dark-matter particle distributions 
whithin each box are projected on successive two-dimensional planes 
up to the redshift of the source. Typically up to  $30$ such planes are
employed to sample the matter density to a source redshift of $ z_s
\sim 1$ and $10^6$ rays are propagated through the N-body
data volume. The computation of the derivatives of the gravitational 
potential in each of these lens planes is performed using FFTs.
This provides the shear tensor at each plane which allows to follow
the deflection of the light rays from one plane to the next.
The Jacobian matrix ${\cal A}$ of the mapping from source to the image
plane determines lensing observables such as the convergence $\kappa$ 
and the shear components $\gamma_1,\gamma_2$, see Eq.(\ref{defA}).
Such an algorithm is also known as
the multiple-lens algorithm (see e.g. \cite{Schneider_et_al1992} for more
detailed discussions).

Various numerical artefacts that determine the resolution of a
ray-tracing simulation include the spatial and mass resolution of
the underlying N-body simulation through which the ray tracing
experiments are being performed as well as the size of the grid
which is used to compute the intermediate projected densities and
the gravitational potential. The finite size of the simulation 
boxes on the other hand
determines the largest angular scales to which we can reliably use
the results from ray-tracing simulations. Depending on the size of
the simulation box and source redshift one can typically construct
a few degree square patches of the sky. To improve the statistics,
the N-body simulation box is rotated and ray tracing experiments
repeated, to generate additional weak lensing sky patches.

Studies using ray tracing simulations were initiated by
\cite{Jain_Seljak_White2000}. They used a $256^3$ adaptive $\rm P^3M$ simulation
outputs from Virgo to perform the ray-tracing simulations. The lens
plane grid used to compute the potential and its derivatives from
the projected matter distribution had a resolution of $2048^2$.
Depending on a specific cosmology these studies generated weak
lensing maps of a few degrees, with resolution down to sub-arc
minutes. Recent ray tracing simulation using multiplane techniques
include the ones presented in~\cite{hamana2000}

\begin{figure}
\begin{center}
\setlength{\unitlength}{1cm}
\begin{picture}(5,8)(0,0)
\put(-.75,-.75){\includegraphics{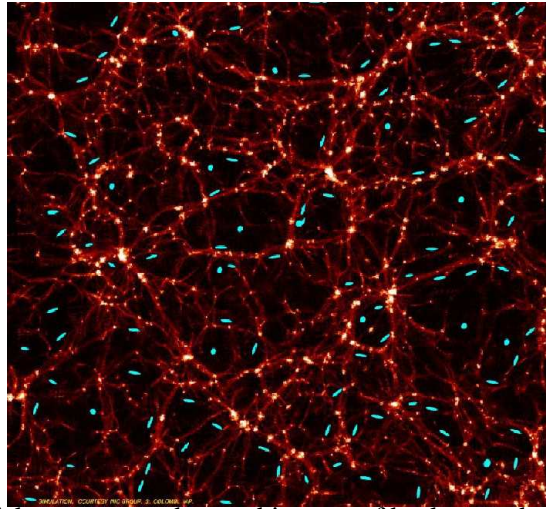}}
\end{picture}
\end{center}
\caption[]{The blue elongated disks represent observed images of
background galaxies. The dark matter filaments from numerical
simulations which were used to simulate the survey are also plotted
to show alignments of ellipticites of observed galaxies with
underlying filamentary structures (Figure courtesy: Stephane
Colombi and Yannick Mellier).} \label{label_fig}
\end{figure}

\subsection{Line-of-sight integration}

Ref.\cite{hcouchman1999} pointed out that several problems can arise 
in weak lensing studies from the use of the multiple-plane approach 
described in the previous Section, especially when the sources
are distributed at relatively high redshifts. In particular,
the projection of the matter distribution onto 
successive 2D planes orthogonal to the mean line of sight clearly
approximates all angular diameter distances by a constant within
a given redshift interval. This may lead to significant errors
if these intervals are too large. Thus, Ref.\cite{hcouchman1999}
introduced a 3D algorithm which computes the second derivatives
of the gravitational potential $\Phi(\bx)$ within the full 3D volume
using FFTs. Then, light rays are followed within the 3D volume
and deflections are taken into account along the 3D grid using the
local derivatives of the gravitational potential and the local angular
distances, which allows to derive the Jacobian matrix $\cA$ between 
source and observer planes (for more detailed discussions
see ~\cite{hcouchman1999,Barber1999,barber2000}).
The variance of the cosmic shear obtained from this algorithm
was compared with analytical predictions and other simulations
in \cite{barber2002}. More detailed comparisons appear in
\cite{Barber_Munshi_Valageas2004,Valageas_Munshi_Barber2005,Munshi_Valageas_Barber2004}
where a good agreement was found with a whole 
range of analytical predictions.

It should be noted that ray-tracing simulations and
line-of-sight integrations still remain costly options to simulate a
reasonable portion of the sky with fairly low resolution. This
limits the number of independent realizations which can be
simulated. However, to probe the small angular scales where
non-linear gravity has not been understood analytically yet, simulations
are the only reliable option against which all analytical
predictions are tested regularly.

\section{Weak Lensing at other wavelengths}
\label{Otherwavelengths}

Current weak lensing surveys mostly rely on statistical studies of
ellipticities of background galaxies at optical wavelengths. However
various authors have considered the possibility of weak lensing
studies in other wavebands, both for individual radio or IR sources
at high redshift and for the fluctuations in the integrated
diffuse emission from unresolved sources
\cite{CoorayAst2004,LiPen04}. It was pointed out that future
facilities at radio wavelengths will even start competing with
space-based optical observations which are limited by their ability
to resolve the shape of distant sources and their small field of
view. Radio surveys by the proposed Square Kilometre Array (SKA)
will be able to make huge progress in this direction by resolving
orders of magnitude more sources \cite{Schneider99}. In addition to
resolved individual sources it was also suggested that integrated
diffuse emission from the first stars and protogalaxies at high
redshift ($z_s= 15-30$) as well as 21cm emission by neutral
intergalactic medium can provide useful arenas for weak lensing
studies. It was realized that such programmes could be very useful
in bridging the gap between weak lensing surveys based on optical
studies of nearby galaxies at a redshift of a few and weak lensing
studies of the CMB at a redshift of $z_s=1100$. Nevertheless
separating galactic contamination from cosmological signal remains a
difficult task.

\subsection{Weak lensing studies in Radio and near IR}

Future radio facilities, such as the proposed 
SKA\footnote{http://www.skatelescope.org/}, will be superior to current
radio observatories by orders of magnitude, particularly in its
field of view, and in sensitivity. In addition, the higher
resolution achieved by SKA will place it in a much better position
than the present generation of radio telescopes. Surveys using SKA
will push radio astronomy in a position where the number density of
radio sources will be comparable to that of the optical sky (where
the number density of useable galaxies is $\sim 30~{\rm arcmin}^{-2}$
up to a redshift of order unity, for ground-based surveys). This is
possible thanks to the fact that SKA will be able to observe 100
times fainter objects than currently achievable; the radio sky at
present is almost literally empty. For effective weak lensing
studies the number density $n$ of sources as well as the mean
redshift of sources $\langle z_s \rangle$ should be as high as
possible. However predictions about these faint radio sources and
their $\langle z_s\rangle$ and $n$ is currently less certain. If the
dominant population consists of normal or star-forming galaxies the
average redshift would be roughly unity. Additional populations of
sources can make the redshift distribution less certain. Another
complexity in the whole scenario is that we do not know to any
accuracy the shape distribution (i.e. the value of
$\sigma_{\epsilon}$) of the faint radio sources. Ideally one would
hope for near-spherical sources with no intrinsic ellipticities
acting as source objects, i.e. with $\sigma_{\epsilon}$ as low as
possible. If the radio sources in SKA are dominated by core-jet type
objects, then $\sigma_{\epsilon}$ can be very high, whereas normal
galaxies have relatively lower values. The field of view (FOV) for
SKA will be large and clearly the PSF will be controllable; this
makes it comparable or probably better than the present generation
of optical surveys. As discussed above, seeing provides a
fundamental limitation for optical telescopes: as they become
fainter, the source galaxies tend to become smaller too. This limits
the number density of sources for which the ellipticity 
can be measured reliably. A situation where point sources
dominate resolved source number counts is already present in some of
the deepest space-based images available to date.  These include the
Hubble Deep Fields North and South and recent images from the
Advanced Camera for Surveys (ACS) on the Hubble Space Telescope
(HST). Clearly the space-based images are much better compared to
their ground-based counterparts, but the FOV of such observations is
limited. Radio observations using future facilities could be highly
productive as they could combine high measurable source density and
large FOV. However, these conclusions will depend to some extent on
the intrinsic properties of the sources, and further assume that
sources are not too elongated and that the average redshift
distribution of these sources is not too shallow.

\subsection{Possibility of 21cm weak lensing studies}

Usually background objects can be broken down to individual sources at
optical and infra-red wavelengths, but the emission of 21cm radiation
\cite{ScottRees90, Madau97,Tozzi00, Iliev02, Furlentto04} from
neutral gas prior to reionization, and the CMB, provide examples of
truly diffuse backgrounds \cite{Peebles70,SunZel70,Silk68,HuDod2002}.

At high redshifts, typically beyond current large-scale surveys,
the unresolved point sources become more dominant. At a source
redshift of $z_s =15 - 30 $, first generations of stars and
protogalaxies start to appear according to current theory 
of galaxy formation. These
point-like objects provide a perfect background for weak lensing
studies due to the large distances that light rays need to travel
from these objects to reach us. Though detection of these objects is
beyond present observational technology, the spatial
fluctuations in the integrated diffuse background emission 
can be potentially interesting for lensing studies if
we assume  the most optimistic emission models.

The analytical formalism for weak lensing studies of diffuse
backgrounds has recently been established \cite{CoorayAst2004}.
Borrowing techniques from studies of weak lensing of the CMB, this
shows that a perturbative approach - typically employed to study CMB
lensing - will still be valid for diffuse background studies. This
is despite the fact that unlike the CMB there is considerable power
at small angular scales due to the lack of damping tail. This
indicates that a perturbative series has a comparatively slow
convergence rate and a larger number of terms need to be included
for realistic calculations.

In contrast to CMB lensing studies \cite{ZaldarriagaSeljak99,
benabed01, Guzik00, HuOkamoto02, Hirata03}, the weak lensing studies
based on diffuse components suffer from the fact that the lensing
modification to the power spectrum is minor at arc-minute angular
scales and the lensing information that one can extract from
low-redshift diffuse background is significantly limited. This is
related to the presence of significant structure in CMB power
spectra, as signified by the acoustics peaks, which is lacking in
the case of diffuse background studies. The presence of a damping
tail in the CMB means that the convolution associated with weak 
lensing effects transfers some power from larger angular scales
to smaller angular scales at the arc-minute range, which makes it
most easily detectable.  Use of polarization information in CMB
also carries much richer lensing information as compared to unpolarized
diffuse background studies. Moreover, at the last
scattering surface where the CMB temperature and polarization
fluctuations are generated they follow a Gaussian pattern. Then
lensing due to the intervening mass distribution imprints a
non-Gaussian footprint which can be effectively used to extract
cosmological information \cite{Bernar97,Kesden02}. There have
been extensive studies in this direction using higher-order
moment-based techniques (see e.g. ~\cite{CoorayKesden03}). By contrast
the non-Gaussianity generated by weak lensing in
diffuse backgrounds such as the 21cm background is unlikely to be
detected in near future. Nevertheless if possible 21cm weak lensing
studies can extend the reconstruction of the integrated matter power
spectrum out to redshifts of 15 to 30, and will bridge the gap
between current and upcoming galaxy lensing studies and the CMB.


\begin{figure}
\begin{tabular}{c}
{\epsfxsize=6.7cm\epsfysize=7cm\epsffile{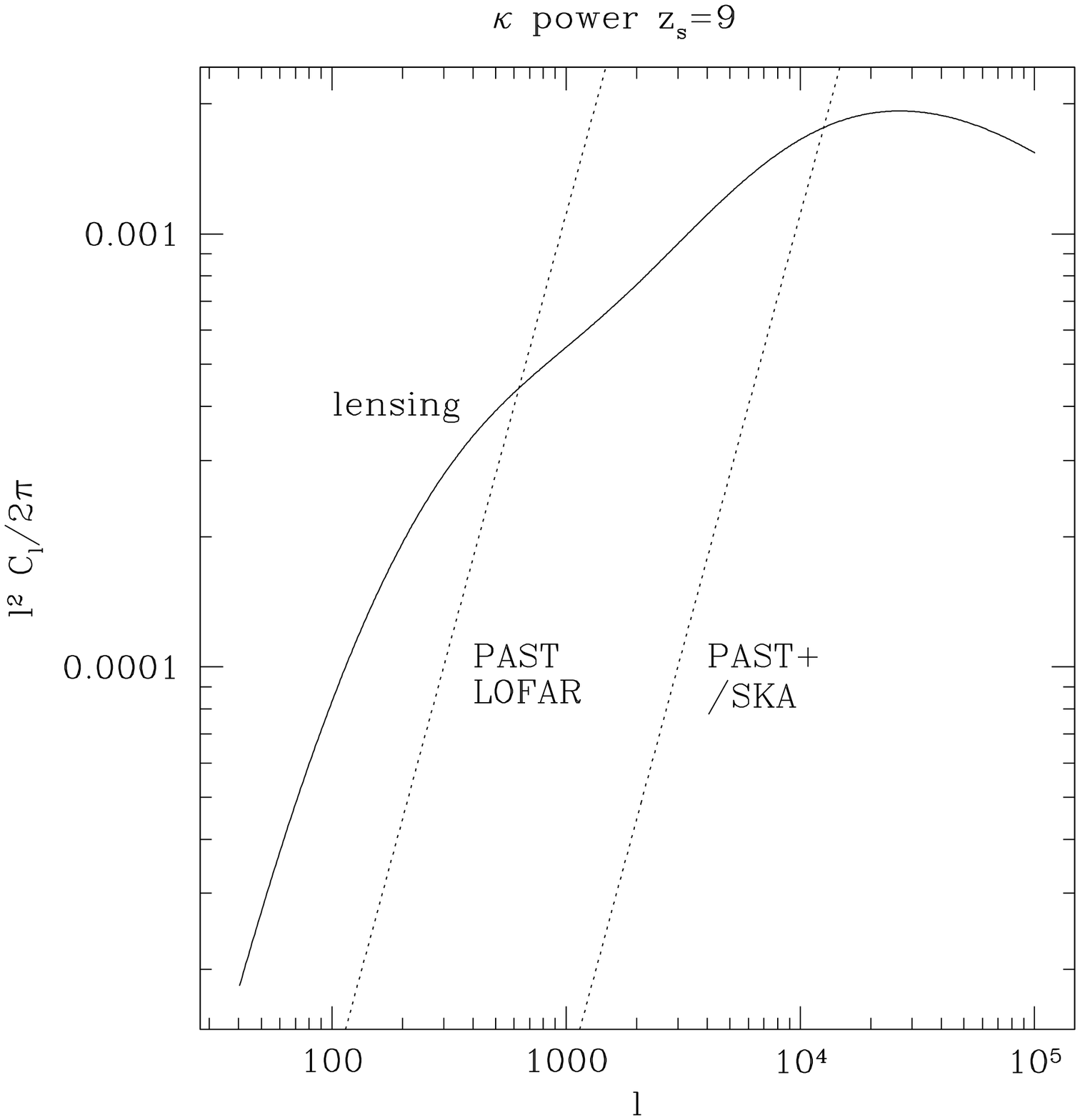}}
{\epsfxsize=6.7cm\epsfysize=7cm\epsffile{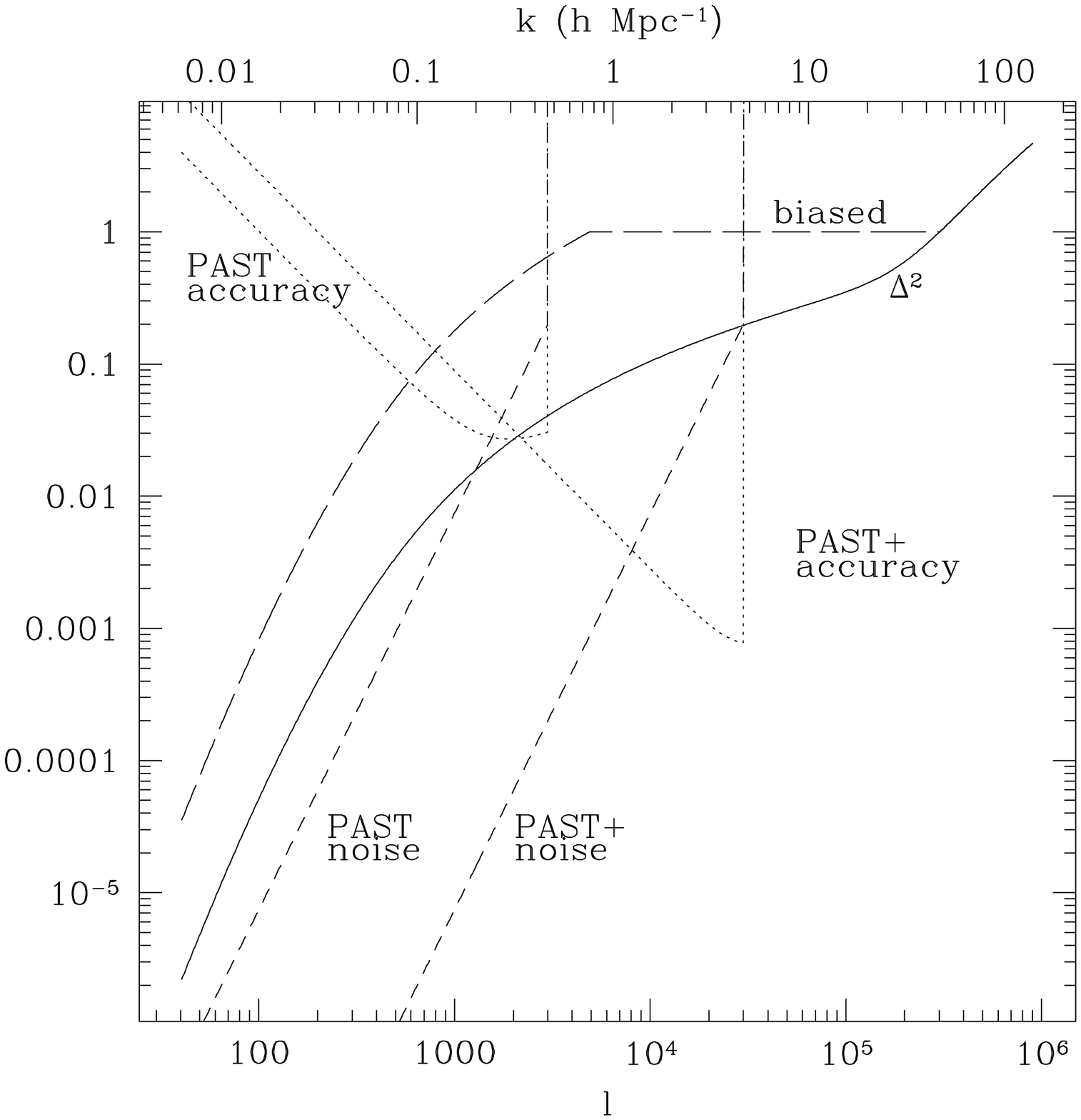}}
\end{tabular}
\caption{In the left panel convergence power spectra for redshift $z_s=9$
are plotted.  The dotted line on the left corresponds to noise level expected from
PAST/LOFAR where as the dotted line on the right corresponds to  noise level expected from second generation PAST+
and the SKA. Measurement errors and accuracy of various surveys in estimating matter power spectra
are compared on the right panel. Solid line corresponds to the matter power spectrum at z=9 as in left figure.
Dot-dashed line corresponds to power spectrum of patchy reionization with a bias factor $b=4$
for more details). Dashed lines correspond to noise levels expected for PAST and second generation PAST+.
Dotted line corresponds to fractional accuracy of power spectrum estimation per logarithmic $l$  bins. (Figure Courtesy Ue Li-Pen)}
\label{UeLiPen}
\end{figure}

\subsection{Using resolved mini-halos for weak lensing studies}

Concentrating on  mini-halos that will contribute to the 21cm
background radiation, \cite{LiPen04} found that instruments such as
PAST, LOFAR would tremendously improve our knowledge of some key
cosmological parameters. Before reionization, most of the baryonic
matter in the Universe is in the form of neutral hydrogen. The first
gravitational bound objects are the collapsed  virialized dark
matter mini-halos where this neutral hydrogen resides.  These
mini-halos \cite{Madau97, Iliev02,Iliev03} provide a fluctuating
background with a characteristic scale that can be used for weak
lensing studies.

For weak lensing studies, one would like
to have a background at a high source redshift that also exhibits
structures at small scales.  Clearly the CMB satisfies the first
criterium but it is smooth at small scales. It was pointed out that the
Universe at the epoch of reionisation might be the most natural
place to look for applications of weak lensing studies. Clearly not only it is at a
very high redshift, but also it emits brightly in the hydrogen
hyperfine transition line and has structures on many scales ranging
from several arcminutes to under a milliarcseconds. Experiments
which will target these particular redshifts and wavelengths are being
planned, including some that are already undergoing construction,
e.g. PAST\footnote
{http://astrophysics.phys.cmu.edu/~jbp/past6.pdf}, LOFAR
\cite{Kassim00}\footnote{http://www.lofar.org/}.  Other experiments
which will have low signal-to-noise include T-REX\footnote
{http://orion.physics.utoronto.ca/sasa/Download/poster/casca\_poster.pdf}
and CATWALK\footnote{ftp://ftp.astro.unm.edu/pub/users/john/AONov03.ppt}.

Presenting a detailed calculation of signal to noise analysis
\cite{LiPen04} argues that weak lensing studies of
epoch-of-reionization gas  can constrain the projected matter power
spectrum to very high accuracy. Use of tomography can further
increase the level of accuracy with which certain cosmological
parameters can be constrained. It was claimed that with such a
technique, the neutrino mass could be constrained with an accuracy
of $0.1$ meV. The inflationary gravity-wave background and
consequently the inflationary dynamics as encoded by the Hubble
parameter during inflation can be constrained with high accuracy
too. These calculations show that such 21cmm weak lensing
observations are an order of magnitude better than those from galaxy
surveys. However such optimistic scenarios will require resolving
each of the $10^{18}$ mini halos that will be observed on the sky.

Clearly several problems arise when one tries to map gravitational
lensing at such an ambitious scale. For example \cite{OhMack2003}
have shown that synchrotron emission from ionised gas can outshine
the 21cm radiation. However, \cite{Furlentto04} pointed out that
such components can be removed by power-law spectra from spatial
fluctuations.



\section{Weak Lensing of the Cosmic Microwave Background}
\label{CosmicMicrowaveBackground}

Observation of cosmic microwave background (CMB) radiation is one of
the cleanest probes of cosmology ~\cite{HuDod2002, Hu2003, Dod2003,
HuWhite2004, Chall2004, DS2006}. However {\em lensing} of CMB
photons by intervening mass clumps can provide additional
information about the structure and dynamics of the Universe.
Besides, lensing of the CMB provides information at larger scales and higher
redshift than can be reached by any other astronomical observations.
Weak
lensing of the CMB is responsible for many observable effects which
have been studied in extensive detail - for a detailed
review see \cite{lewischallinor2006}.  Calculations of these effects
have been made both for temperature and polarisation anisotropies
~\cite{BlanSch1987,ColeEfsth1989,Linder1990,Seljak1996,MS1999}.

\subsection {Effect of weak lensing on the temperature and polarisation power-spectrum}

Lensing broadens the acoustic peaks and enhances power at small
angular scales. These features are non degenerate with the standard
cosmological parameters. This allows a determination of the lensing
amplitude by comparing the observed spectra with CMB spectra of
different mass fluctuations.The magnitude of distortion is sensitive
to the level of  mass fluctuations as a function of redshift and
scale. This in turn depends on the background cosmology. In this way
lensing of CMB can be used as a window to probe the fluctuations in
the dark matter distribution over a huge range of redshifts and length
scales.

Lensing generates a non zero B-mode polarization from a purely
E-mode. (Note that here we are referring to E and B modes in the CMB
polarisation map, not in the shear map).   B-modes generated during
inflation by tensor mode perturbations or gravitational waves
produce a distinct spectrum which can be used to discriminate
between different classes of inflationary models. Lensing of E-modes
produces B-modes which dominate the more primordial signal on small
scales, and which are considered mainly as a source of confusion in
polarisation experiments. Detection of a primordial B-mode signal in
the presence of lensing therefore pushes observational strategies
towards favouring larger sky coverage
\cite{KamionKosowStebb1997,SeljakZal1997}. A low sky coverage and
presence of boundaries  also causes additional confusion by
introducing mixing of E and B modes. It is however useful to note
that as the inflationary B-mode signal is mainly significant on
large angular scales experiments with low resolution can also be
very promising. High resolution B-mode detection experiments will
typically employ a ``delensing'' step in data reduction to
effectively restore the unlensed sky from the lensed data.

Typically the power spectrum of the lensed map is computed by a series expansion
in the deflection angle. A perturbative expansion in the deflection angle
is a transparent way to understand most of the lensing effects. Non-perturbative
evaluations of lensing effects are carried out by considering the correlation
function \cite{Seljak1996,ChallinorLewis2005,CoorayHu2002,Zal2000,CoorayAst2004,MandelZal2005}.

\begin{figure}
\setlength{\unitlength}{.5cm}
\begin{picture}(12,11.5)(0,0)
\epsfig{file=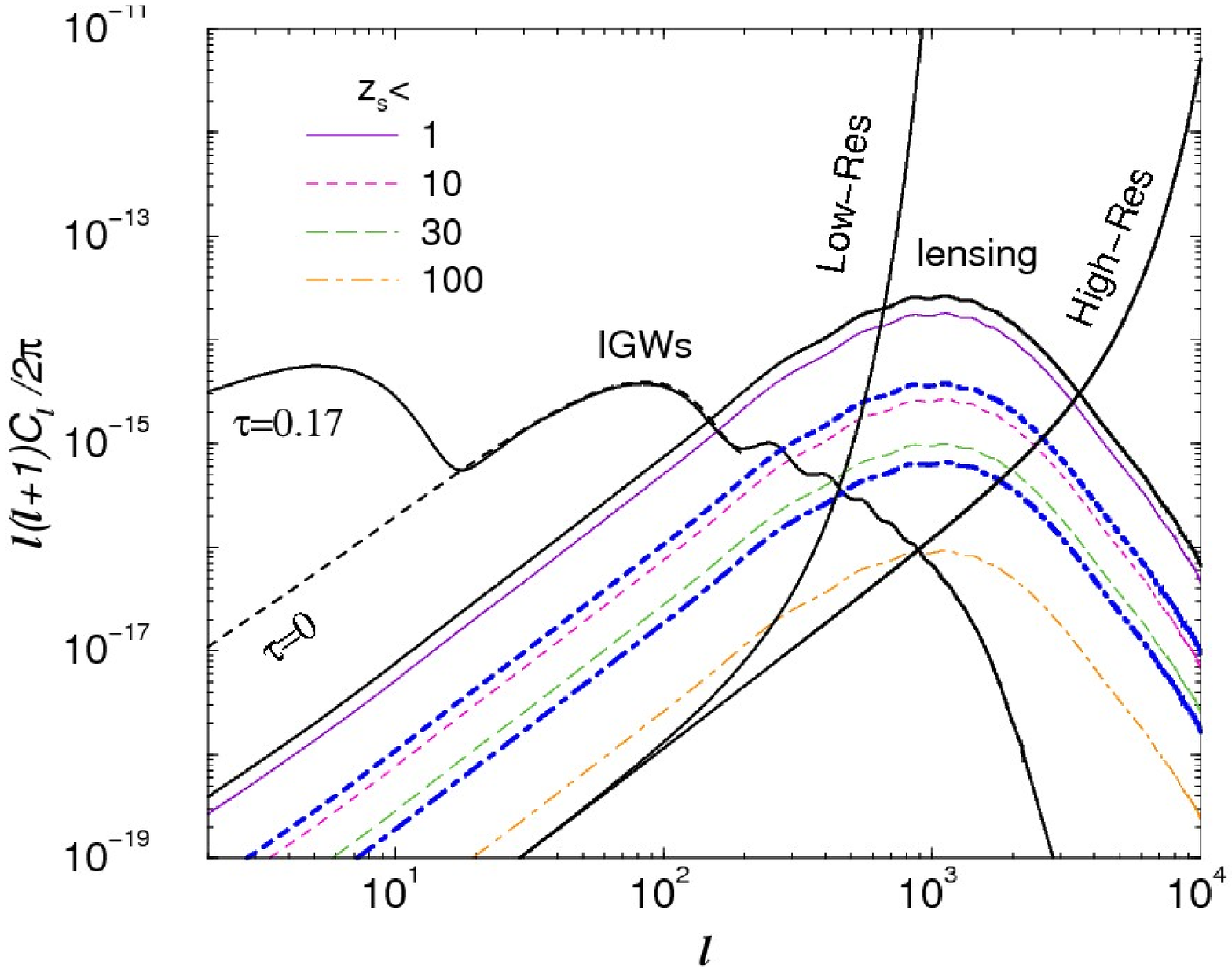,height=5cm,width=7cm}
\epsfig{file=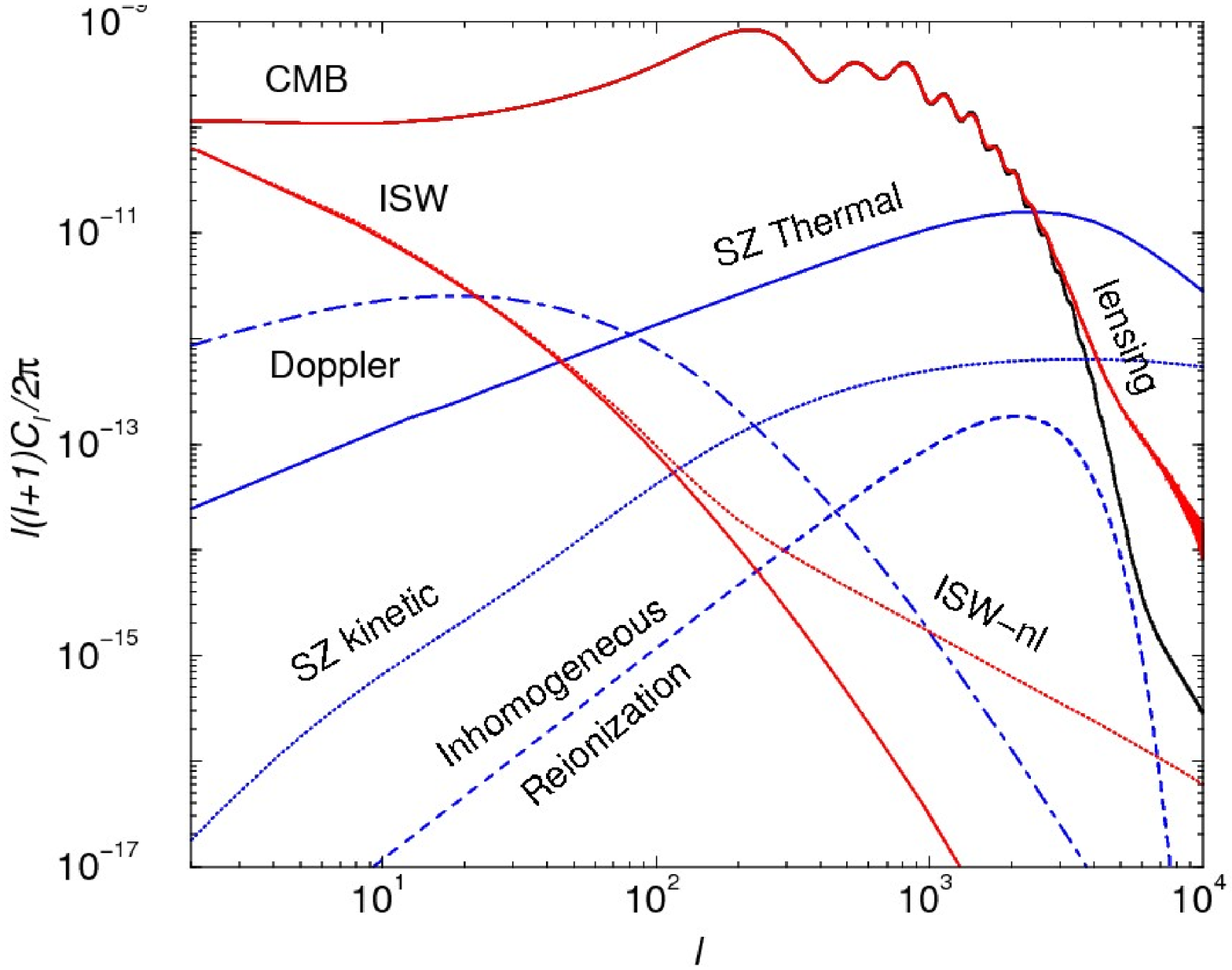,height=5cm,width=7cm}
\end{picture}
\caption{Left Panel: Power spectrum for the temperature
anisotropies in the fiducial $\Lambda$CDM model
with $\tau=0.1$. In the case of temperature, the curves show the
local universe contributions to CMB due to
gravity (ISW and lensing) and scattering (Doppler, SZ effects,
patchy reionization). See, Cooray,Baumann \& Sigurdson (2004) for a
review on large scale structure contributions to temperature anisotropies (Figure Courtesy: Asantha Cooray).
Right Panel: CMB B-mode polarization. The curve labeled `IGWs'
is the IGW contribution with a tensor-to-scalar ratio of 0.1 with (solid line; $\tau=0.17$) and without (dashed line)
reionization. The curve labeled `lensing' is the total lensing confusion to B-modes.
Thin lines show the residual B-mode lensing contamination for removal of with lensing out to $z_{s}$.
See, Sigurdson \& Cooray (2005) for details.}
\end{figure}


\subsection{Non-Gaussianity in the CMB induced by Weak Lensing}

Primordial perturbations produced during inflation are very nearly
Gaussian and the linear evolution of these perturbations up to the last-scattering
surface does not generate any non-Gaussianity. Weak lensing
by the intervening matter distribution however does introduce
non-Gaussianity in maps of the CMB sky \cite{Zal2000}.

All odd-order correlation functions vanish for a Gaussian random
field. On the other hand weak lensing generates a non-zero contribution through
correlations between the large-scale temperature and lensing
potential. This cross-correlation which is due to the Integrated
Sachs-Wolfe (ISW) effect at large angular scales can therefore be a
very useful tool to probe the growth of perturbations and the
expansion history of the Universe
\cite{Hu2002,SelkakZald1999,GoldSper1999,Giovi03,Giovi05,Gold06}.
This implies that such studies can be very useful tools for studying
the dark energy equation of state \cite{Stompor99} or the neutrino
mass \cite{Les06}. These correlations are negligible on smaller
scales. Besides non-linear evolution of the matter distribution as well as
late time non-linear effects including the Sunyaev-Zeldovich (SZ)
effect also contribute to the bispectrum. The
four-point correlation function is the lowest order non-zero
correlation function which does not vanish in the absence of any the
cross-correlation between the low redshift mass distribution and the
CMB temperature distribution. The non-Gaussianity studies involving
polarisation fields are very similar to the temperature case, at
least for the E-mode polarisation. Various combinations of
four-point correlation functions involving E- and B- mode
polarisations have been studied in the literature
\cite{Bernar97,Hu01}. For future high resolution all-sky
polarisation surveys such studies will be feasible.

\subsection{Weak lensing effects as compared to other secondary anisotropies}

The thermal Sunyaev-Zeldovich (tSZ) effect, which is caused by the
scattering of photons from hot electrons in clusters, is a dominant anisotropy
contribution on small scales. Due to a very characteristic frequency
spectrum such a component can however be readily separated from
primordial anisotropies. Another secondary contamination, the
kinetic Sunyaev-Zeldovich (kSZ) effect, has the same frequency
dependence as the primary anisotropies, so is harder to separate.
Similarly, lensing does not cause any frequency shift in an
otherwise perfect blackbody spectrum of primary CMB. Thus non-linear
sources, such as kSZ are potential sources of confusion, when trying
to understand the effect of lensing on the CMB. Current uncertainties in the
reionization history and topology of reionization patches also make it
more difficult to model. For polarisation spectra the kinetic SZ is
sub-dominant, simplifying the situation to some extent
\cite{Valageas01,Gibli97,ChallinorFord00, Shimon06}.

\subsection{Lensing of the CMB by individual sources}
On small scales, the CMB lacks power.  As mentioned earlier, most of
the power is generated by secondary anisotropies and transfer of
power from larger scales to smaller scales due to lensing. This
raises the possibility of detecting individual cluster-mass objects
in CMB maps by their effect of lensing on the CMB. Unlike other
probes such as the SZ which depends on baryon physics, such studies
can constrain the physical mass distribution of individual objects
directly \cite{LewisKing05,dodStark03,Cooray03b}.

\subsection{Future Surveys}
In the future, satellite experiments such as
Planck\footnote{http://www.rssd.esa.int/index.php?project=Planck} (to
be launched in 2008) will be in a good position to detect the effect
of lensing in temperature power spectra.In the case of polarisation,
experiments such as CLOVER\footnote
{http://www.astro.cf.ac.uk/groups/instrumentation/projects/clover/}
and QUIET\footnote {http://quiet.uchicago.edu}, which are being
planned, will have detection of B-mode from inflation as their
primary science driver \cite{liddlelyth2000, MukhaFeldBrand1992,
SteinTurok2002, LindeMukhaSasa2005}.  Ongoing experiments such as
QUaD\footnote{http://www.astro.cf.ac.uk/groups/instrumentation/projects/quad/},
may reach the required sensitivity, but for all of these, lensing is
the main source of confusion on small scales for such experiments.



\section{\label{External-data-sets} Weak lensing and External data sets: Independent and Joint Analysis}

Weak lensing is a very powerful probe of the projected dark matter
clustering, and in principle it is very good at constraining the
cosmological parameters playing a dominant role in the structure
growth. The main limitation comes from the strong degeneracy between
the dark matter power spectrum normalisation $\sigma_8$ and the
matter density $\Om$ (see the introduction of Section~\ref{Non-Gaussianities}
and Eq.(\ref{kappa})). These two parameters are degenerate
because any change in $\Om$ can be balanced by an appropriate
change in $\sigma_8$ slowing down or accelerating the growth rate.
In \cite{Bernardeau_et_al1997} it was shown that the shear variance
at scale $\theta_s$ scales roughly as
\begin{equation}
\langle \bargamma^2\rangle_{\theta_s} \propto \sigma_8^2\Om^{1.5}
z_s^{1.5} \theta_s^{-{(n+2)\over 2}}. \label{scaling}
\end{equation}
This equation, which is the first term of the perturbation series on
the mass density contrast, assumes a power-law power spectrum with
constant slope $n$ and a single source redshift $z_s$. It has a very
limited application but it has a pedagogical value in that it shows
that the parameter degeneracy between $\sigma_8$ and $\Om$
extends to the source redshift as well. Therefore, weak lensing in
2D can become a high precision cosmology tool only if it is combined
with another cosmology probe and provided we have a {\it good}
knowledge of the source redshift distribution. The simultaneous
observation of non-linear and linear scales in lensing surveys shows
some features in the projected mass power spectrum which helps to
lift this degeneracy \cite{JS97}. Alternative approaches which can also
lift degeneracies were discussed in Section~\ref{ch4:3Dweaklensing}, using 3D
information, and in Section~\ref{Non-Gaussianities},
using higher-order correlations beyond the shear variance.
However, like any other
cosmology probe, this is not enough for weak lensing to be a high
precision cosmology tool alone. In this Section we review the
advantages of combining lensing with different probes and we outline
the gain on cosmological constraints in the future missions.

\subsection{With CMB, Supernovae and Baryon Acoustic Oscillations to probe Cosmology}

\begin{figure}
\centerline{\psfig{figure=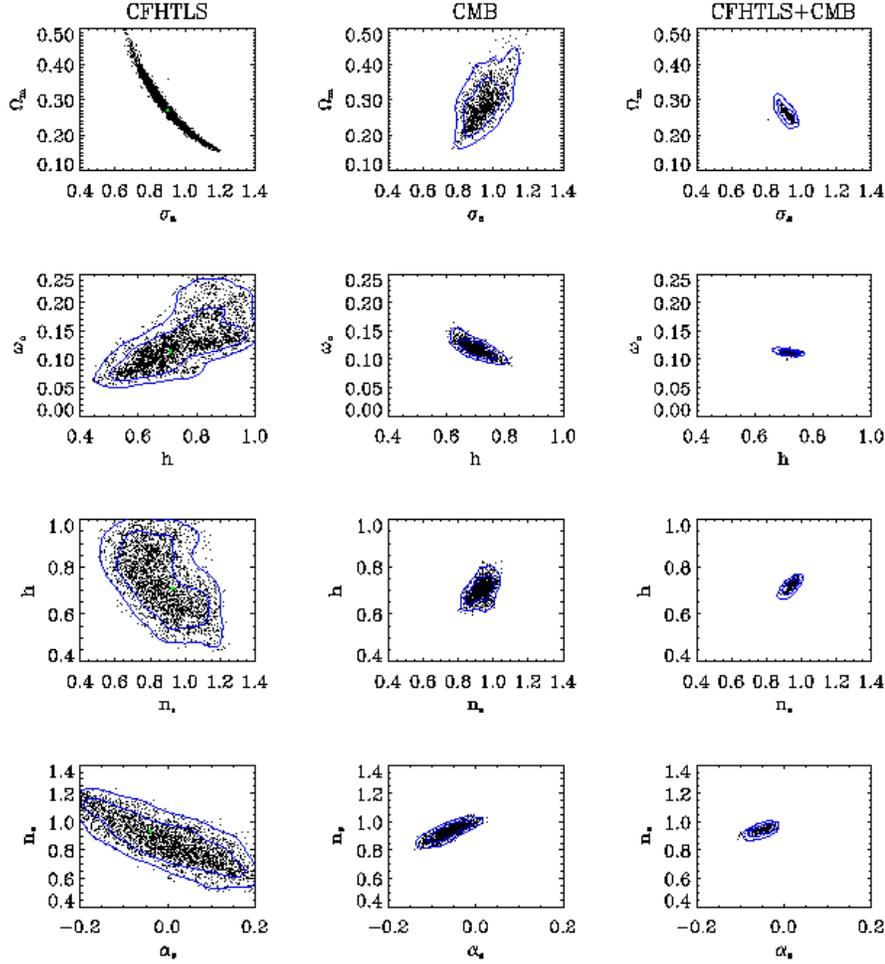,height=13cm} }
\caption[]{Figure showing the most orthogonal parameter degeneraties
between CMB (WMAP1) and weak lensing (CFHTLS). The set of parameters
is $\Om$, $\sigma_8$, $\omega_c=\Omega_c h^2$, $n_s$,
$\alpha_s$ and $h$ (from Tereno et al. 2004). \label{cfhtlscmb}}
\end{figure}

Weak lensing alone provides a measure of $\Om^{0.7} \sigma_8$,
provided that the redshift of the sources is known. This means that
weak lensing best performance is in the measurement of the amplitude
of the projected mass power spectrum. For this reason, it is very
powerful at breaking the parameters degeneracies seen in other
cosmology probes, which are usually sensitive to other combinations of
cosmological parameters. It is known, for instance, that a precise
measurement of $\sigma_8$ and $\Om$ can be obtained from the
combination of lensing and CMB \cite{vanWaerbeke_et_al2002,contaldi2003}.
Figure~\ref{cfhtlscmb} shows the set of parameters to be combined
for an optimal joint CMB-lensing analysis \cite{tereno2004}. This
result, obtained for the CFHTLS and WMAP1 surveys, is not modified
for a different choice of lensing and CMB data sets. Note that the
best improvement is obtained on parameters which the shape of the
dark matter power spectrum is the most sensitive to: the mass
density $\Om$, the power spectrum normalisation $\sigma_8$, the
reduced Hubble constant $h$, the primordial power spectrum slope $n_s$ and
the running spectral index $\alpha=-{\rm d}\ln(n_s)/{\rm d} \ln k$.

\begin{figure}
\centerline{\psfig{figure=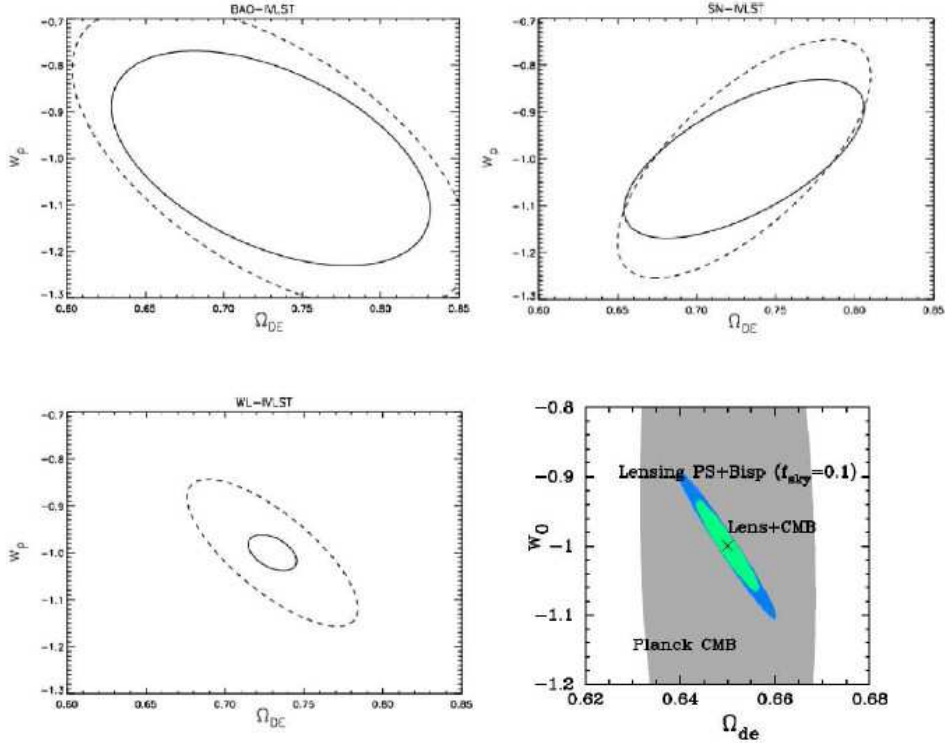,height=10cm}}
 \caption[]{Panels showing the measurement forecast of the dark energy
 equation of state parameter $w_p$ and dark energy density
 $\Omega_{DE}$ from the baryon oscillations (top-left panel),
 supernovae Ia (top-right panel) and weak lensing (bottom-left) from
 the DETF report (2006).  $w_p$ is the equation of state parameter
 of the Dark Energy parameter at an intermediate `pivot' redshift,
 usually a little less than 0.5.
 Solid and dashed lines are for optimistic and pessimistic surveys
 respectively. The surveys characteristics are described in the
 text. The bottom-right panel shows the constraints from the PLANCK
 CMB experiment with a lensing survey (Takada \& Jain 2005).
\label{forecast}}
\end{figure}

For the lensing signal alone, an improvement of a factor $\sim 3$ in
the parameter errors is expected if the lensing signal is combined
for different source redshift slices instead of measured from the
broad source distribution \cite{Takada_Jain2004}. This is the well known
tomography technique which requires the measurement of photometric
redshifts (see Section~\ref{tomography}).
Unfortunately, for these parameters, the improvement
cannot be further increased with a larger number of source slices:
one can show that the signal-to-noise saturates for a maximum of 3-5
redshift slices \cite{TW04}, which is a consequence of the fact that the
lensing selection window is a rather flat function of redshift where the
signal is the strongest. Another source of improvement for the
lensing constraints is the use of higher-order statistics
(\cite{Takada_Jain2004}, see Section~\ref{Non-Gaussianities} of this review): the probe of the
non-linear regime of structure formation helps considerably in the
determination of the precise moment when large scale structures
become non-linear and at which scale. This event is a strong
function of the cosmological parameters, in particular of dark
energy \cite{MB04,BDW}. Some measurement of high order statistics
have been done \cite{BMVW,Pen_et_al2003b}, but this area of research is
still in its infancy.

The main science driver for cosmology has become the measurement of
the Dark Energy equation of state. The Dark Energy Task Force report
(DETF) \cite{DETF} and the ESA-ESO working group report
\cite{ESAESO} are summaries of where cosmology is heading to for the
next decade: there is a consensus among cosmologists that the goal
of measuring the dark energy parameters can only be achieved from
the joint analysis of several cosmological probes. Figure
\ref{forecast} shows the dark energy parameters forecast for the
future experiments (which includes a pessimistic and optimistic
cases). Most of the panels on this figure are extracted from the
DETF report, and they show the various degeneracies for different
cosmological probes. The three projects that DETF has considered are
the following:

\begin{itemize}
\item{} Baryon Oscillations: $20000$ square degrees, ground-based
survey. Photometric redshifts cover the 0.2-3.5 range, and their
precision is $0.01$ for the optimistic case and $0.05$ for the
pessimistic one.
\item{} Supernovae: ground-based survey with 300000 supernovae.
The photometric redshift accuracy is $0.01(1+z)$ for the optimistic
case and $0.05(1+z)$ for the pessimistic.
\item{} Weak Lensing: $20000$ square degrees, ground-based survey.
The shear calibration is $f_{cal}=0.01$ and photometric redshift
accuracy $\sigma_z=0.01(1+z)$ for the pessimistic case.
$f_{cal}=0.001$ and $\sigma_z=0.001(1+z)$ for the optimistic case.
\end{itemize}
Note that these surveys are all providing predictions of very
accurate determination of the dark energy equation of state $w_p$
and energy density $\Ode$, far better than any joint
analysis would do today \cite{spergel06}. Here $w_p$ is the equation of
state at an intermediate reshift, typically around 0.4. The bottom
right panel on Figure~\ref{forecast} is from \cite{Takada_Jain2004} and shows
the joint constraints on $w_p$ and $\Ode$ from the future
CMB PLANCK mission and a lensing survey covering $4000$ square
degrees, combining the two and three-points statistics. Figure
\ref{forecast} shows that the supernovae, CMB and lensing (or BAO)
have very different degeneracies in the dark energy parameter space,
which offers an optimal complementarity in the combination of
cosmology probes. It is believed that only the combination of all
probes together will provide convincing constraints regarding the
dark energy, although each individual probe seems to predict  very
accurate results. The main limitation being the systematics, it is
indeed important to have more constraints than parameters we wish to
measure: weak lensing for instance may appear as the most powerful
probe. However, the lensing signal is rather featureless compared to
CMB and BAO, which makes it more vulnerable to systematics like
shear calibration and photometric redshift inaccuracies
(see Section~\ref{ch:6}). The future of cosmology certainly
lies in a joint analysis of
surveys, and lensing surveys play a particular role in the sense
that this is the only probe which can provide an unbiased
measurement of the dark matter fluctuations amplitude.
Note that the
constraints on quintessence models can be improved by the combination of
various cosmology probes, such as lensing and SNeIa as demonstrated in
\cite{Schmid06}.

The success of the joint analysis is subject to the validity of our
Cold Dark Matter and Dark Energy picture. Given the unknown nature
of these ingredients, it is possible that our description might be
incomplete or wrong. We must therefore keep an open mind and
consider alternative interpretations of the data. The modified
gravity theory proposed by \cite{milgrom} lacks a solid physical
motivation and it appears to be unable to explain the halo
flattening seen in weak lensing data \cite{HYG04}. Nevertheless,
this study initiated the idea that a modified theory of General
Relativity could solve the dark matter and dark energy problems.
These models include the string-motivated braneworld scenarios
\cite{dvali2000}. Ref.\cite{UB01} suggested that modified gravity can be
tested directly from a verification of the Poisson equation
\begin{equation}
\nabla^2\Phi=4\pi G\rho.
\end{equation}
The gravitational potential $\Phi$ can be measured from weak lensing
by large-scale structures, and the mass density $\rho$ from the
galaxy distribution, which at large enough scale can be assumed to
be an unbiased tracer of the mass distribution, as shown in \cite{verde02}
from the 2dF Galaxy Redshift Survey.

A general problem with modified theories of gravity is the
significantly increased level of degeneracy due to a larger number
of degrees of freedom. As pointed out by \cite{HL06}, it is
necessary to measure independently the growth rate of structures and
the Universe expansion rate in order to reduce the degeneracy to a
reasonable level. Modified gravity also might not be the cause of
all the dark components of the Universe. The investigation of a
realistic mix of dark component and modified gravity in \cite{HL06}
shows that neglecting the modified gravity severely biases the dark
energy parameters, even with a joint Supernovae, CMB and weak
lensing constraints. To make things worse, the addition of inflation
parameters introduces quantum correction, the running spectral
index, to the primordial mass power spectrum which is a
scale-dependent effect. Yet another degree of complication comes
from the small-scale amplitude prediction which is highly
non-linear. Dark Matter itself could also be self-interacting as
suggested by \cite{CMH92}, resulting in density-dependent
observational effects \cite{AZ06}.

The present situation is that the simple Big-Bang scenario (i.e. no
alternative theories) will indeed be accurately constrained by a
joint analysis of different surveys. A general scenario including
alternative theories might be harder to constrain. Whether or not
this is possible with the next generation of surveys remains to be
demonstrated, but it is clear that weak lensing by large scale
structure plays a central role in being sensitive to the ``dark
matter" whether it is real dark matter or modified gravity.

\subsection{With Galaxy Surveys to probe bias}

Weak lensing is the only reliable technique able to probe the dark
matter distribution up to redshift of a few using optical surveys,
or at even higher redshifts from lensing of the CMB and the 21cm
line. The peculiar velocity field, which is also an unbiased tracer
of the matter distribution, is no longer accurate at distances
larger than a few hundred Megaparsecs. The combination of lensing
with the galaxy distribution is therefore a unique way of
constraining the relative amount of matter with respect to light,
the so-called bias. This can be done out to reasonably high
redshift, giving astronomers access to the dark side of galaxy
formation over a wide range of its evolutionary history.

\subsubsection{Galaxy biasing}
\label{Galaxy biasing}

An apparent limitation of lensing surveys is that the mass is only
seen in projection, impeding a 3-dimensional probe of the bias. A
method for alleviating this problem was proposed by \cite{vw98} and
\cite{schneider98}. It is an alternative to the 3D mass
reconstruction technique discussed in Section~\ref{ch4:3Dweaklensing},
which was designed
specifically for potential or density measurements. We assume a
population of foreground galaxies with a known narrow redshift
distribution $n_{\rm f}(z)$ centered on $z_{\rm f}$ (in units of arcmin$^{-2}$),
from which we want
to measure the mass-to-light ratio. The lensing signal is measured
from a background galaxy population which can have a broad redshift
distribution $n_{\rm b}(z)$ (normalized to unity), while the technique
discussed in Section~\ref{ch4:3Dweaklensing} requires an
estimate for all redshifts. The cross-correlation of the lensing
signal with the foreground galaxy density distribution provides a
measurement of the bias at redshift $z_{\rm f}$. The scale dependence of
the bias can be obtained by filtering the lensing signal and the
galaxy density with an aperture filter, which is a narrow band
filter (Fig.~\ref{kernels}).

Assuming that the galaxy number density contrast $\delta_{\rm gal}$ is proportional
to the matter density contrast $\delta$ with the proportionality
factor defined as the bias parameter $b$, the lensing-galaxy density
cross-correlation at scale $\theta_s$ is given by \cite{vw98}:
\begin{equation}
\langle \Map(\theta_s){\cal N}(\theta_s)\rangle
= b \pi \theta_s^2 \int_0^{\chirad_{\rm max}} \d \chirad \,
\hat{n}_{\rm f}(\chirad) \frac{\wh(\chirad)}{\De^2(\chirad)}
\int \frac{\d \ell}{2\pi} \, \ell \,
P\left({\ell\over\De(\chirad)};\chirad\right) W_{\Map}^2(\ell\theta_s) ,
\label{MN_def}
\end{equation}
see Eqs.(\ref{kappa})-(\ref{Map}),(\ref{Powerkappa}) and (\ref{Map_Pkappa})
in Sections~\ref{ch:2} and \ref{ch2:2Pointstatistics}. Here the weight
$\wh(\chirad)$ along the line of sight depends on the redshift distribution
$n_{\rm b}(z)$ of the background sources as in Eq.(\ref{kappanz})
whereas $\hat{n}_{\rm f}(\chirad)$ is the redshift distribution of the foreground
galaxies (with $\hat{n}_{\rm f}(\chirad)\d\chirad=n_{\rm f}(z)\d z$) of
``number counts'' ${\cal N}$ with respect to the aperture of radius $\theta_s$.
$P(k;\chirad)$ is the time-evolving 3-D matter power spectrum (where
$\chirad$ is a parameterization of the redshift along the line of sight)
and $W_{\Map}^2(\ell\theta_s)$ is the square of the Fourier transform of the
aperture filter, which peaks at some effective wavelength $\ell_{\rm
eff}\sim 5/\theta_s$, see Fig.~\ref{kernels}. For a narrow foreground redshift
distribution, the function $\hat{n}_{\rm f}(\chirad)$ peaks at some radial
comoving distance $\chirad_{\rm f}(z_{\rm f})$. Eq.(\ref{MN_def}) shows that
the cross-correlation is dominated by
$P\left(k_{\rm eff};\chirad_{\rm f}\right)$ where
$k_{\rm eff}\sim 5/(\De(\chirad_{\rm f})\theta_s)$. In order to extract
the bias parameter $b$, we need to define the variance of the number density
fluctuations for the foreground galaxies:
\begin{equation}
\langle {\cal N}^2(\theta_s)\rangle = b^2 (\pi\theta_s^2)^2 \int\d\chirad
\frac{\hat{n}_{\rm f}^2(\chirad)}{\De^2(\chirad)}
\int \frac{\d \ell}{2\pi} \, \ell \,
P\left({\ell\over\De(\chirad)};\chirad\right) W_{\Map}^2(\ell\theta_s) ,
\label{N2_def}
\end{equation}
and define the ratio $R$
\begin{equation}
R\equiv{\langle M_{\rm ap}(\theta_{\rm c}){\cal N}(\theta_{\rm
c})\rangle\over \langle {\cal N}^2(\theta_{\rm c})\rangle}.
\end{equation}
It was shown in \cite{vw98} that this ratio is nearly independent of
scale for all cosmologies and any dark matter power spectrum, unless
the bias parameter is scale-dependent (this statement was shown to
be valid in the non-linear regime as well). The bias at angular
scale $k_{\rm eff}^{-1}$ and redshift $z_{\rm f}$ can therefore be
measured from weak lensing data.

\begin{figure}
\centerline{\psfig{figure=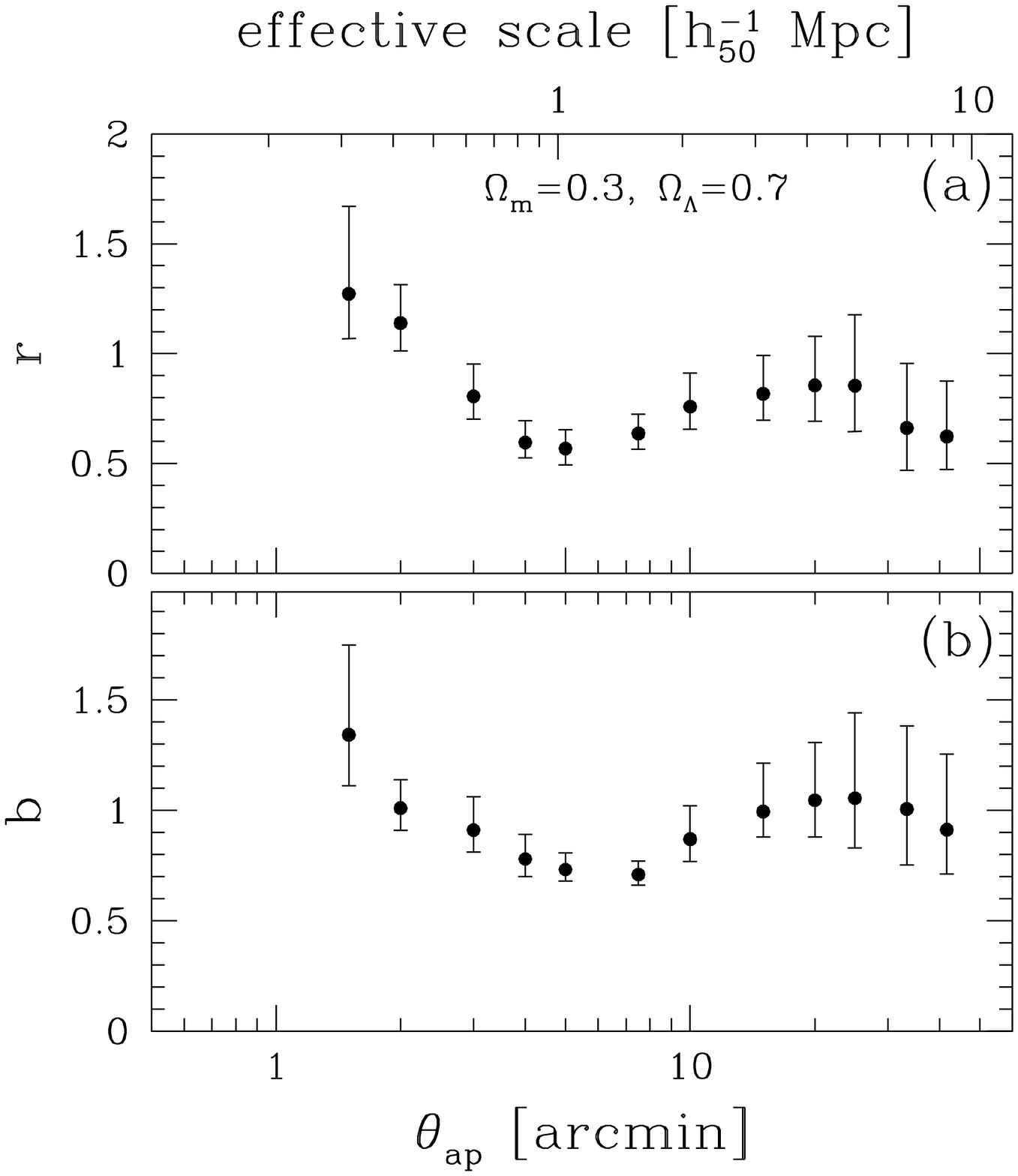,height=7cm}
\psfig{figure=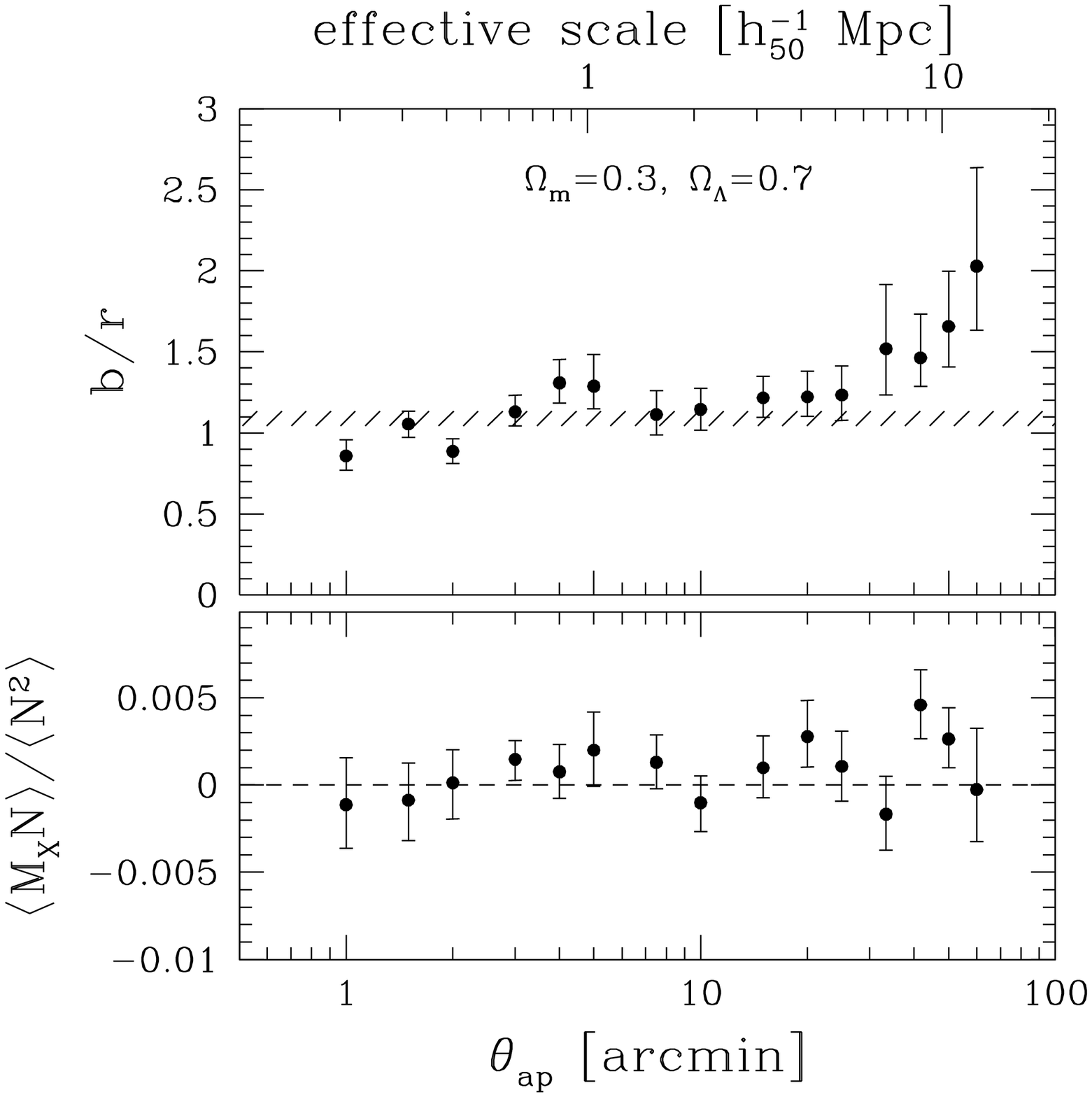,height=7cm}} \caption[]{Galaxy
biasing constraints from the VIRMOS and RCS weak lensing surveys
(Hoekstra et al. 2002). The left panel shows the cross-correlation
coefficient $r$ and biasing $b$ as function of scale. The right
panel, top plot, shows the ratio $b/r$ and its prediction for a
concordance model $\Omega_m=0.3$, $\Omega_\Lambda=0.7$. The bottom
plot shows the most convincing test of residual systematics obtained
by rotating the lensed galaxies by $45$ degrees, which is supposed
to cancel the lensing signal. \label{bias}}
\end{figure}

Refs.\cite{Hetal2002} \cite{simon2006} have performed the only
application of this technique. \cite{Hetal2002} measured the bias
from a combination of the VIRMOS \cite{vanWaerbeke_et_al2002} and RCS
\cite{Hoekstra_et_al2002} surveys, which is shown on Figure \ref{bias}. This
analysis shows a significant scale dependence of the bias, although
its calibration is still uncertain due to incomplete knowledge of
the source redshift distribution. With the GaBoDS surveys,
\cite{simon2006} found $b=0.8\pm 0.1$ and $r=0.6\pm 0.2$, which is
consistent with \cite{Hetal2002}.

\subsubsection{Galaxy-galaxy Lensing}

\begin{figure}
\centerline{\psfig{figure=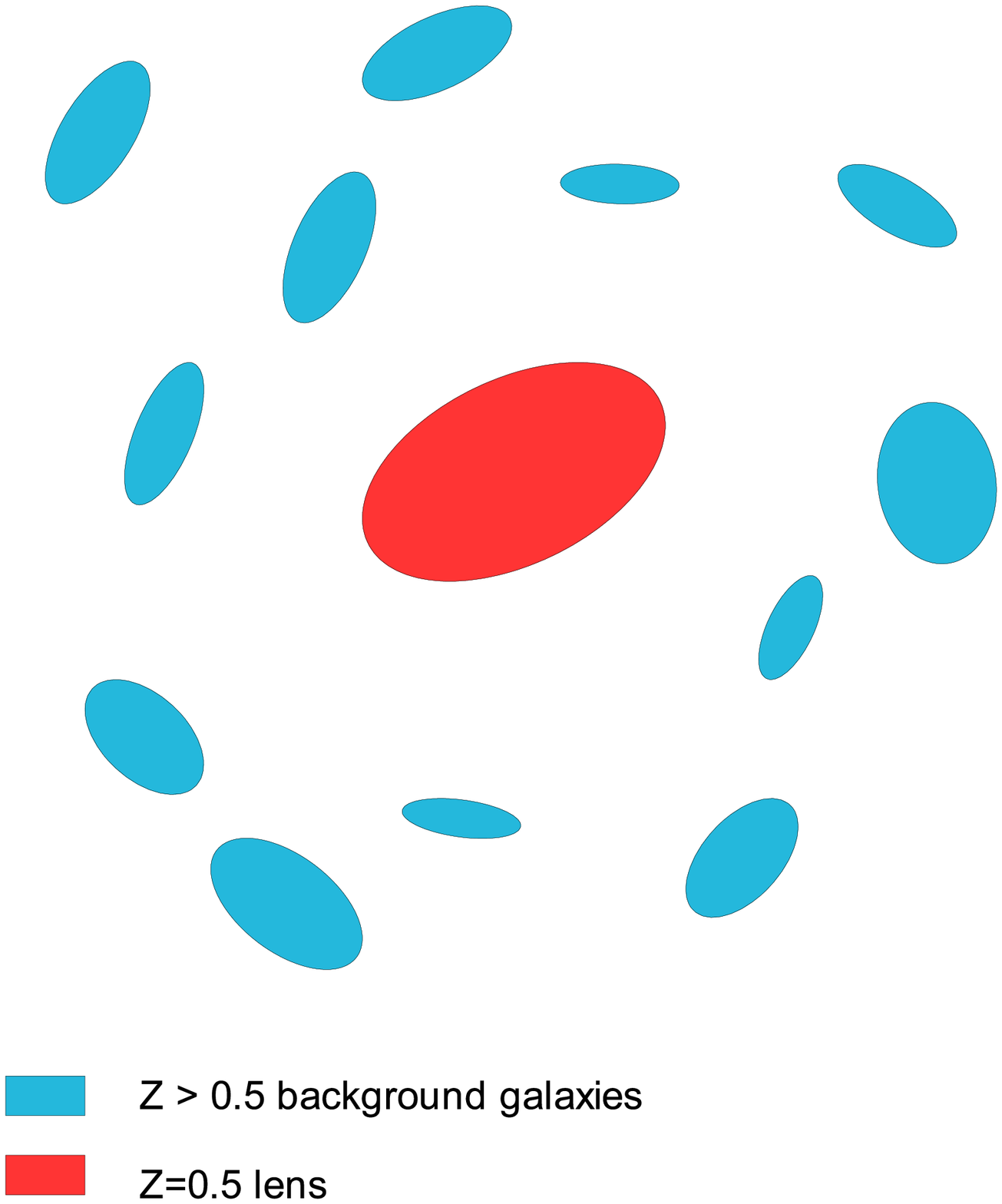,height=7cm}
\psfig{figure=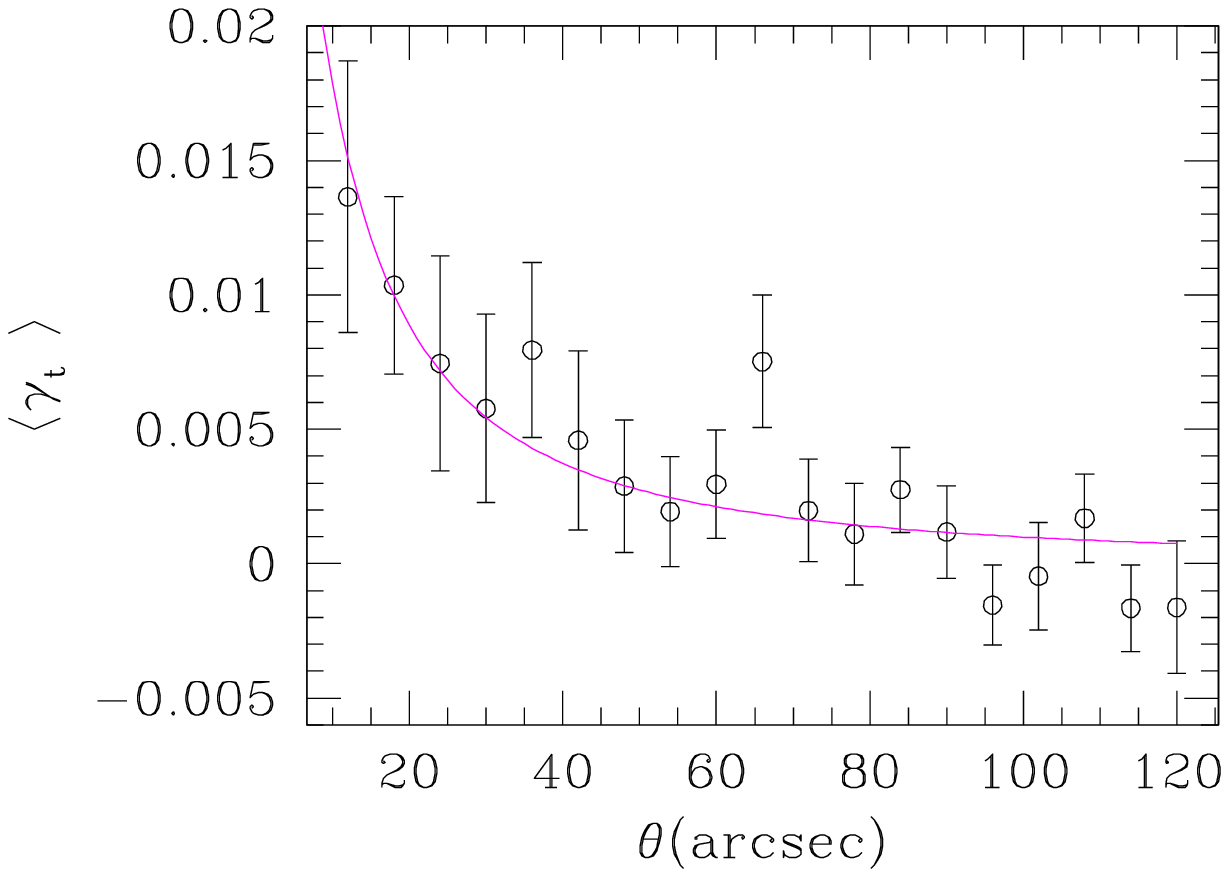,height=6cm}} \caption[]{Left panel: schematic
illustration of galaxy-galaxy lensing. Right panel: Galaxy-galaxy
lensing signal from the GEMS survey (Heymans et al. 2006). The solid
line shows the best fit NFW mass profile, assuming an average lens
redshift of $z_l=0.65$. \label{galgallens}}
\end{figure}

The probe of galaxy biasing can be extended down to galactic halo
scales with a technique called galaxy-galaxy lensing. First proposed
by \cite{tyson84} and then formalised by \cite{BBS}, the idea is to
cross-correlate the shape of distant lensed galaxies with foreground
galaxies. As shown in the left panel of Figure \ref{galgallens}, the
background galaxies lensed by a foreground galactic halo are
preferentially tangentially aligned with respect to the foreground
galaxy. The situation is identical to lensing by a cluster of
galaxies, but the lensing by individual galaxies is much weaker,
since the amplitude of the effect scales as the mass of the lens. To
measure the galaxy-galaxy lensing, one must therefore stack the
lensing signal behind a large number of foreground lenses. The
average shear as function of angular distance from the center gives
an estimate of the average halo profile around the foreground
galaxies (see right panel in Figure \ref{galgallens}).

The tangential shear of a lensed galaxy at position angle $\theta$
with respect to the lens on the sky is given by
\begin{equation}
\gamma_t=-e_1~\cos(2\theta)+e_2~\sin(2\theta).
\end{equation}
It is straightforward to show that, when averaged over all position
angles, the mean tangential ellipticity is unchanged if a constant is
added to the galaxy ellipticity $\vec e=(e_1,e_2)$. For this reason,
galaxy-galaxy lensing is robust against an imperfect Point Spread
Function (PSF) anisotropy correction, as long as the latter is a
slowly-varying function of position. The lensing signal is therefore
relatively easy to measure, even if the amplitude of the signal is
low.

Given that the foreground galaxies span a large range of velocity
dispersion and luminosity, scaling relations are needed in order to
calibrate the expected lensing signal to the same {\it fiducial}
galaxy with luminosity $L_\star$ and size $s_\star$. If
$\sigma_\star$ is the velocity dispersion of the fiducial galaxy,
the scaling relations are:
\begin{equation}
s=s_\star\left({\sigma\over \sigma_\star}\right)^2 \ \ \ ; \ \ \
{L\over L_\star}=\left({\sigma\over \sigma_\star}\right)^\eta,
\label{scaling2}
\end{equation}
where the latter corresponds to the Tully-Fisher or Faber-Jackson
relation for spiral and elliptical galaxies respectively if
$\eta=1/4$. $(L,s,\sigma)$ are the luminosity, scale and velocity
dispersion of one foreground galaxy. The lens mass model chosen by
\cite{BBS} is a truncated isothermal sphere (TIS), which is also
frequently used by several authors \cite{hudson98,HYG04}. The mass
density of the TIS is given by
\begin{equation}
\rho(r)={\sigma^2~s^2\over 2\pi G r^2(r^2+s^2)}, \label{rhofct}
\end{equation}
whose total mass $M_{tot}$ is \cite{HYG04}:
\begin{equation}
M_{tot}=7.3\times 10^{12}h^{-1}~M_\odot~\left({\sigma\over 100 {\rm
kms}^{-1}}\right)^2\left({s\over 1 {\rm Mpc}}\right).
\end{equation}
The tangential shear is calculated from the mass model (e.g.
Eq.\ref{rhofct}), and can be compared to the data provided that an
estimate of the lens and source redshifts is know. In general, the
free parameters measured with galaxy-galaxy lensing are the fiducial
velocity dispersion $\sigma_\star$ and truncation radius $s_\star$
of an $L_\star$ galaxy. The apparent magnitude of each foreground
lens is used to estimate its absolute luminosity $L$ (which may
require an appropriate $k$-correction to be included), then the
velocity dispersion $\sigma$ is obtained from the scaling relations.
The maximum likelihood technique developed by \cite{SR97} ensures an
optimal analysis, in particular if individual photometric redshifts
can be obtained. Note that simple mass models such as the singular
isothermal sphere (SIS) do not have a truncation radius:
\begin{equation}
\rho(r)={\sigma^2\over 2\pi G r^2},
\end{equation}
in which case one could constrain the scaling relations,
Eq.(\ref{scaling2}), like the parameter $\eta$, in addition to the
fiducial velocity dispersion $\sigma_\star$. \cite{hudson98} and
\cite{K06} measured a $\eta$ parameter very close to the
Tully-Fisher and Faber-Jackson relation.

At low redshift, the most extensive galaxy-galaxy lensing analysis
was performed by \cite{GS02} on the SDSS data set. They found a
Virial mass $M_{\rm 200}=5-10\times 10^{11}~h^{-1}M_\odot$ for
$L_\star=10^{10}~h^{-1} L_\odot$, depending on the galaxy color and
morphological type. In the redshift range $0.2-0.7$ a Virial mass of
$M_{\rm 200}=4-8\times 10^{11}~h^{-1}M_\odot$, the less massive
corresponding to the bluest galaxies \cite{heymans2004,K06}.
\cite{sheldon04} measured the galaxy-mass bias from the SDSS
galaxy-galaxy lensing signal, and found a constant bias over
correlation coefficient ratio $b/r=1.3\pm0.3$ for $\Om=0.27$,
which is in agreement with the cosmic shear study presented in
Section~\ref{Galaxy biasing} (see Figure \ref{bias}).

\begin{figure}
\centerline{\psfig{figure=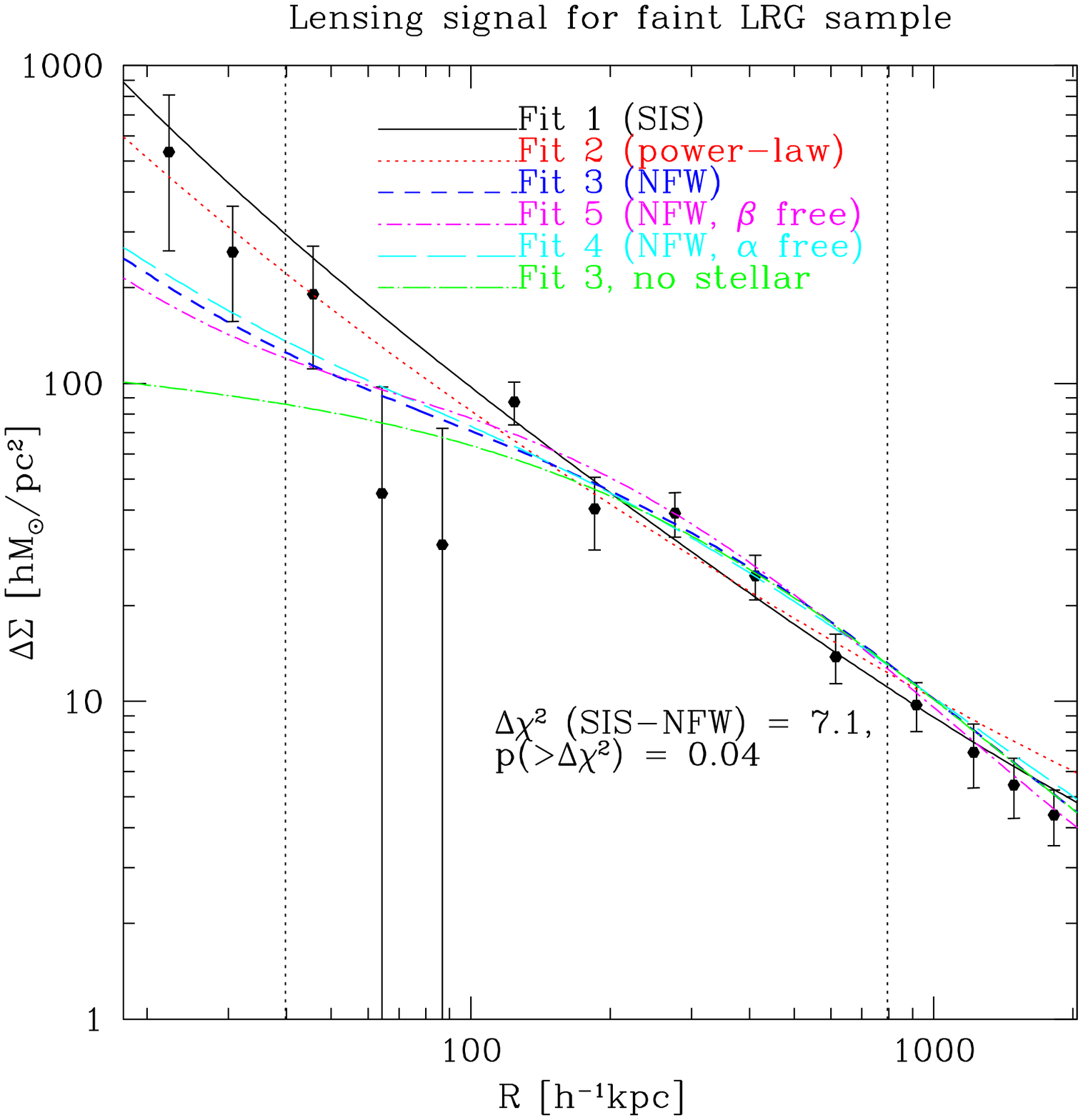,height=8cm}
\psfig{figure=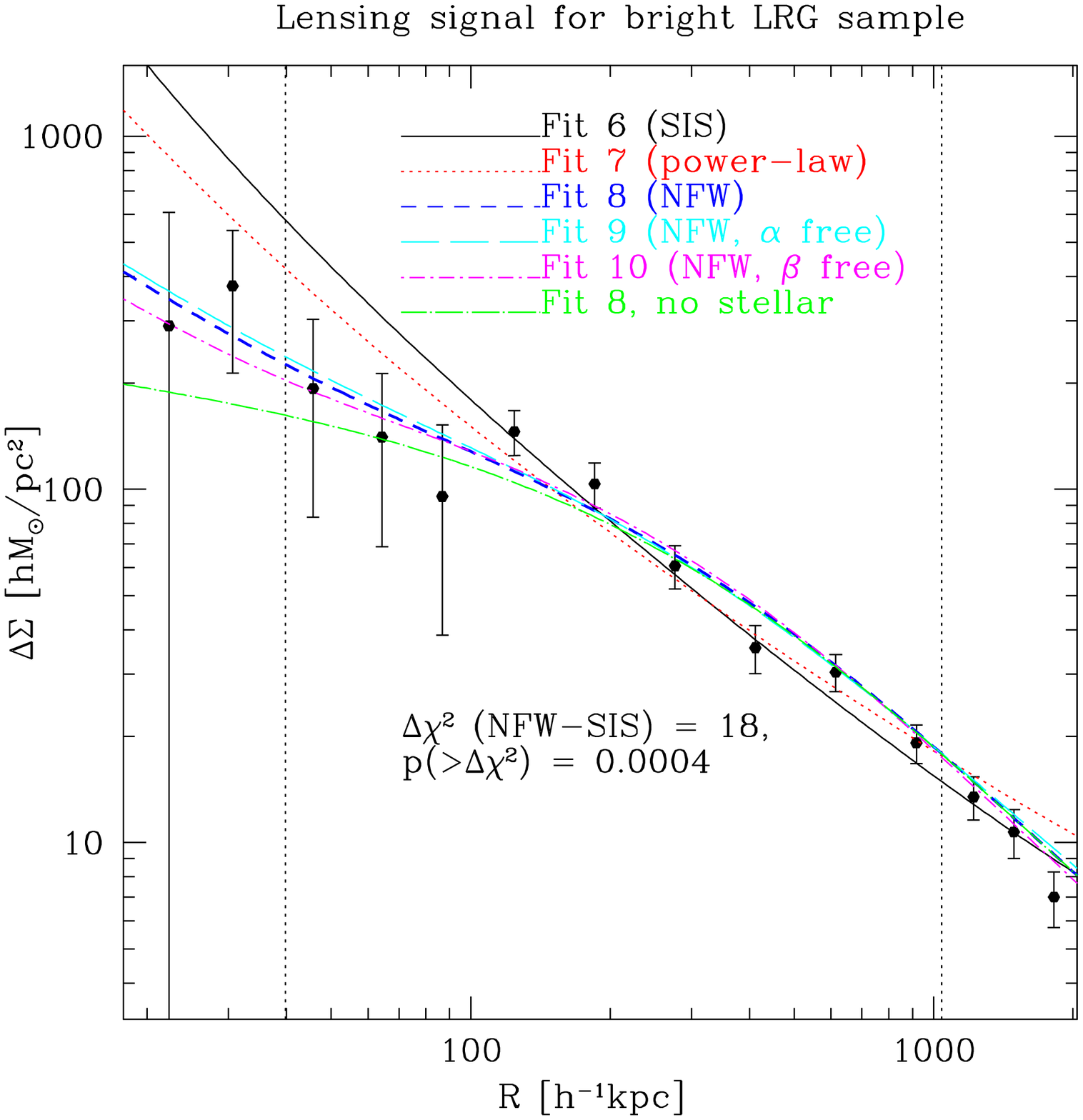,height=8cm} } \caption[]{Galaxy-galaxy
lensing signal for the Luminous Red Galaxies (LRG), used as tracers
of galaxy groups and small clusters (Mandelbaun et al. 2006). 43335
LRG from the Sloan Digital Sky Survey were used. The left and right
panels correspond to faint ($M_r>-22.3$) and bright ($M_r<-22.3$)
samples respectively. \label{galgallens2}}
\end{figure}

Recent studies use the Navarro-Frenk-White density profile as the
parametric mass model, which is particularly well suited for
galaxy-galaxy lensing in more massive structures such as galaxy
groups \cite{Hoekstra_Yee_Gladders2002} and galaxy clusters \cite{NK97,GS99}. In
\cite{rachel06a} the authors measured the lensing around Luminous
Red Galaxies (LRG) from the Sloan Digital Sky Survey. The LRG sample
was split in a bright and faint sample at $M_{\rm cut}=-22.3$. LRG
are particularly good foreground targets for probing larger mass
halos because they are known to be present in the core of groups and
cluster of galaxies. Figure~\ref{galgallens2} shows the measured
signal against several mass models. It is particularly interesting
to note that the flattening of the dark halo profile is clearly
visible, leading to a concentration parameter of $c\sim 5-7$, in
perfect agreement with Cold Dark Matter predictions \cite{NFW97}.

\cite{heymans2004} marginally measured a halo over lens ellipticity
ratio of $e_h/e_g\sim 0.8\pm 0.2$, consistent with the Cold Dark
Matter Scenario \cite{DC91}. A significant, but contradictory,
measurement is shown in \cite{rachel06b}, who found $e_h/e_g=0.1\pm
0.06$ for red galaxies and $0.8\pm 0.4$ for blue galaxies. The
baryon fraction can also be measured by galaxy-galaxy lensing from a
comparison of the lensing to the stellar mass. \cite{HH05} and
\cite{CH06} find a virial to baryon mass-to-light ratio between 50
and 100, and they show that massive early type galaxies have a
stellar to total baryon fraction of $\sim 10\%$, which indicates
that these galaxies are not efficient at producing stars.

\subsection{With Sunyaev-Zeldovich studies to probe small scale baryonic physics}

The energy of the CMB photons is boosted by scattering on the free
electrons contained in the plasma of the Intra Cluster Medium (ICM).
It is the source of the Sunyaev-Zeldovich (SZ) effect, which shifts
the spectral energy distribution of CMB photons for the
lines-of-sight containing hot, ionized, gas. The SZ effect is a
probe of the hot baryon distribution, which can be compared to
stellar and lensing mass distributions in order to learn about the
cluster physics. If one assume that the gas is isothermal, SZ is a
measure of the electronic density $n_e(\vtheta,z)$ projected along
the line of sight:
\begin{equation}
y(\vtheta)={kT\over m_e c^2}\sigma_T \int {\rm d}z ~n_e(\vtheta,z),
\end{equation}
where $\sigma_T$ is the Thompson scattering cross-section and $T$
the temperature of the gas. The SZ effect is therefore independent
on the cluster redshift, as opposed to the lensing effect which has
its maximum sensitivity at mid distance between the observer and the
sources. Lensing and SZ are complementary because they are both
linear in the density: for this reason they probe the same regions
of galaxy clusters, while X-rays for instance scale as the density
squared, which is more sensitive to the cluster core.

One approach of the combined SZ and lensing data sets consists of
working on individual clusters. Both data sets can be used to
predict the X-ray emission of the cluster \cite{dore01}, which
provides a direct test of the cluster dynamical equilibrium. The
3-dimensional halo shape (assuming axial symmetry) can be
reconstructed from the combination of SZ, lensing and X-ray
observations \cite{zaroubi98,PB06}. \cite{SVJ} have shown that large
lensing and SZ surveys (a few hundred square degrees) would be able
to detect the evolution of the mass-SZ luminosity relation with
redshift, which in turn is a measure of the cluster baryonic
physics. This is particularly important for the understanding of the
source of energy maintaining the temperature of the intra-cluster
plasma at high temperature. The other approach is statistical: the
goal is to use the cluster number counts to probe cosmology
\cite{matthias01}, the lensing data is used to measure the cluster
masses. but this requires a full sky survey with a mass sensitivity
down to a few $10^{14}~M_\odot$ to provide interesting constraints.

One should remember that cluster masses derived from lensing are
subject to large noise and projection effects
\cite{hoekstra01,white02}, although recent studies seem to show that
the discrepancy between dynamical, lensing and X-ray masses is
rather small \cite{hicks06}. The ideas developed around the
combination of SZ and lensing yet remain to be put into practice:
instruments such as the Cosmic Background Interferometer, Atacama
Cosmology Telescope, Arcminute Imager and South Pole Telescope, all with angular
resolution of the order of a few arcminutes, are very promising for
this purpose. The understanding of cluster physics from SZ and
lensing is not without any consequences for cosmological studies:
statistical lensing for instance probes the projected mass power
spectrum down to arbitrarily small angular scale (if the statistical
noise is low enough). At the cluster scale, typically one arcminute,
we know that baryon cooling and heating modify the cluster
gravitational potential well, and therefore the dark matter
distribution itself. \cite{ZK04} and \cite{white04} have shown how
the cluster physics could affect the power spectrum by $\sim 10\%$
below one arcminute. Taking into account this effect might be
necessary for the next generation of high precision weak lensing
studies.

\subsection{Weak lensing of supernovae and effects on parameter estimation}

Type Ia supernovae (SNeIa), which are believed to be the
thermonuclear explosion of an accreting white dwarf, are standard
candles with a small intrinsic dispersion around their average
luminosity. Moreover, this dispersion can be further reduced to
about $0.12$ mag using an empirical relation between the peak
magnitude and the width of the light curve \cite{RPK}. This makes
SNeIa excellent tools for observational cosmology. In particular, by
measuring their apparent magnitude we can derive their distance from
us and obtain the redshift-distance relation up to $z\sim 1$ which
provides useful constraints on cosmological parameters \cite{GP95}.
This method has supplied the main contribution to the discovery of
the present acceleration of the Universe \cite{R98,P99} and it is
the basis of future proposals to probe the nature of dark energy
through its equation of state (e.g. the SuperNova Acceleration
Probe, \cite{Aldering2005}). On the other hand SNeIa, like all
radiation sources, are affected by gravitational lensing effects
which can magnify or demagnify their observed luminosity. This is a
source of noise for studies which intend to measure the
redshift-distance relation but this effect may also be used by
itself to constrain cosmology in the same manner as gravitational
lensing distortion of distant galaxies.

There are several important differences between SNeIa and galaxies
as probes of gravitational lensing effects. Firstly, since SNeIa are
point sources one uses the magnification (associated with the
convergence $\kappa$) rather than the shear $\gamma$ which we
focussed on in previous Sections. In particular, the magnification
$\mu$ can be written as: \beq \mu =
\frac{1}{(1-\kappa)^2-|\gamma|^2} \;\;\;\; \mbox{whence} \;\; \;\;
\mu\simeq 1+2\kappa \;\;\;\; \mbox{for} \;\;\;\; |\kappa| \ll 1,
|\gamma|\ll 1 . \label{mu} \eeq Second, whereas weak gravitational
lensing only modifies the observed ellipticities of galaxies at
$z_s=1$ by less than $10\%$, so that one needs many galaxies to
extract the signal, the magnification of a type Ia supernova at
$z_s=1$ by gravitational lensing is of the same order as the
intrinsic magnitude dispersion. Hence the signal-to-noise ratio is
larger for SNeIa but since we have many more observed galaxies than
SNeIa, accurate weak gravitational lensing effects have only been
measured from galaxy ellipticities so far. Third, in order to derive
the coherent shear on large scales one must cross-correlate the
observed ellipticities of many distant galaxies over a window radius
$\theta_s$ of the order of a few arcmin (since galaxies are not
exactly spherical). This leads to observables such as the
aperture-mass $\Map$ or the mean shear over scale $\theta_s$ which
probe matter density fluctuations over scales of the order of
$\De\theta_s$. On the contrary, the magnification of each SNeIa only
probes the density fluctuations along its line-of-sight, which
corresponds to no smoothing ($\theta_s=0$). Then, by measuring the
probability distribution of observed SNeIa magnitudes (or its
variance) rather than cross-correlating different SNeIa, one can
probe the statistics of the convergence $\kappa$ with $\theta_s=0$.
This means that the signal is dominated by scales where
$\ell^2 P_{\kappa}(\ell)$ is maximum, which are set by the matter
power-spectrum. For CDM power-spectra this corresponds for sources
at $z_s=1$ to wavenumbers $k \sim 10 h$ Mpc$^{-1}$ whereas smoothing
galaxy ellipticites over $1$ arcmin mainly probes smaller
wavenumbers $k \sim 1 h$ Mpc$^{-1}$ \cite{Valageas2000}. Thus, weak
lensing magnification of SNeIa allows us to probe density
fluctuations on smaller scales than with galaxy ellipticities. This
also implies that non-Gaussianities are more important for SNeIa
gravitational lensing distortions. On the other hand, this
difference between the scales probed by galaxy shear maps and SNeIa
magnitude distortions means that both effects are only weakly
correlated so that weak-lensing shear maps obtained from surrounding
galaxies are not very efficient to correct the SNeIa luminosities
\cite{Dalal03}.

The rms magnification of SNeIa was computed by analytical means in
\cite{Frieman1997} who found that this would not significantly
decrease the accuracy of the distance-redshift relation at $z_s<0.5$
used to measure the current acceleration of the universe but it
could have a significant effect at higher redshifts $z_s>1$, in
agreement with the numerical simulations performed in
\cite{wambsganss1997}. In particular, although future surveys covering
more than a few square degrees will be unaffected, pencil beams
surveys ($<1$ deg$^2$) suffer significant contamination
\cite{CoorayHH2006}.
\begin{figure}
\begin{tabular}{c}
{\epsfxsize=7cm\epsfysize=7cm\epsffile{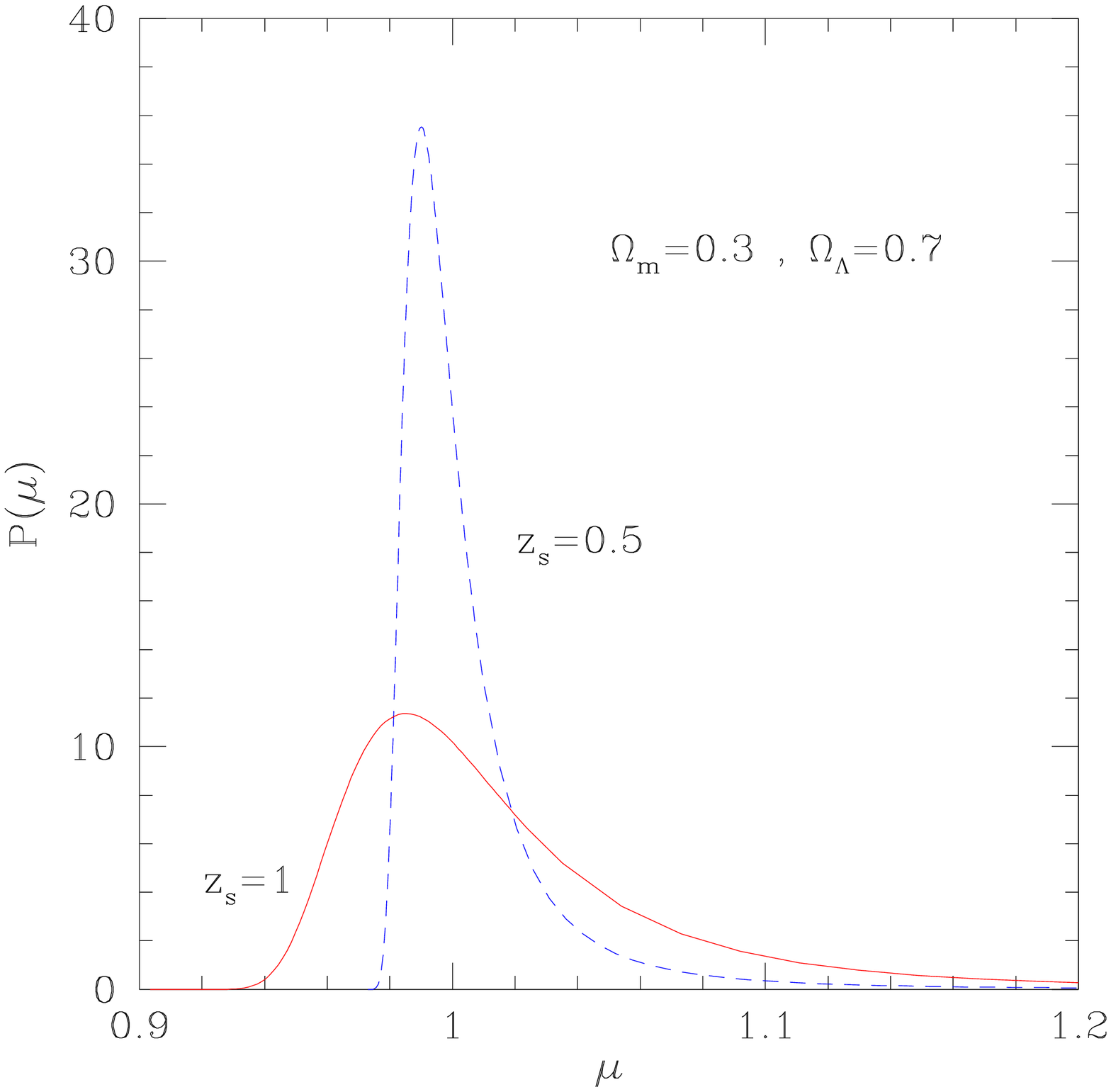}}
{\epsfxsize=7.2cm\epsfysize=7.2cm\epsffile{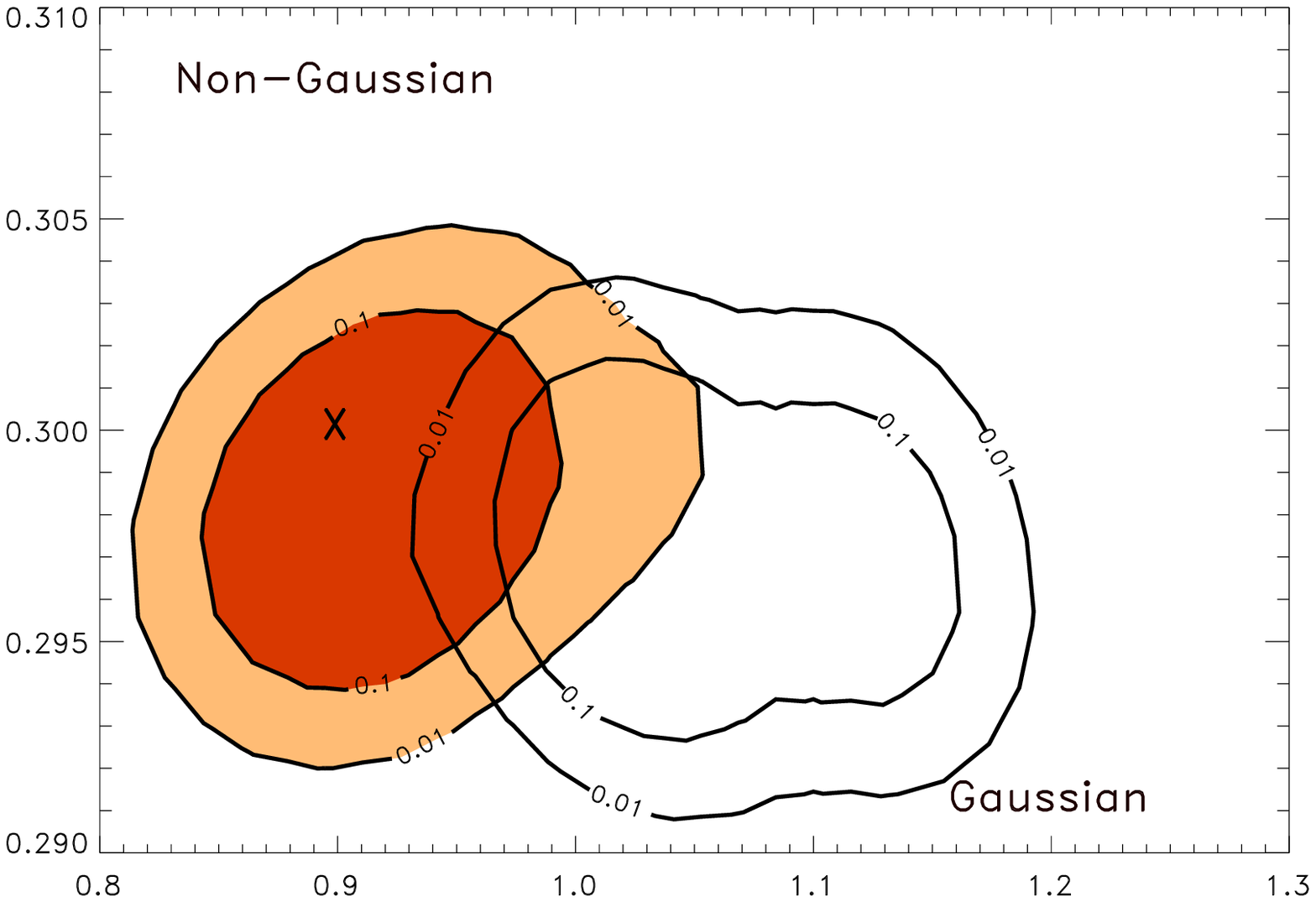}}
\end{tabular}
\caption{ {\it Left panel:}  the probability distribution $\cP(\mu)$
of the magnification within a $\Lambda$CDM universe for SNeIa at
redshifts $z_s=0.5$ or $z_s=1$. From Valageas (2000a). {\it Right
panel:} the constraints on cosmological parameters obtained from the
dispersion of SNeIa magnitudes from a SNAP-like survey. The two sets
of contours correspond to analyzing the data assuming the true
non-Gaussian distribution (shaded) or a Gaussian distribution
(unshaded). From Dodelson \& Vallinotto (2006). } \label{fig:SNeIa}
\end{figure}
Turning to the use of SNeIa as a tool to detect gravitational
lensing, \cite{Metcalf99} computed analytically the rms
magnification of SNeIa to find that in a $\Lambda$CDM universe one
needs at least $2400$ SNeIa at $z_s=0.5$ (or $110$ SNeIa at $z_s=1$)
to detect weak lensing from their observed magnitudes. The PDF of
the magnification $\cP(\mu)$ was computed from a hierarchical model
in \cite{Valageas2000} as well as its impact on the measure of
cosmological parameters through the distance-redshift relation. In
particular, this study shows the strong non-Gaussianity of the
magnification with an extended high-$\mu$ tail which follows the
high-density tail of the underlying matter density field, as can be
seen from left panel of Fig.~\ref{fig:SNeIa}. Then, the
high-luminosity tail of observed SNeIa could be used to detect weak
lensing since it should be significantly enhanced by gravitational
lensing. However, the number of observed SNeIa is still too small to
draw definite conclusions about a possible detection of weak lensing
effects by this method \cite{wang2005}.

On the other hand, since the amplitude of the gravitational lensing
contribution to the dispersion of observed SNeIa magnitudes depends
on cosmological parameters it could be used to constrain cosmology
\cite{CoorayHH2006}. Thus, \cite{DodelsonVallinotto06} found that
$2000$ SNeIa in the redshift range $0.5<z_s<1.7$ should be able to
constrain up to $5\%$ the amplitude $\sigma_8$ of the matter
power-spectrum. However, they point out that one needs to take into
account the non-Gaussianity of the weak-lensing magnification
distribution in order to obtain correct estimates, as seen in right
panel of Fig.~\ref{fig:SNeIa}.

In order to eliminate the systematic errors associated with
uncertainties on the intrinsic SNeIa luminosity distribution and its
possible dependence on redshift, it is possible to cross-correlate
SNeIa observed magnitudes with foreground galaxies, as advocated in
\cite{Metcalf01}. Indeed, in the absence of gravitational lensing
this cross-correlation would vanish. Using a halo model
\cite{Metcalf01} found that a gravitational lensing signal should be
detected with $\sim 250$ SNeIa at $z_s=1$. This will be within the
reach of future experiments (e.g. the SNAP satellite should observe
thousands of SNeIa up to $z\sim 1.7$) but current surveys are too
small to detect a correlation between SNeIa magnitudes and galaxy
overdensities \cite{MenardDalal05}.

\section{Summary and outlook}
\label{Conclusion}

Ten years ago, the detection of weak lensing by large scale
structures was only a dream. The progress accomplished, summarized
in this review, are remarkable. The main reason for this progress in
essentially the development of the Charge Coupled Device detectors
which had a major impact on the accuracy of galaxy shape
measurement. Adequate detectors are probably more important than
larger telescopes; weak lensing needs wide field of view and well
sampled Point Spread Function. One should remember for instance that
the very first tentative to measure distorted galaxy shapes
\cite{tyson84} failed because of inadequate technology detectors.

In this review we mostly focussed on the statistical lensing, what
one can learn on cosmology from the shear or convergence field,
rather then focussing on individual lenses like clusters of
galaxies. Like most of the cosmology probes, the information is
mainly encoded in the power spectrum. We have shown how different
cosmological parameters affect the projected mass (convergence or
shear) and how it can be measured. In particular a combination of
$\sigma_8$ and $\Om$ appears well constrained from current
lensing surveys. Statistical lensing can also be used to probe the
biasing between dark matter and light, which provides clues on the
assembly of galaxies inside their hosting halos. Future surveys
should be able to provide accurate measurement of the biasing
history as function of scale. A particularly interesting aspect of
lensing is that it can probe the non-linear regime without any
assumption on how light traces mass, and therefore go beyond the
traditional power spectrum analysis. This property gives access to a
different sensitivity to the cosmological parameters, an important
feature to help breaking the parameter degeneracy between $\Om$
and $\sigma_8$, but it also probes the history of the gravitational
collapse. In that respect, high-order lensing statistics can be used
to test the role of gravity during the collapse. The tomography
technique, which has just started to be applied to lensing data, is
very promising in probing the Dark Energy equation of state. The
redshift slicing of the lensed sources, or 3D analysis of the 
shear field, are the only ways to measure
the growth rate of structures (from the power spectrum and
bi-spectrum) which is very sensitive to the dark energy content of
the Universe. An interesting alternative to this is the
3-dimensional reconstruction of the mass distribution which in
addition will give us the distribution of the dark matter in space.
The combination of statistical lensing with other cosmology surveys
in other wavelengths was also shown to be very important for three
reasons i) the sensitivity to the cosmological parameters is
different and sometime orthogonal ii) the systematics can be
drastically reduced (e.g. 21 cm observation which offers perfect
source redshift measurement) iii) this is the only way to probe the
physics of the lenses beyond the simple mass-light relation.

It appears that ray-tracing techniques are an essential ingredient
of any precision weak lensing study. In particular this would be the
only way to address many of the complicated higher-order lensing
effects such as multiplanes deflections, source clustering,
intrinsic and intrinsic-shear alignment which cannot be modeled with
high precision. This is particularly relevant for future lensing
surveys. Moreover, a proper assessment of the cosmic variance
associated with the survey geometry can also be done from numerical
simulation, which is also important for non-linear scales (typically
less than half a degree) where prediction from semi-analytical
models is challenging.

In 2006, weak lensing by large structures has just began to reach a
status of scientific maturity: the first successes have demonstrated
the feasibility of the technique and now begins an era of thorough
weak lensing studies. The situation is similar to the Cosmic
Microwave Background research in the pre-COBE era, before 1992: many
independent groups have now measured the lensing fluctuation
amplitude, and the next generation of lensing surveys will provide
full sky, or nearly full sky, coverage. Table~\ref{tabcs} summarizes
the ongoing and future lensing surveys. The largest ongoing effort
is the Canada France Hawaii Telescope Legacy Survey (CFHTLS). This
survey represents a transition because this is the last one which
requires a percent accuracy of galaxy shape measurement in order to
be fully scientifically exploited. One percent accuracy is what we
are currently capable of. All future surveys require a sub-percent
precision level, which is still beyond our capability, as
demonstrated by STEP. In fact, below the percent accuracy, many
other effects will complicate the lensing measurement and analysis.

All recent studies show that the only way to quantify these effects,
such as intrinsic alignment, shear-intrinsic alignment and source
clustering, is to have a redshift estimate of each galaxy. This is
only doable with photometric redshifts, which poses additional
challenges: assuming one can get enough colors to obtain accurate
photometric redshifts, there is no spectroscopic survey to help
calibrating objects fainter than $I\sim 24$. The magnitude
calibration of brighter sources could be problematic, especially
from the ground due to zero-point fluctuations for different
wavelengths. An absolute zero-point variation is not a problem, but
an explicit dependence with color could jeopardize photometric
redshift estimates. It appears that the only way to obtain unbiased
photometric redshift is to have uniform magnitude calibration and to
conduct a deep spectroscopic survey in order to validate the
technique at faint magnitude/high redshift. Both might only be
doable from space where photometry is stable. The tunable laser
project \cite{Albert} is an interesting solution for the zero-point
issue, while the satellite GAIA \cite{GAIA}, successor of Hypparcos,
could help with absolute astrometry calibration. Interestingly, it
seems that the shape measurement problem is likely to be solved in
the next two or three years, thanks to the Shear TEsting Program
\footnote{http://www.physics.ubc.ca/~heymans/step.html}. The main
limitation of weak lensing by large scale structures might therefore
not be shape measurement anymore but multicolor photometric
calibration, a very old astronomical problem!

\begin{table}[t]
\caption{List of forthcoming lensing surveys (adapted from Peacock
et al. 2006). Surveys are sorted in three groups separated by a
line. The top group show the current lensing surveys, the middle
group shows survey starting in one year at the latest, and the group
at the bottom is essentially not funded or partially funded
projects.}
\label{tabcs}
\bigskip
\begin{tabular}{lcccc}
\hline
\\
Survey & Telescope & Sky coverage & Filters & depth \\
\hline
Deep Lens Survey & CTIO & 7x4 ~$deg^2$ & BVRz' & R=25  \\
CFHTLS-Wide & CFHT & 170 ~$deg^2$ & ugriz & $i_{AB}$=24.5  \\
RCS2 & CFHT & 1000 ~$deg^2$ & grz & $i_{AB}$=22.5  \\
\hline
KIDS & VST & 1500 ~$deg^2$ & ugriz & $i_{AB}$=22.9  \\
Pan-STARRS & PS1 & 30000 ~$deg^2$ & grizy & $i_{AB}$=24  \\
VIKING & VISTA & 1500 ~$deg^2$ & zYJHK & $i_{AB}$=22.9  \\
Dark Energy Survey & CTIO & 5000 ~$deg^2$ & griz & $i_{AB}$=24.5  \\
\hline
DarkCam & VISTA & 10000 ~$deg^2$ & ugriz & $i_{AB}$=24  \\
HyperCam & SUBARU & 3500 ~$deg^2$ & TBD & TBD  \\
SNAP & Space & 300/2000 ~$deg^2$ & Narrow band (0.35-1.6) & TBD  \\
LSST & 6m ground & 20000 ~$deg^2$ & Narrow band (0.35-1.2) & $i_{AB}$=27  \\
DUNE & Space & 20000 ~$deg^2$ & TBD & $i_{AB}$=25.5  \\
\hline
\end{tabular}
\end{table}

\subsection*{Acknowledgements}

We thank Stephane Colombi, Asantha Cooray, Scott Dodelson, Wayne Hu, Bhuvnesh Jain, 
Rachel Mandelbaum, Yannick Mellier, Ue Li-Pen, Elisabetta Semboloni, Masahiro Takada
and Alberto Vallinotto for use of their figures.
It is a pleasure to thank Catherine Heymans, Sanaz Vafaei and
Jonathan Benjamin for their help with some of the figures and
tables. DM would like to thank Martin Kilbinger, Jerry Ostriker, Lindsay King,
Tony Tyson and members of Cambridge Planck Analyis Center for many 
useful discussions. LVW is supported by NSERC, CIAR and CFI. DM is supported 
by PPARC.

\appendix

\section{Analytical modeling of gravitational clustering and weak-lensing
statistics}
\label{Analytical modeling of gravitational clustering and weak-lensing
statistics}

In order to derive the properties of weak lensing observables, like the
shear $\gamma$, one first needs to specify the properties of the
underlying density field. We briefly describe in this appendix two such models
which can be used to predict weak lensing statistics, the hierarchical models
presented for instance in \cite{Valageas2004} and
the halo model described in detail in \cite{Cooray_Sheth2002}.

\subsection{From density to weak-lensing many-body correlations}
\label{From density to weak-lensing many-body correlations}

At lowest-order weak-lensing observables can be written as linear functionals
of the density field, as seen in Eqs.(\ref{kappa})-(\ref{kappal}).
Therefore, their many-body connected correlation functions can be directly written
in terms of the connected correlations $\xi_p$ of the 3D matter density field 
defined by \cite{peebles1980}:
\beq
\xi_p(\bx_1,\ldots,\bx_p;z) = \lag \delta(\bx_1,z) \ldots \delta(\bx_p,z) \rag_c .
\label{xip}
\eeq
Note that for a Gaussian field we have $\xi_p=0$ for $p\geq3$.
For a weak-lensing observable $\barX$ defined as in Eq.(\ref{Xx}) this gives
in real space \cite{Valageas2000}:
\beq
\lag\barX^p\rag_c = \int_0^{\chirad_s} \d\chirad \; \wh^p
\int_{-\infty}^{\infty} \prod_{i=2}^{p} \d\chirad_i \int \prod_{i=1}^{p}
\d{\vtheta}_i \; U_X({\vtheta}_i) \;
\xi_p\left( \bea{l} 0 \\ \De {\vtheta}_1 \ea ,
\bea{l} \chirad_2 \\ \De {\vtheta}_2 \ea , \ldots ,
\bea{l} \chirad_p \\ \De {\vtheta}_p \ea ; z \right) ,
\label{cumX}
\eeq
and in Fourier space \cite{Valageas2004}:
\beq
\lag \barX^p \rag_c = (2\pi)^{-2p-1} \int \d\chirad \; \wh^p
\int \prod_{j=1}^p \d\kperpj \; W_X(\kperpj\De\theta_s) \;
\lag \delta(\vec{k}_{\perp 1}) \ldots \delta(\vec{k}_{\perp p})\rag_c .
\label{cumXk1}
\eeq
Here we used Limber's approximation $k\simeq k_{\perp}$ as for 
Eq.(\ref{Powerkappa}) and the longitudinal Dirac factor 
$\delta_D(k_{\parallel 1}+\ldots+k_{\parallel p})$ has been factorized out
of the correlation $\lag \delta .. \delta\rag_c$. 
Next, from the correlation functions one can obtain the full probability
distribution function (PDF). Indeed, if we define the generating function
$\varphi_X(y)$ of the cumulants $\lag \barX^p\rag_c$ as:
\beq
\varphi_X(y) = \sum_{p=2}^{\infty} \frac{(-1)^{p-1}}{p!} \; S^{(X)}_p \; y^p
\hspace{0.3cm} \mbox{with} \hspace{0.3cm}
S^{(X)}_p = \frac{\lag \barX^p\rag_c}{\lag \barX^2\rag_c^{\; p-1}} ,
\label{phiSp}
\eeq
where we used $\lag \barX\rag=0$, then one can show that the PDF $\cP_X(\barX)$ is
given by the inverse Laplace transform:
\beq
\cP_X(\barX)= \int_{-i\infty}^{i\infty} \frac{\d y}{2\pi i \lag \barX^2\rag_c}
\; e^{[\barX y - \varphi_X(y)]/\lag \barX^2\rag_c} .
\label{PXphi}
\eeq
Thus, in order to compute $\cP_X(\barX)$ one only needs to derive $\varphi_X(y)$,
which may be written in terms of the density field from
Eq.(\ref{cumX}) or Eq.(\ref{cumXk1}).\

For the smoothed convergence $\barkappa$ the cumulants $\lag\barkappa^p\rag_c$
correspond to averages of the density cumulants over cylindrical cells
along the line of sight and they can be obtained with a good accuracy from
the 3D density cumulants averaged over spherical cells. If we also use a
mean-redshift approximation one obtains 
\cite{Valageas2000,Valageas2000b,Barber_Munshi_Valageas2004}:
\beq
\varphi_{\kappa}(y) \simeq |\kappa_{\rm min}|^2 \varphi_{\delta}
\left(\frac{y}{|\kappa_{\rm min}|}\right) \;\;\;\; \mbox{with} \;\;\;\;
\kappa_{\rm min} = - \int\d\chirad \; \wh ,
\label{phikappa}
\eeq
where we introduced the minimum value $\kappa_{\rm min}$ of the convergence,
which corresponds to an empty line of sight ($\delta=-1$,
see Eq.(\ref{kappanz})). This gives:
\beq
\cP_{\kappa}(\barkappa) \simeq \frac{1}{|\kappa_{\rm min}|} \cP_{\delta}
\left(\delta\rightarrow\frac{\barkappa}{|\kappa_{\rm min}|},
\xib_2 \rightarrow\frac{\lag\barkappa^2\rag}{|\kappa_{\rm min}|^2}\right) ,
\label{Pkappa}
\eeq
where $\varphi_{\delta}$ and $\cP_{\delta}$ are the generating function and the
PDF of the 3D matter density contrast at the mean redshift and
scale probed by the smoothed convergence $\barkappa$ (they depend on the angular
radius $\theta_s$ and the galaxy distribution $n(z_s)$).
Thus, within this simple approximation the PDF of the projected density field
$\barkappa$ is directly expressed in terms of the PDF of the underlying 3D
density contrast.

\subsection{Hierarchical models}
\label{Hierarchical models}

For more intricate observables like the shear or the aperture-mass which
involve compensated filters one cannot perform approximations such as
Eqs.(\ref{phikappa})-(\ref{Pkappa}) and it is not possible to approximate
high-order cumulants or the PDF of weak-lensing observables in terms of
those of the smoothed density contrast.
Therefore, one needs to specify the detailed angular behavior of the
many-body correlation functions $\xi_p(\bx_1,\ldots,\bx_p)$.
A simple prescription is provided by the general class of ``tree-models''
defined by the hierarchical property \cite{Schaeffer84,Groth77}:
\beq
\xi_p(\bx_1, \ldots ,\bx_p) = \sum_{(\alpha)} Q_p^{(\alpha)} \sum_{t_{\alpha}}
\prod_{p-1} \xi_2(\bx_i,\bx_j)
\label{tree}
\eeq
where $(\alpha)$ is a particular tree-topology connecting the $p$ points
without making any loop, $Q_p^{(\alpha)}$ is a parameter associated with the
order of the correlations and the topology involved, $t_{\alpha}$ is a
particular labeling of the topology, $(\alpha)$, and the product is made over
the $(p-1)$ links between the $p$ points with two-body correlation functions.
Then, as seen in \cite{Valageas2000b} the 2D correlations $\omega_p$ involved in
weak-lensing cumulants such as (\ref{cumX}) exhibit the same tree-structure,
with:
\beq
\omega_p({\vtheta}_1,\ldots,{\vtheta}_p;z) =
\int_{-\infty}^{\infty} \prod_{i=2}^{p} \d\chirad_i \;
\xi_p\left( \bea{l} 0 \\ \De {\vtheta}_1 \ea , \ldots ,
\bea{l} \chirad_p \\ \De {\vtheta}_p \ea ; z \right) .
\label{omp}
\eeq
In order to perform numerical computations we need to specify the weights
$Q_p^{(\alpha)}$. Thus, the ``minimal tree-model'' corresponds to the
specific case where the weights $Q_p^{(\alpha)}$ are given by
\cite{Bernardeau92}:
\beq
Q_p^{(\alpha)} = \prod_{\mbox{vertices of } (\alpha)} \nu_q
\label{mintree}
\eeq
where $\nu_q$ is a constant weight associated to a vertex of the tree
topology with $q$ outgoing lines. The advantage of this minimal tree-model
is that it is well-suited to the computation of the cumulant generating
functions as defined in Eq.(\ref{phiSp}).
Indeed, for an arbitrary real-space filter, $F(\bx)$, which
defines the random variable $\dum$ as:
\beq
\dum = \int \d\bx \; F(\bx) \; \delta(\bx) \hspace{0.4cm} \mbox{and}
\hspace{0.4cm} \xidum= \lag \dum^2 \rag ,
\label{mub}
\eeq
it is possible to obtain a simple implicit expression for the
generating function, $\phidum(y)$, see \cite{Bernardeau92,Jannink87}:
\beqa
{\displaystyle \phidum(y)} & = & {\displaystyle y \int \d\bx \; F(\bx) \;
\left[ \zeta_{\nu}[ \tau(\bx)] - \frac{\tau(\bx) \zeta'_{\nu}[\tau(\bx)]}{2}
\right] }
\label{implicitphi} \\
{\displaystyle \tau(\bx) } & = & {\displaystyle -y \int \d\bx' \; F(\bx')
\; \frac{\xi_2(\bx,\bx')}{\xidum} \; \zeta'_{\nu}[\tau(\bx')] }
\label{implicittau}
\eeqa
where the function $\zeta_{\nu}(\tau)$ is defined as the generating function
for the coefficients $\nu_p$:
\beq
\zeta_{\nu}(\tau) = \sum_{p=1}^{\infty} \frac{(-1)^p}{p!} \; \nu_p \; \tau^p
\hspace{0.4cm} \mbox{with} \hspace{0.4cm} \nu_1 = 1.
\label{zetanu}
\eeq
Since the 2D correlations $\omega_p$ obey the same minimal tree-model we can
perform the resummation (\ref{implicitphi})-(\ref{implicittau}) which yields
\cite{Bernardeau00,Barber_Munshi_Valageas2004}:
\beq
\varphi_X(y) = \int_0^{\chirad_s} \d\chirad \; \frac{\lag \barX^2\rag_c}{\overline{\omega}_{2X}} \;
\varphi_{\rm cyl.}\left( y \wh \frac{\overline{\omega}_{2X}}{\lag \barX^2\rag_c} ; z \right) ,
\label{phiXtree}
\eeq
where we introduced the 2D generating function $\varphi_{\rm cyl.}$
associated with the 2D correlations $\omega_p$, given by the resummation:
\beqa
{\displaystyle \varphi_{\rm cyl.}(y)} & = & {\displaystyle y
\int \d{\vtheta} \; U_X({\vtheta}) \;
\left[ \zeta_{\nu}[ \tau({\vtheta})]
- \frac{\tau({\vtheta}) \zeta'_{\nu}[\tau({\vtheta})]}{2}
\right] }
\label{phiom} \\
{\displaystyle \tau({\vtheta}) } & = & {\displaystyle -y
\int \d{\vtheta}' \; U_X({\vtheta}') \;
\frac{\omega_2({\vtheta},{\vtheta}';z)}{\overline{\omega}_{2X}(z)} \;
\zeta'_{\nu}[\tau({\vtheta}')] }
\label{tauom}
\eeqa
Here we introduced the angular average $\overline{\omega}_{2X}$ of the 2D correlation
$\omega_2$, associated with the filter $U_X$:
\beq
\overline{\omega}_{2X}(z) = \int \d{\vtheta}_1 \d{\vtheta}_2 \;
U_X({\vtheta}_1) U_X({\vtheta}_2) \;
\omega_2({\vtheta}_1,{\vtheta}_2;z) .
\label{omb2X}
\eeq
Thus, one obtains in this way the generating function $\varphi_X(y)$
which yields in turn the PDF $\cP_X(\barX)$ from Eq.(\ref{PXphi}).

A second simple model is the ``stellar model'' introduced in
\cite{Valageas2004} where we only keep the stellar diagrams
in Eq.(\ref{tree}). Thus, the $p-$point connected correlation $\xi_p$ of
the density field can now be written as:
\beq
\xi_p(\bx_1, \ldots ,\bx_p) = \frac{\tilde{S}_p}{p} \; \sum_{i=1}^p \prod_{j\neq i}
\xi_2(\bx_i,\bx_j) .
\label{stellar}
\eeq
The advantage of the stellar-model (\ref{stellar}) is that it leads to
very simple calculations in Fourier space. Indeed, Eq.(\ref{stellar}) reads
in Fourier space:
\beq
\lag \delta(\bk_1) \ldots \delta(\bk_p) \rag_c = \frac{\tilde{S}_p}{p} \;
(2\pi)^3 \delta_D(\bk_1+\ldots+\bk_p) \sum_{i=1}^p \prod_{j\neq i} P(k_j) .
\label{stellark}
\eeq
Using the standard exponential representation of the Dirac distribution
gives \cite{Valageas2004}:
\beq
\lag \barX^p \rag_c = \tilde{S}_p \int_0^{\chirad_s} \d\chirad \; \wh^p \int \d{\vtheta}
\; U_X({\vtheta}) \; I_X(\chirad,{\vtheta})^{p-1} ,
\label{cumXk2}
\eeq
where we introduced:
\beq
I_X(\chirad,{\vtheta}) = \int \frac{\d\kperp}{(2\pi)^2} \; e^{-i\kperp.\De\vtheta} \; 
W_X(\kperp\De\theta_s)  \; P(k_{\perp};z) .
\label{IX}
\eeq
Then, using Eq.(\ref{phiSp}) we obtain:
\beqa
\varphi_X(y) = \int_0^{\chirad_s} \d\chirad \int \d{\vtheta}
\; U_X({\vtheta}) \frac{\lag \barX^2\rag_c}{I_X(\chirad,{\vtheta})}
\; \varphi_{\delta}\left( y \wh \frac{I_X}{\lag \barX^2\rag_c} ; z \right) ,
\label{phiXstellar}
\eeqa
which directly gives $\varphi_X(y)$ in terms of the cumulant generating
function of the 3D density field $\varphi_{\delta}$.

\subsection{Halo models}
\label{Halo models}

An alternative to the hierarchical models presented in
Section~\ref{Hierarchical models} is provided by the halo model where the
matter density field is described as a collection of halos
\cite{McClelland,Scherrer,Sheth97}.
Then, the density correlation functions are obtained through a convolution
over the halo density profiles. One also needs to specify the many-body
correlations of the halos themselves as well as their multiplicity function.
Thus, the density field is written as the superposition of the halo profiles:
\beq
\rho(\bx) = \sum_i m_i u_{m_i}(\bx-\bx_i) \;\;\; \mbox{with} \;\;\;
\int\d\bx \; u_m(\bx) = 1 ,
\label{rhohalo}
\eeq
where we introduced the normalized density profile $u_m(x)$ of halos of mass
$m$. Then, the density two-point correlation function is expressed as the sum
of correlations within a single halo (1-halo term, both points are within the
same halo) and between different halos (2-halo term, the two points are
within two different halos):
\beqa
\lefteqn{ \xi_2(\bx_1,\bx_2) = \int\d m \; n(m) \left(\frac{m}{\rhob}\right)^2
\int\d\bx' u_m(\bx_1-\bx') u_m(\bx_2-\bx')
+ \int\d m_1 n(m_1) \frac{m_1}{\rhob} }\nonumber \\
&& \times \!\!\int\!\!\!\d m_2 n(m_2) \frac{m_2}{\rhob}
\!\!\int\!\!\!\d\bx_1' u_{m_1}(\bx_1-\bx_1')
\!\!\int\!\!\!\d\bx_2' u_{m_2}(\bx_2-\bx_2') \xi_h(\bx_1',\bx_2';m_1,m_2)
\label{xi2halo}
\eeqa
where $n(m)$ is the halo mass function and $\xi_h$ is the halo two-point
correlation \cite{Takada_Jain2003a,Takada_Jain2003b}. 
In a similar fashion one can write
the $n$-point density correlation as a sum of 1-halo up to $n$-halo terms.
In practice, it is more convenient to work in Fourier space (to simplify
the convolution products and to take advantage of the statistical
homogeneity and isotropy of the system). Thus, the density power-spectrum
reads \cite{Seljak00,Ma00,Scoccimarro01}:
\beq
P(k) = I_2^0(k,k) + \left[ I_1^1(k) \right]^2 P_L(k)
\label{Pdeltahalo}
\eeq
with:
\beq
I_{\mu}^{\beta}(k_1,\ldots,k_{\mu}) = \int\d m \; n(m) b(m)^{\beta}
\left(\frac{m}{\rhob}\right)^{\mu} u_m(k_1) \ldots u_m(k_{\mu}) ,
\label{Ideltahalo}
\eeq
where $u_m(k)$ is the Fourier transform of the halo density profile.
In Eq.(\ref{Pdeltahalo}) we assumed that the halo-halo power-spectrum
can be written as $P_h(k;m_1,m_2)=b(m_1)b(m_2)P_L(k)$ where $P_L$ is the
linear matter power-spectrum and $b(m)$ is the bias parameter which describes how
the halo distribution is biased with respect to the dark matter density field.
Indeed, the 2-halo term dominates at large scales which are described by the
quasi-linear theory (hence it is sufficient to use $P_L$) whereas the 1-halo
term dominates a small non-linear scales.
This also ensures that one recovers the results of standard linear theory
at large scales.
To complete the halo model one needs to specify the halo density profile,
which is often taken from \cite{NFW97}, the halo mass
function, taken for instance from \cite{Sheth99,Press74}, and the biasing of the halo distribution
as from \cite{Mo96}. Then, the correlation functions of weak-lensing
observables can be derived from Eqs.(\ref{cumX})-(\ref{cumXk1}).
For instance, the 1-halo contribution to the angular two-point correlation
of the convergence field reads \cite{Takada_Jain2003a}:
\beq
\lag\kappa({\vtheta}_1)\kappa({\vtheta}_2)\rag_{1h} =
\int\d\chirad \; \frac{\wh^2}{\De^2} \int\d m \; n(m)
\left(\frac{m}{\rhob}\right)^2 \int\d l \frac{l}{2\pi} |u_m(k)|^2
J_0(l\theta) ,
\label{xikappa1h}
\eeq
where $k=l/\De(\chirad)$. Similar expressions give the 1-halo contribution to
higher-order convergence correlation functions whereas 2-halo terms
involve the linear density power-spectrum, 3-halo terms involve the density
bispectrum and so on. This approach has been used to estimate the
low-order correlations of weak-lensing observables up to fourth-order
(the kurtosis of the shear, see \cite{Takada_Jain2002}) 
but no attempt has been made to predict the full PDF $\cP_X(\barX)$ 
yet (however see \cite{Taruya03} for a model for the tails of the matter
density PDF).


\begin{thebibliography}{100}

\bibitem{Albert}
J. Albert, W. Burgett, J. Rhodes, \eprint{astro-ph/0604339}

\bibitem{DETF}
A. Albrecht, et al., {\em Dark Energy Task Force Report}, (2006), \eprint{astro-ph/0609591}.

\bibitem{Aldering2005}
G. Aldering, \nar{2005}{49}{346}

\bibitem{AmbValeWhite}
A. Amblard, C. Vale and M.J. White, \nas{2004}{9}{687}, \eprint{astro-ph/0403075}.

\bibitem{babul1991}
A. Babul and M.H. Lee, \mnras{1991}{250}{407}

\bibitem{Bacon_Refregier_Ellis2000}
D.J. Bacon, A. Refregier, R. Ellis, \mnras{2000}{318}{625}

\bibitem{Bacon_et_al2003}
D.J. Bacon, R. Massey, A. Refregier, R. Ellis, \mnras{2003}{344}{673}

\bibitem{BT03}
D.J. Bacon, A.N. Taylor, \mnras{2003}{344}{1307}

\bibitem{Bacon2006}
D.J. Bacon, D.M. Goldberg, B.T.P. Rowe, A.N. Taylor, \mnras{2006}{365}{414}

\bibitem{Barber1999}
A.J. Barber, P.A. Thomas, H.M.P. Couchman, \mnras{1999}{310}{453}

\bibitem{barber2000}
A.J. Barber, P.A. Thomas, H.M.P. Couchman, C.J. Fluke, \mnras{2000}{319}{267}

\bibitem{barber2002}
A.J. Barber, \mnras{2002}{335}{909}

\bibitem{Barber_Munshi_Valageas2004}
A.J. Barber, D. Munshi, P. Valageas, \mnras{2004}{347}{667}

\bibitem{bartelmann1996}
M. Bartelmann, \aap{1996}{313}{697}

\bibitem{BS99}
M. Bartelmann, P. Schneider, \aap{1999}{345}{17}

\bibitem{matthias01}
M. Bartelmann, \aap{2001}{370}{754}

\bibitem{Bartelmann_Schneider2001}
M. Bartelmann, P. Schneider, \phr{2001}{340}{291}

\bibitem{BDW}
M. Bartelmann, M. Doran, C. Wetterich, \aap{2006}{454}{27}

\bibitem{benabed01}
K. Benabed, F. Bernardeau and L. Van Waerbeke, \prd{2001}{63}{043501}

\bibitem{jonben06}
J. Benjamin, et al., (2006), in preparation

\bibitem{Bernardeau92}
F. Bernardeau, R. Schaeffer, \aap{1992}{255}{1}

\bibitem{Bernardeau_et_al1997}
F. Bernardeau, L. Van Waerbeke, Y. Mellier, \aap{1997}{322}{1}

\bibitem{Bernar97}
F. Bernardeau, \aap{1997}{324}{15}, \eprint{astro-ph/9611012}.

\bibitem{Bernardeau1998}
F. Bernardeau, \aap{1998}{338}{375}

\bibitem{Bernardeau00}
F. Bernardeau, P. Valageas, \aap{2000}{364}{1}

\bibitem{BMVW}
F. Bernardeau, Y. Mellier, L. Van Waerbeke, \aap{2002}{389}{L28}

\bibitem{Bernardeau_et_al2003}
F. Bernardeau, L. Van Waerbeke, Y. Mellier, \aap{2003}{397}{405}

\bibitem{Bernardeau2005}
F. Bernardeau, \prl{2005}{}{}, submitted, \eprint{astro-ph/0409224}.

\bibitem{BJ2002}
G. Bernstein, M. Jarvis, \aj{2002}{123}{583}

\bibitem{BJ04}
G. Bernstein, B. Jain, \apj{2004}{600}{17}

\bibitem{BlanSch1987}
A. Blanchard and J. Schneider, \aap{1987}{184}{1}, ADS.

\bibitem{Blandford_et_al1991}
R.D. Blandford, A.B. Saust, T.G. Brainerd, J.V. Villumsen, \mnras{1991}{251}{600}

\bibitem{BM1995}
H. Bonnet, Y. Mellier, \aap{1995}{303}{331}

\bibitem{bradac2006}
M. Bradac, D. Clowe, A. Gonzales, et al., \apj{2006}{652}{937}

\bibitem{BBS}
T.G. Brainerd, R.D. Blandford, I. Smail, \apj{1996}{466}{623}

\bibitem{bridle98}
S.L. Bridle, M.P. Hobson, A.N. Lasenby, R. Saunders, \mnras{1998}{299}{895}

\bibitem{bridle2001}
S.L. Bridle, (2001)

\bibitem{brown}
M. Brown, A.N. Taylor, N.C. Hambly, S. Dye, \mnras{2002}{333}{501}

\bibitem{Brown_et_al2003}
M. Brown, A.N. Taylor, D.J. Bacon, M.E. Gray, S. Dye, K. Meisenheimer, C. Wolf, \mnras{2003}{341}{100}

\bibitem{CMH92}
E.D. Carlson, M.E. Machacek, L.J. Hall, \apj{1992}{398}{43}

\bibitem{castro2005}
P.G. Castro, A.F. Heavens, T.D. Kitching, \prd{2005}{72}{3516}

\bibitem{catelan2001}
P. Catelan, M. Kamionkowski, R. Blandford, \mnras{2001}{320}{L7}

\bibitem{ChallinorFord00}
A. Challinor, M. Ford and A. Lasenby, \mnras{2000}{312}{159}, \eprint{astro-ph/9905227}.

\bibitem{Chall2004}
A. Challinor, (2004), \eprint{astro-ph/0403344}.

\bibitem{ChallinorLewis2005}
A. Challinor and A. Lewis, \prd{2005}{71}{103010}, \eprint{astro-ph/0502425}.

\bibitem{cheval01}
M. Chevallier, D. Polarski, \ijmpd{2001}{10}{213}

\bibitem{clowe2006}
D. Clowe, M. Bradac, A. Gonzales, et al., \apj{2006}{648}{L109}

\bibitem{clowe2006b}
D. Clowe, et al., \aap{2006}{451}{395}

\bibitem{ColeEfsth1989}
S. Cole and G. Efstathiou, \mnras{1989}{239}{195}

\bibitem{contaldi2003}
C. Contaldi, H. Hoekstra, A. Lewis, \prl{2003}{90}{1303}

\bibitem{CoorayHu2002}
A. Cooray and W. Hu, \apj{2002}{574}{19}, \eprint{astro-ph/0202411}.

\bibitem{Cooray_Sheth2002}
A. Cooray, R. Sheth, \phr{2002}{372}{1}

\bibitem{Cooray03b}
A.R. Cooray, \apj{2003}{596}{L127}, \eprint{astro-ph/0305515}.

\bibitem{CoorayKesden03}
A. Cooray and M. Kesden, \nas{2003}{8}{231}

\bibitem{CoorayAst2004}
A.R. Cooray, \nas{2004}{9}{173}, \eprint{astro-ph/0309301}.

\bibitem{CoorayBauSigr2004}
A. Cooray, D. Baumann and K. Sigurdson, (2004), \eprint{astro-ph/0410006}

\bibitem{CoorayHH2006}
A. Cooray, D. Huterer, D.E. Holz, \prl{2006}{96}{1301}

\bibitem{hcouchman1999}
H.M.P. Couchman, A.J. Barber, P.A. Thomas, \mnras{1999}{308}{180}

\bibitem{critten2001}
R. Crittenden, et al., \apj{2000}{559}{552}

\bibitem{crittendenetal2002}
R. Crittenden, P. Natarajan, U.-L. Pen, T. Theuns, \apj{2002}{568}{20}

\bibitem{croft2000}
R. Croft, C. Metzler, \apj{2000}{541}{561}

\bibitem{Dalal03}
N. Dalal, D.E. Holz, X. Chen, J.A. Frieman, \apj{2003}{585}{L11}

\bibitem{Dod2003}
S. Dodelson, {\em Modern Cosmology}, Academic Press, (2003), ISBN:0122191412

\bibitem{dodStark03}
S. Dodelson and G.D. Starkman, (2003), \eprint{astro-ph/0305467}.

\bibitem{DodelsonVallinotto06}
S. Dodelson, A. Vallinotto, \prd{2006}{74}{3515}

\bibitem{Dolney_Jain_Takada_2004}
D. Dolney, B. Jain, M. Takada, \mnras{2004}{352}{1019}

\bibitem{dore01}
O. Dore, F. Bouchet, Y. Mellier, R. Teyssier, \aap{2001}{375}{14}

\bibitem{DC91}
J. Dubinski, R. Carlberg, \apj{1991}{378}{496}

\bibitem{dvali2000}
G. Dvali, G. Gabadadze, M. Porrati, \phlb{2000}{485}{208}

\bibitem{Dyson_Eddington_Davidson1920}
F.W. Dyson, A.S. Eddington, C. Davidson, \phil{1920}{220A}{291}

\bibitem{Einstein1915}
A. Einstein, {\em Sitzungber. preuss. Akad. Wiss.}, (1915), 831.

\bibitem{erben01}
T. Erben, et al., \aap{2001}{366}{717}

\bibitem{fahlman94}
G. Fahlman, N. Kaiser, G. Squires, D. Woods, \apj{1994}{437}{56}

\bibitem{Forero-Romero}
J. E. Forero-Romero, J. Blaizot, J. Devriendt, et al., \eprint{astro-ph/0611932}. 

\bibitem{FM94}
B. Fort, Y. Mellier, \aarv{1994}{5}{239}

\bibitem{Frieman1997}
J. Frieman, \coa{1997}{}{}, \eprint{astro-ph/9608068}.

\bibitem{Furlentto04}
S.R. Furlanetto, A. Sokasian, L. Hernquist, \mnras{2004}{347}{187}

\bibitem{GS2006}
R. Gavazzi, G. Soucail, \eprint{astro-ph/0605591}

\bibitem{GS99}
B. Geiger, P. Schneider, \mnras{1999}{302}{118}

\bibitem{Gibli97}
M. Gibilisco, \ass{1997}{249}{189}, \eprint{astro-ph/9701203}.

\bibitem{GAIA}
G. Gilmore, et al., \eprint{astro-ph/9805180}.

\bibitem{Giovi03}
F. Giovi, C. Baccigalupi and F. Perrotta, \prd{2003}{68}{123002}, \eprint{astro-ph/0308118}.

\bibitem{Giovi05}
F. Giovi, C. Baccigalupi and F. Perrotta, \prd{2005}{71}{103009}, \eprint{astro-ph/0411702}.

\bibitem{Gold06}
B. Gold, \prd{2005}{71}{063522}, \eprint{astro-ph/0411376}.

\bibitem{GoldSper1999}
D.M. Goldberg and D.N. Spergel, \prd{1999}{59}{103002}, \eprint{astro-ph/9811251}.

\bibitem{GN02}
D. Goldberg, P. Natarajan, \apj{2002}{564}{65}

\bibitem{GB05}
D. Goldberg, D. Bacon, \apj{2005}{619}{741}

\bibitem{GP95}
A. Goobar, S. Perlmutter, \apj{1995}{450}{14}

\bibitem{Groth77}
E. Groth, P.J.E. Peebles, \apj{1977}{217}{385}

\bibitem{Guzik00}
J. Guzik, U. Seljak and M. Zaldarriaga, \prd{2000}{62}{043517}

\bibitem{GS02}
J. Guzik, U. Seljak, \mnras{2002}{335}{311}

\bibitem{hamana2000}
T. Hamana, H. Martel, T. Futamase, \apj{2000}{529}{56}

\bibitem{Hamana_et_al2002}
T. Hamana, S.T. Colombi, A. Thion, J.E. Devriendt, Y. Mellier, F. Bernardeau, \mnras{2002}{330}{365}

\bibitem{Hamana_et_al2003}
T. Hamana, S. Miyazaki, K. Shimasaku, H. Furusawa, M. Doi, et al., \apj{2003}{597}{98}

\bibitem{Hammerle_et_al2002}
H. H\"{a}mmerle, J.M. Miralles, P. Schneider, T. Erben, R.A. Fosbury, \aap{2002}{385}{743}

\bibitem{Hetal2000}
A.F. Heavens, A. Refregier, C.E.C. Heymans, \mnras{2000}{319}{649}

\bibitem{H03}
A.F. Heavens, \mnras{2003}{343}{1327}

\bibitem{Hetal06}
A.F. Heavens, T.D. Kitching, A.N. Taylor, \mnras{2006}{}{}, in press, \eprint{astro-ph/0606568}.

\bibitem{Hetal2006}
M. Hetterscheidt, et al., \aap{2005}{442}{43}

\bibitem{HH03}
C.E.C. Heymans, A.F. Heavens, \mnras{2003}{339}{711}

\bibitem{heymans2004}
C.E.C. Heymans, M.L. Brown, A.F. Heavens, K. Meisenheimer, A.N. Taylor, C. Wolf, \mnras{2004}{347}{895}

\bibitem{STEP1}
C.E.C. Heymans, et al., \mnras{2006}{368}{1323}

\bibitem{Hetal06b}
C.E.C. Heymans, M. White, A.F. Heavens, C. Vale, L. Van Waerbeke, \mnras{2006}{371}{750}.

\bibitem{CH06}
C.E.C. Heymans, et al., \mnras{2006}{371}{L60}

\bibitem{hicks06}
A.K. Hicks, E. Ellingson, H. Hoekstra, H.K.C. Yee, \apj{2006}{652}{232}

\bibitem{HS03}
C. Hirata, U. Seljak, \mnras{2003}{343}{459}

\bibitem{Hirata03}
C. Hirata, U. Seljak, \prd{2003}{67}{043001}

\bibitem{HS04}
C. Hirata, U. Seljak, \prd{2004}{204}{3056}

\bibitem{Hetal98}
H. Hoekstra, et al., \apj{1998}{504}{636}

\bibitem{hoekstra01}
H. Hoekstra, \aap{2001}{370}{743}

\bibitem{Hoekstra_et_al2002}
H. Hoekstra, H.K.C. Yee, M. Gladders, L. Felipe Barrientos, P.B. Hall, L. Infante, \apj{2002}{572}{55}

\bibitem{Hoekstra_Yee_Gladders2002}
H. Hoekstra, H.K.C. Yee, M. Gladders, \apj{2002}{577}{595}

\bibitem{Hetal2002}
H. Hoekstra, L. Van Waerbeke, M. Gladders, Y. Mellier, H. Yee, \apj{2002}{577}{604}

\bibitem{H2004}
H. Hoekstra, \mnras{2004}{347}{1337}

\bibitem{HYG04}
H. Hoekstra, H.K.C. Yee, M. Gladders, \apj{2004}{606}{67}

\bibitem{HH05}
H. Hoekstra, B.C. Hsieh, H.K.C. Yee, H. Lin, M.D. Gladders, \apj{2005}{635}{73}

\bibitem{hoekstra2006}
H. Hoekstra, et al., \apj{2006}{647}{116}

\bibitem{HuWhite1996}
W. Hu and M.J. White, \aap{1996}{315}{33}, \eprint{astro-ph/9507060}.

\bibitem{hu1999}
W. Hu, \apj{1999}{522}{21}

\bibitem{hu2000}
W. Hu, \prd{2000}{62}{3007}

\bibitem{wHu2000}
W. Hu, \apj{2000}{529}{12}, \eprint{astro-ph/9907103}.

\bibitem{Hu01}
W. Hu, \prd{2001}{64}{083005}, \eprint{astro-ph/0105117}.

\bibitem{huwhite}
W. Hu, M. White, \apj{2001}{554}{67}

\bibitem{Hu2002}
W. Hu, \prd{2002}{65}{023003}, \eprint{astro-ph/0108090}.

\bibitem{HuOkamoto02}
W. Hu, T. Okamoto, \apj{2002}{574}{566}

\bibitem{HuDod2002}
W. Hu, S. Dodelson, \araa{2002}{40}{171}, \eprint{astro-ph/0110414}.

\bibitem{Hu2003}
W. Hu, \annp{2003}{303}{203}, \eprint{astro-ph/0210696}.

\bibitem{HK03}
W. Hu, C. Keeton, \prd{2003}{66}{3506}

\bibitem{HuWhite2004}
W. Hu, M. White, \scam{2004}{290}{44}, http://background.uchicago.edu/whu/Papers/HuWhi04.pdf .

\bibitem{hudson98}
M. Hudson, S. Gwyn, H. Dahle, N. Kaiser, \apj{1998}{503}{531}

\bibitem{HL06}
D. Huterer, E. Linder, (2006), \eprint{astro-ph/0608681}.

\bibitem{huterer2006}
D. Huterer, M. Takada, G. Bernstein, B. Jain, \mnras{2006}{366}{10}

\bibitem{ilbert06}
O. Ilbert, et al., \aap{2006}{457}{841}

\bibitem{Iliev02}
I.T. Iliev, P.R. Shapiro, A. Ferrara, H. Martel, \apjl{2002}{572}{L123}

\bibitem{Iliev03}
I.T. Iliev, E. Scannapieco, H. Martel, P.R. Shapiro, \mnras{2003}{341}{81}

\bibitem{JS97}
B. Jain, U. Seljak, \apj{1997}{484}{560}

\bibitem{Jain_Seljak_White2000}
B. Jain, U. Seljak, S.D.M. White, \apj{2000}{530}{547}

\bibitem{JVW2000}
B. Jain, L. Van Waerbeke, \apj{2000}{530}{L1}

\bibitem{JT03}
B. Jain, A. N. Taylor, \prl{2003}{91}{1302}

\bibitem{JJB06}
B. Jain, M. Jarvis, G. Berstein, \jcap{2006}{02}{001}

\bibitem{Jannink87}
G. Jannink, J. Des Cloiseaux, {\em Les polym\`eres en solution}, Les editions de physique, Les Ulis, France (1987)

\bibitem{jarros1990}
M. Jarosszy\'nsky, C. Park, B.Pacz\'nisky  and  J.R. Gott, \apj{1990}{365}{22}

\bibitem{Jarvis_et_al2002}
M. Jarvis, G.M. Bernstein, P. Fisher, D. Smith, B. Jain, J.A. Tyson, D. Wittman, \apj{2002}{125}{1014}

\bibitem{Jarvis_et_al2004}
M. Jarvis, G. Bernstein, B. Jain, \apj{2004}{338}{}

\bibitem{jing2001}
Y.P. Jing, \mnras{2002}{335}{L89}

\bibitem{Kaiser1992}
N. Kaiser, \apj{1992}{388}{272}

\bibitem{KS93}
N. Kaiser, G. Squires, \apj{1993}{404}{441}

\bibitem{K94}
N. Kaiser, G. Squires, G. Fahlman, D. Woods, T. Broadhurst, {\em Wide Field Spectroscopy and the Distant Universe}, Proceedings of the 35th Herstmonceux Conference, held July 4-8, (1994).

\bibitem{Kaiser95}
N. Kaiser, \apj{1995}{439}{L1}

\bibitem{KSB95}
N. Kaiser, G. Squires, T. Broadhurst, \apj{1995}{449}{460}

\bibitem{KS96}
N. Kaiser, G. Squires, \apj{1996}{473}{65}

\bibitem{Kaiser1998}
N. Kaiser, \apj{1998}{498}{26}

\bibitem{K2000}
N. Kaiser, \apj{2000}{537}{555}

\bibitem{Kaiser_Wilson_Luppino2000}
N. Kaiser, G. Wilson, G.A. Luppino, (2000), \eprint{astro-ph/0003338}

\bibitem{KamionKosowStebb1997}
M. Kamionkowski, A. Kosowsky and A. Stebbins, \prd{1997}{55}{7368}, \eprint{astro-ph/9611125}.

\bibitem{Kassim00}
N.E. Kassim, T.J.W. Lazio, W.C. Erickson, P.C. Crane, R.A. Perley., B. Hicks, (2000). In: Proc. SPIE, {\em Radio Telescopes}, Harvey R. Butcher, Ed., Vol. 4015, p. 328-340.

\bibitem{Kesden02}
M.H. Kesden, A. Cooray and M. Kamionkowski, \prd{2002}{66}{083007}, \eprint{astro-ph/0208325}.

\bibitem{Kilbinger_Schneider2005}
M. Kilbinger, P. Schneider, \aap{2005}{442}{69}

\bibitem{KS02}
L. King, P. Schneider, \aap{2002}{396}{411}

\bibitem{king2005}
L. King, \aap{2005}{441}{47}

\bibitem{kitching2006}
T.D. Kitching, A.F. Heavens, A.N. Taylor, M.L. Brown, K. Meisenheimer, C. Wolf, M.E. Gray, D.J. Bacon, (2006), \eprint{astro-ph/0610284}.

\bibitem{K06}
M. Kleinheinrich, et al., \aap{2006}{455}{441}

\bibitem{KK99}
K. Kuijken, \aap{1999}{352}{355}

\bibitem{KK06}
K. Kuijken, \aap{2006}{456}{827}

\bibitem{lee1990}
M.H. Lee and B. Pacz\'nisky, \apj{1990}{357}{32}

\bibitem{Les06}
J. Lesgourgues, et al., \prd{2006}{73}{045021}, \eprint{astro-ph/0511735}.

\bibitem{LewisKing05}
A. Lewis and L. King, (2005), \eprint{astro-ph/0512104}.

\bibitem{lewischallinor2006}
A. Lewis, A. Challinor, \phr{2006}{429}{1}

\bibitem{liddlelyth2000}
A. Liddle and D. Lyth, {\em Cosmological Inflation And Large-Scale Structure}, Cambridge University Press, (2000)

\bibitem{Limber1954}
D.N. Limber, \apj{1954}{119}{655}

\bibitem{LindeMukhaSasa2005}
A. Linde, V. Mukhanov and M. Sasaki, \jcap{2005}{0510}{002}, \eprint{astro-ph/0509015}.

\bibitem{Linder1990}
E.V. Linder, \mnras{1990}{243}{353}

\bibitem{LK97}
G. Luppino, N. Kaiser, \apj{1997}{475}{20}

\bibitem{Ma00}
C.-P. Ma, J.N. Fry, \apj{2000}{543}{503}

\bibitem{ma2006}
Z. Ma, W. Hu, D. Huterer, \apj{2006}{636}{21}

\bibitem{McClelland}
J. McClelland, J. Silk, \apj{1977}{217}{331}

\bibitem{Madau97}
P. Madau, A. Meiksin, M.J. Rees, \apj{1997}{475}{429}

\bibitem{MandelZal2005}
K.S. Mandel and M. Zaldarriaga, (2005), \eprint{astro-ph/0512218}.

\bibitem{Mandel05}
R. Mandelbaum, et al., \mnras{2005}{361}{1287}

\bibitem{rachel06b}
R. Mandelbaum, et al., \mnras{2006}{370}{1008}

\bibitem{rachel06a}
R. Mandelbaum, et al., \mnras{2006}{372}{758}

\bibitem{Maoli_et_al2001}
R. Maoli, L. van Waerbeke, Y. Mellier, P. Schneider, B. Jain, et al., \aap{2001}{368}{766}

\bibitem{MB04}
L. Marian, G. Bernstein, \prd{2006}{73}{l3525}

\bibitem{marshall}
P.J. Marshall, M.P. Hobson, S.F. Gull, S.L. Bridle, \mnras{2002}{335}{1037}

\bibitem{massey2005}
R. Massey, A. Refregier, D.J. Bacon, R.S. Ellis, M.L. Brown, \mnras{2005}{359}{1277}

\bibitem{MR2005}
R. Massey, A. Refregier, \mnras{2005}{363}{197}

\bibitem{massey2006}
R. Massey, et al., (2006), submitted to Nature.

\bibitem{STEP2}
R. Massey, et al., (2006), \eprint{astro-ph/0608643}.

\bibitem{MRRBB2006}
R. Massey, B. Rowe, A. Refregier, D.J. Bacon, J. Berge, (2006), \eprint{astro-ph/0609795}.

\bibitem{Mellier1999}
Y. Mellier, \araa{1999}{37}{127}

\bibitem{Mellier04}
Y. Mellier, G. Meylan, (eds), {\em Impact of Gravitational Lensing on Cosmology}, IAU S225, Cambridge university press, (2004)

\bibitem{MenardDalal05}
B. M\'enard, N. Dalal, \mnras{2005}{358}{101}

\bibitem{MS1999}
R.B. Metcalf and J. Silk, \apj{1997}{489}{1}, \eprint{astro-ph/9708059}.

\bibitem{Metcalf99}
B. Metcalf, \mnras{1999}{305}{746}

\bibitem{Metcalf01}
B. Metcalf, \mnras{2001}{327}{115}

\bibitem{milgrom}
M. Milgrom, \apj{1983}{270}{365}

\bibitem{Miralda-Escude1991}
J. Miralda-Escude, \apj{1991}{380}{1}

\bibitem{miya2002}
S. Miyazaki, et al., \apj{2002}{580}{L97}

\bibitem{Mo96}
H.J. Mo, S.D.M. White, \mnras{1996}{282}{347}

\bibitem{MukhaFeldBrand1992}
V. Mukhanov, H. Feldman and R. Brandenberger, \phr{1992}{215}{203}

\bibitem{Munshi_Valageas_Barber2004}
D. Munshi, P. Valageas, A.J. Barber, \mnras{2004}{350}{77}

\bibitem{munshivalageas2005a}
D. Munshi, P. Valageas, \mnras{2005}{356}{439}

\bibitem{munshivalageas2005b}
D. Munshi, P. Valageas, \mnras{2005}{360}{1401}

\bibitem{narayan1991}
R. Narayan and R.D. Blandford, \anyas{1991}{647}{}

\bibitem{NK97}
P. Natarajan, J.P. Kneib, \mnras{1997}{287}{833}

\bibitem{NFW97}
J.F. Navarro, C.S. Frenk, S.D.M. White, \apj{1997}{490}{493}

\bibitem{OhMack2003}
S.P. Oh, K.J. Mack, \mnras{2003}{346}{871}

\bibitem{OkaHu03}
T. Okamoto, W. Hu, \prd{2003}{67}{083002}.

\bibitem{Peacock_Dodds1996}
J.A. Peacock, S.J. Dodds, \mnras{1996}{280}{L19}

\bibitem{ESAESO}
J.A. Peacock, et al., (2006), \eprint{astro-ph/0610906}.

\bibitem{Peebles70}
P.J.E. Peebles and J.T. Yu, \apj{1970}{162}{815}

\bibitem{peebles1980}
P.J.E. Peebles, {\em The Large-Scale Structure of the Universe}, Princeton University Press, Princeton (1980).

\bibitem{Pen_et_al2003b}
U.-L. Pen, T. Zhang, L. Van Waerbeke, Y. Mellier, P. Zhang, J. Dubinski, \apj{2003}{592}{664}

\bibitem{LiPen04}
U.-L. Pen, \nas{2004}{9}{417}

\bibitem{P99}
S. Perlmutter, et al., \apj{1999}{517}{565}

\bibitem{Press74}
W. Press, P. Schechter, \apj{1974}{187}{425}

\bibitem{PB06}
E. Puchwein, M. Bartelmann, \aap{2006}{455}{791}

\bibitem{Refregier_Rhodes_Groth2002}
A. Refregier, J. Rhodes, E. Groth, \apj{2002}{572}{L131}

\bibitem{Refregier2003}
A. Refregier, \araa{2003}{41}{645}

\bibitem{AR2003}
A. Refregier, \mnras{2003}{338}{35}

\bibitem{RB2003}
A. Refregier, D. Bacon, \mnras{2003}{338}{48}

\bibitem{Rhodes_Refregier_Groth2001}
J. Rhodes, A. Refregier, E. Groth, \apj{2001}{552}{L85}

\bibitem{RPK}
A.G. Riess, W.H. Press, R.P. Kirshner, \apj{1996}{473}{88}

\bibitem{R98}
A.G. Riess, et al., \aj{1998}{116}{1009}

\bibitem{Sachs1961}
R. Sachs, \prs{1961}{A264}{309}

\bibitem{Schaeffer84}
R. Schaeffer, \aap{1984}{134}{L15}

\bibitem{Scherrer}
R.J. Scherrer, E. Bertschinger, \apj{1991}{381}{349}

\bibitem{sch98}
J. Schmalzing, K.M. Gorski, \mnras{1998}{297}{355}

\bibitem{Schmid05}
C. Schmid, J.-P. Uzan, A. Riazuelo, (2005), \eprint{astro-ph/0412120}.

\bibitem{Schmid06}
C. Schmid, I. Tereno, J.-P. Uzan, et al., \eprint{astro-ph/0603158}.

\bibitem{weiss1988}
P. Schneider and A. Weiss, \apj{1988}{330}{1}

\bibitem{Schneider_et_al1992}
P. Schneider, J. Ehlers, E.E. Falco, {\em Gravitational Lenses}, Springer-Verlag, Berlin, (1992)

\bibitem{schneiderseitz1995}
P. Schneider, S. Seitz, \aap{1995}{294}{411}

\bibitem{Schneider1996}
P. Schneider P., \mnras{1996}{283}{837}

\bibitem{SR97}
P. Schneider, H.W. Rix, \apj{1997}{474}{25}

\bibitem{schneider98}
P. Schneider, \apj{1998}{498}{43}

\bibitem{schneideretal1998}
P. Schneider, L. Van Waerbeke, B. Jain, G. Kruse, \mnras{1998}{296}{873}

\bibitem{Schneider99}
P. Schneider, {\em Perspectives on Radio Astronomy, Scientific Imperatives at cm and m wavelengths}, Proceedings of a workshop in Amsterdam, April 7-9, (1999).

\bibitem{SEF}
P. Schneider, J. Ehlers, E. Falcon, {\em Gravitational Lenses}, Springer (1999)

\bibitem{schneider2002a}
P. Schneider, L. Van Waerbeke, Y. Mellier, \aap{2002}{389}{729}

\bibitem{schneider2002b}
P. Schneider, L. Van Waerbeke, M. Kilbinger, Y. Mellier, \aap{2002}{396}{1}

\bibitem{Schneider2003}
P. Schneider, \aap{2003}{408}{829}

\bibitem{Schneider_Lombardi2003}
P. Schneider, M. Lombardi, \aap{2003}{397}{809}

\bibitem{Schneider2005}
P. Schneider, {\em Gravitational Lensing: Strong, Weak and Micro}, Lecture Notes of the 33rd Saas-Fee Advanced Course, G. Meylan, P. Jetzer \& P. North (eds.), Springer-Verlag: Berlin, (2006), p.273, \eprint{astro-ph/0509252}

\bibitem{Schneider_et_al2005}
P. Schneider, M. Kilbinger, M. Lombardi, \aap{2005}{431}{9}

\bibitem{sem06}
E. Semboloni, et al., \aap{2006}{452}{51}

\bibitem{Scoccimarro01}
R. Scoccimarro, R.K. Sheth, L. Hui, B. Jain, \apj{2001}{546}{652}

\bibitem{ScottRees90}
D. Scott and M. J. Rees, \mnras{1990}{247}{510}

\bibitem{DS2006}
D. Scott, G. Smoot, 2006, \eprint{astro-ph/0601307}.

\bibitem{scrantonetal2005}
R. Scranton, et al., \apj{2005}{633}{589}

\bibitem{SVJ}
C. Sealfon, L. Verde, R. Jimenez, \apj{2006}{649}{118}

\bibitem{Seitz_et_al1994}
S. Seitz, P. Schneider, J. Ehlers, \cqg{1994}{11}{2345}

\bibitem{SSB98}
S. Seitz, P. Schneider, M. Bartelmann, \aap{1998}{337}{325}

\bibitem{Seljak1996}
U. Seljak, \apj{1996}{463}{1}, \eprint{astro-ph/9505109}.

\bibitem{SeljakZal1997}
U. Seljak and M. Zaldarriaga, \prl{1997}{78}{2054}, \eprint{astro-ph/9609169}.

\bibitem{SelkakZald1999}
U. Seljak and M. Zaldarriaga, \prd{1999}{60}{043504}, \eprint{astro-ph/9811123}.

\bibitem{Seljak00}
U. Seljak, \mnras{2000}{318}{203}

\bibitem{Semboloni07}
E. Semboloni, L. Van Waerbeke, C. Heymans, et al., \eprint{astro-ph/0606648}.

\bibitem{sheldon04}
E. Sheldon, et al., \aj{2004}{127}{2544}

\bibitem{Sheth97}
R.K. Sheth, B. Jain, \mnras{1997}{285}{231}

\bibitem{Sheth99}
R.K. Sheth, G. Tormen, \mnras{1999}{308}{119}

\bibitem{Shimon06}
M. Shimon et al., (2006), \eprint{astro-ph/0602528}

\bibitem{Silk68}
J. Silk, \apj{1968}{151}{459}

\bibitem{simon2006}
P. Simon, M. Hetterscheidt, M. Schirmer, T. Erben, P. Schneider, C. Wolf, K. Meisenheimer, \aap{2006}{}{}, in press.

\bibitem{smithetal2001}
D.R. Smith, G.M. Bernstein, P. Fischer, M. Jarvis, \apj{2001}{551}{643}

\bibitem{Smith_et_al2003}
R.E. Smith, J.A. Peacock, A. Jenkins, S.D.M. White, C.S. Frenk, et al., \mnras{2003}{341}{1311}

\bibitem{spergel06}
D. Spergel, et al., (2006), \eprint{astro-ph/0603449}.

\bibitem{starck}
J.-L. Starck, S. Pires, A. Refregier, \aap{2006}{451}{1139}

\bibitem{SteinTurok2002}
P.J. Steinhardt and N. Turok, \prd{2002}{65}{126003}, \eprint{hep-ph/0111098}.

\bibitem{Stompor99}
R. Stompor and G. Efstathiou, \mnras{1999}{302}{735}, \eprint{astro-ph/9805294}.

\bibitem{SunZel70}
R.A. Sunyaev, Ya.B. Zel'dovich, \ass{1970}{7}{3}

\bibitem{SunZel1972}
R. Sunyaev and Y. Zel'dovich, \casp{1972}{4}{173}

\bibitem{Takada_Jain2002}
M. Takada, B. Jain, \mnras{2002}{337}{875}

\bibitem{Takada_Jain2003a}
M. Takada, B. Jain, \mnras{2003}{340}{580}

\bibitem{Takada_Jain2003b}
M. Takada, B. Jain, \mnras{2003}{344}{857}

\bibitem{Takada_Jain2004}
M. Takada, B. Jain, \mnras{2004}{348}{897}

\bibitem{TW04}
M. Takada, M. White, \apj{2004}{601}{L1}

\bibitem{Taruya_et_al2002}
A. Taruya, M. Takada, T. Hamana, I. Kayo, T. Futamase, \apj{2002}{571}{638}

\bibitem{Taruya03}
A. Taruya, T. Hamana, I. Kayo, \mnras{2003}{339}{495}

\bibitem{taylor2001}
A.N. Taylor, (2001), \eprint{astro-ph/0111605}.

\bibitem{tayloretal2006}
A.N. Taylor, T.D. Kitching, D.J. Bacon, A.F. Heavens, \mnras{2006}{}{} in press, \eprint{astro-ph/0606416}.

\bibitem{tereno2004}
I. Tereno, O. Dore, L. Van Waerbeke, Y. Mellier, \aap{2005}{429}{383}

\bibitem{Tozzi00}
P. Tozzi, P. Madau, A. Meiskin and M. J. Rees, \apj{2000}{528}{597}

\bibitem{turner1984}
E.L. Turner, J.P. Ostriker and J,R. Gott, \apj{1984}{284}{1}

\bibitem{tyson84}
J. Tyson, F. Valdes, J. Jarvis, A. Mills, \apj{1984}{281}{L59}

\bibitem{UB01}
J.P. Uzan, F. Bernardeau, \prd{2001}{64}{3004}

\bibitem{Valageas2000}
P. Valageas, \aap{2000}{354}{767}

\bibitem{Valageas2000b}
P. Valageas, \aap{2000}{356}{771}

\bibitem{Valageas01}
P. Valageas, A. Balbi and J. Silk, \aap{2001}{367}{1}, \eprint{astro-ph/0009040}.

\bibitem{Valageas2004}
P. Valageas, A.J. Barber, D. Munshi, \mnras{2004}{347}{654}

\bibitem{Valageas_Munshi_Barber2005}
P. Valageas, D. Munshi, A. Barber, \mnras{2005}{356}{386}

\bibitem{vw98}
L. Van Waerbeke, \aap{1998}{334}{1}

\bibitem{vw1999}
L. Van Waerbeke, F. Bernardeau, Y. Mellier, \aap{1999}{342}{15}

\bibitem{vw2000}
L. Van Waerbeke, \mnras{2000}{313}{524}

\bibitem{vwetal2000}
L. Van Waerbeke, Y. Mellier, T. Erben, J.C. Cuillandre, et al., \aap{2000}{358}{30}

\bibitem{vw2001}
L. Van Waerbeke, T. Hamana, R. Scoccimarro, S. Colombi, F. Bernardeau, \mnras{2001}{322}{918}

\bibitem{vanWaerbeke_et_al2001a}
L. van Waerbeke, Y. Mellier, M. Radovich, E. Bertin, M. Dantel-Fort, et al., \aap{2001}{374}{757}

\bibitem{vanWaerbeke_et_al2002}
L. van Waerbeke, Y. Mellier, R. Pell\'{o}, U.-L. Pen, H.J. McCracken, B. Jain, \aap{2002}{393}{369}

\bibitem{LVWM2003}
L. van Waerbeke, Y. Mellier, (2003), \eprint{astro-ph/0305089}.

\bibitem{vwetal06}
L. Van Waerbeke, M. White, H. Hoekstra, C. Heymans, \apj{2006}{26}{91}

\bibitem{verde02}
L. Verde, A.F. Heavens, W.J. Percival, \mnras{2002}{335}{432}

\bibitem{wambsganss1995}
J. Wambsganss, R. Cen, J.P. Ostriker, E.L. Turner, \sci{1995}{268}{274}

\bibitem{wambsganss1997}
J. Wambsganss, R. Cen, G. Xu, J.P. Ostriker, \apjl{1997}{475}{81}

\bibitem{wambsganss1998}
J. Wambsganss, R. Cen, J.P. Ostriker, \apj{1998}{494}{29}

\bibitem{Wang_et_al2002}
Y. Wang, D. Holz, D. Munshi, \apj{2002}{572}{15}

\bibitem{wang2005}
Y. Wang, (2005), \eprint{astro-ph/0406635}.

\bibitem{white02}
M. White, L. Van Waerbeke, J. Mackey, \apj{2002}{575}{640}

\bibitem{white04}
M. White, \apj{2004}{22}{211}

\bibitem{Wittman_et_al2000}
D.M. Wittman, J. Tyson, D. Kirkman, I. Dell'Antonio, G. Bernstein, \nat{2000}{405}{143}

\bibitem{ZaldarriagaSeljak99}
M. Zaldarriaga, U. Seljak, \prd{1999}{59}{123507}

\bibitem{Zal2000}
M. Zaldarriaga, \prd{2000}{62}{063510}, \eprint{astro-ph/9910498}.

\bibitem{Zaldarriaga_Scoccimarro2003}
M. Zaldarriaga, R. Scoccimarro, \apj{2003}{584}{559}

\bibitem{zaroubi98}
S. Zaroubi, G. Squires, Y. Hoffman, J. Silk, \apj{1998}{500}{L87}

\bibitem{ZK04}
H. Zhan, L. Knox, \apj{2004}{616}{L75}

\bibitem{ZHS05}
J. Zhang, L. Hui, A. Stebbins, \apj{2005}{635}{806}

\bibitem{ZP06}
P. Zhang, U.L. Pen, \prl{2005}{95}{241302}

\bibitem{AZ06}
A. Zhitnitsky, \prd{2006}{74}{043515}

\end{thebibliography}
\end{document}